\raggedbottom
% TeX MACRO file containing standard abreviations and useful macros
% plus commented out changes to the default parameters enabling different
% size pages and other type setting requirements to be altered
%
% *** OVERRIDE DEFAULT PAGE SIZE AND TYPE-SETTING PARAMETERS  ***
% double spaced lines
% \baselineskip=20pt
% do not print page numbers
%\nopagenumbers
% the extra vertical space between paragraphs and the size of the paragraph
% indentation
%\parskip 0pt
%\parindent 1pc
% pagesize and offset of the page centre from its normal position
%\topskip=0pt
%\hoffset=2.0cm
%\hsize 15.2cm
%\voffset=1.1cm
%\vsize 21.6cm
%
%  ** FONT DEFINITIONS **
%  (see tex$:fonts.lis and tex$pixel directory for all availabe fonts)
% giant sive

% huge size

% big size

% standard normal size
%\font\rm=cmr10     scaled \magstep0
%\font\bf=cmbx10    scaled \magstep0
%\font\it=cmti10    scaled \magstep0
% small  size

% funny fonts

%
% **  Standard astronomical symbol abbreviations **

\def \h-{{\rm h}^{-1}}

\def \w{w(\theta)}
\def \etal{{\it et al.\/} } 
\def \rms{ {\it rms } }
\def \hmpc{{\rm h}^{-1} {\rm Mpc}}

\def \bj{{\rm b_J}}

%
% ** with long names in TeX or none at all**

\def\bk{\break}
% produces <~ or >~ signs
\def\spose#1{\hbox to 0pt{#1\hss}}
\def\simlt{\mathrel{\spose{\lower 3pt\hbox{$\mathchar"218$}}
     \raise 2.0pt\hbox{$\mathchar"13C$}}}
\def\simgt{\mathrel{\spose{\lower 3pt\hbox{$\mathchar"218$}}
     \raise 2.0pt\hbox{$\mathchar"13E$}}}
% 
%
% *** REFERENCES ***
% macro to deal with long references

% journal abreviations for use in MNRAS

%
%
% *** CORRECTION OF AMERICAN SPELLINGS OF TEX COMMANDS ***

\def\ref{\parskip=0pt\par\noindent\hangindent\parindent
    \parskip =2ex plus .5ex minus .1ex}
%
%        \font\first=cmr10  scaled\magstep1
        
%\first

%--------------------------------------
%% set up page style
%%\nopagenumbers

\magnification = \magstep1

\vsize = 9.5truein
\hsize = 6.6 truein

\baselineskip 15pt
\parskip 7pt

%% some fonts
\font\bigb=cmbx10 scaled\magstep1

%% some useful definitions 

\def\hmpc{ \, {\rm h}^{-1} {\rm Mpc}}

\def\br{{\bf r}}

\def\bk{{\bf k}}

\def\bj{{\rm b_J}}

%
% produces <~ or >~ signs
\def\spose#1{\hbox to 0pt{#1\hss}}
\def\simlt{\mathrel{\spose{\lower 3pt\hbox{$\mathchar"218$}}
     \raise 2.0pt\hbox{$\mathchar"13C$}}}
\def\simgt{\mathrel{\spose{\lower 3pt\hbox{$\mathchar"218$}}
     \raise 2.0pt\hbox{$\mathchar"13E$}}}
% Journal abbreviations
%

\newcount\sectno
\newcount\chapno
\def\brk{\break}

%--------------------------------------

\hfuzz 50pt 

\centerline{ \bigb The APM Galaxy Survey III: An Analysis of}
\centerline {\bigb  Systematic Errors in the Angular Correlation Function}
\centerline{\bigb and Cosmological Implications}

\vskip 0.2in
{\leftskip 0.7in \noindent
{\raggedright
{\bf S.J. Maddox\footnote{$^1$}{\rm 
Present address: The Royal Greenwich Observatory, 
Madingley Road, Cambridge, CB3 0EZ, UK}, 
G. Efstathiou, 
and W.J. Sutherland}\brk
Department of Physics, \brk
Nuclear and Astrophysics Laboratory, \brk
Keble Road, Oxford, \brk
OX1 3RH, England \brk}
\par}

\vskip 0.2in

\centerline{ABSTRACT}

{ %\second
\narrower\smallskip\noindent
\baselineskip 10pt
We present measurements of the angular two-point galaxy correlation
function, $\w$, from the APM Galaxy Survey.  The performance of
various estimators of $\w$ is assessed by analyzing simulated galaxy
catalogues. We use these tests, and analytic arguments, to select
estimators which are least
affected by large-scale gradients in the galaxy counts correlated with
the survey boundaries.  An error analysis of the plate matching procedure
in the APM Galaxy Survey shows that residual plate-to-plate errors do
not  bias our estimates of $\w$ by more than $\sim 1 \times 10^{-3}$. 
A direct comparison
between our photometry and external CCD photometry of over $13,000$
galaxies from the Las Campanas Deep Redshift Survey shows that the
{\it rms} error in the APM plate zero points lies in the range
$0.04$--$0.05$ magnitudes, in agreement with our previous estimates.
The comparison with the CCD photometry sets tight limits on any
variation of the magnitude scale with right ascension. We find no
evidence for any systematic errors in the survey correlated with the
date of scanning and exposure.  We estimate the effects on $\w$ of 
atmospheric extinction and obscuration by dust in our Galaxy and conclude
that these are negligible.  There is no evidence for any correlations
between the errors in the survey and limiting magnitude except
at the faintest magnitudes of the survey, $b_J > 20$, where 
star-galaxy classification begins to break down introducing
plate-to-plate variations in the completeness of the survey.  We 
use our best estimates of the systematic errors in the survey
to calculate corrected  estimates of $\w$. Deep redshift surveys are used
to determine the selection function of the APM Galaxy Survey, {\it
i.e.} the probability that a galaxy at redshift $z$ is included in the
sample at a given magnitude limit. The selection function is applied
in Limber's equation to compute how $\w$ scales as a function of
limiting magnitude. Our estimates of $\w$ are in excellent agreement
with the scaling relation, providing further evidence that systematic
errors in the APM survey are small.  We explicitly remove large-scale
structure by applying filters to the APM galaxy maps and conclude that
there is still strong evidence for more clustering at large scales
than predicted by the standard scale-invariant cold dark matter 
(CDM) model. We compare the APM $\w$ and the three dimensional power 
spectrum derived
by inverting $\w$, with the predictions of scale-invariant CDM models.
We show that the observations require $\Gamma = \Omega_0 h$ in the
range $0.2$--$0.3$ and are incompatible with the value $\Gamma=0.5$
of the standard CDM model. \smallskip}

\vfill\eject

\noindent
{\bf 1. Introduction} 

\vskip 0.15 truein

The APM Galaxy Survey is a machine measured
survey of over $2$ million galaxies in the southern
sky to a magnitude limit $\bj = 20.5$. It was constructed by
scanning $185$ UK Schmidt telescope (UKST) survey plates,
covering $4300$ square degrees, with the Automatic Plate
Measuring (APM) machine at Cambridge. The construction
of the survey and the star-galaxy separation algorithm
have been described in Paper I of this series (Maddox \etal
1990a). The photometric accuracy and plate matching algorithm
have been described in Paper II (Maddox \etal 1990b).

One of the main aims in constructing the APM Galaxy Survey
has been to determine the
two-point galaxy correlation function $\xi ( r)$, or equivalently
the power spectrum $P(k)$ (see Baugh and Efstathiou 1993, 1994), 
at large spatial scales. 
The two-point spatial correlation function is the Fourier 
transform of the power spectrum 
$$ \eqalignno{
\xi(r) &= {V \over (2 \pi)^3} \int 
P(k)\;
e^{-i\bk.\br} \, d^3k,  &(1)}
$$
where for a point process of mean density $\overline n$,
$$
\eqalignno{
P(k) &=  \langle \vert \delta_{\bk} \vert^2 \rangle
-  {1 \over \overline n V}, &(2)}
$$
and  $\delta_{\bk}$ is the Fourier transform 
$$ \eqalignno{
\delta_{\bk} &= {1 \over \overline n V}
\sum_i 
e^{-i\bk.\br_i}, &(3)}
$$
where the sum in equation (3) extends over all points located
at positions ${\bf r}_i$ (see Peebles 1980 \S 41). 
In equations (1)-(3) we assume
that the distribution is periodic in a large
box of volume $V$. For a Gaussian random field, 
phases of different Fourier
modes are uncorrelated ({\it ie} $\left<\delta_{\bk} \delta_{\bk^\prime}
\right> = P({\b k}) \delta({\b k} - {\b k^\prime})$) and so
the two-point correlation function provides a
{\it complete} statistical description of the density field. 
The two-point galaxy correlation function is, therefore, an especially
important measure of large-scale structure, particularly if the
galaxy distribution can be described by a Gaussian random field,
as expected in many models of the origin of structure ({\it e.g.}
theories of inflation in the early universe, 
Kolb and Turner 1990). Even if the
fluctuations are non-Gaussian, the two-point correlation function
can provide an important test of theoretical models and is the 
most easily measured of a hierarchy of $N-$point correlation functions
(see Peebles 1980 \S34-35).

The three dimensional two-point function, $\xi({r}),$ can be inferred
from the angular two-point function, $w(\theta)$, via Limber's
formula (Limber 1954, equation 31 below).
Analyses of several samples of galaxies ({\it e.g.} Peebles 1974,
see also Section 4.1) have shown
that at small angles the angular correlation function is a power
law,
$$\eqalignno{
w(\theta) & = A \theta^{1-\gamma}, & (4)\cr}$$ 
where $\gamma \approx 1.7,$ and $A$ depends on the sample depth.  In an
earlier paper, we presented measurements of $w(\theta)$ from the APM
Galaxy Survey (Maddox \etal 1990c) which show  that $w(\theta)$
 steepens and lies below the 
extrapolation of the power law (4) at large angles.
At a magnitude limit of $b_J = 20$, the break occurs at an
angle $\theta \approx 2^\circ$, corresponding to a physical
scale of $15 \hmpc$\footnote*{Throughout this paper  $h$
denotes  the Hubble constant
$H_0$ in units of $100{\rm km}{\rm s}^{-1}{\rm Mpc}^{-1}$;
$\Omega_0$ is the mean mean mass density of the universe
at the present epoch divided by the critical 
density of the Einstein de-Sitter model.}
at the median depth of the survey. However, even 
though we observe a break in $w(\theta)$, our results
provide strong evidence for more clustering at large
scales than predicted by the standard $\Omega_0=1$ 
scale-invariant cold dark matter model (see Efstathiou 1995a
for a recent review and Section 5 of this paper).

Measurements of clustering on scales $\simgt 10
\hmpc$ are
extremely difficult because  $w(\theta)$ is
small and hence sensitive to non-uniformities in the galaxy
catalogue. For example, the results from the APM Galaxy Survey disagree
with Groth and Peebles (1977; hereafter GP77) determination of $w(\theta)$
from the Lick sample (Shane and Wirtanen 1967, Seldner \etal 1977), which is
the largest galaxy catalogue comparable to the APM Galaxy Survey.  GP77
found a much sharper break from the power law on scales $\theta \simgt
3^\circ $ (see Section 4.2.3).  The APM 
angular correlation function is in better 
agreement with that measured from the Edinburgh-Durham
Southern Galaxy Catalogue, a machine measured sample similar
to the APM Survey but covering about one third of the area of sky
(Collins \etal 1992, Nichol \& Collins 1993, see Section 4.2.2).
It is unfortunate that the results from the machine 
measured catalogues and the Lick survey disagree, for it implies
that one or more samples  are affected by systematic errors.
The uniformity of the Lick catalogue has been a subject of some
controversy (see {\it e.g.} Geller \etal 1984, de Lapparent \etal 1986,
Groth and Peebles 1986a,b), but because it was
constructed by eye, and the information is extremely limited 
(consisting of counts of galaxies in $10^\prime \times 10^\prime$
cells),  only a restricted
set of checks can be applied to test its uniformity.
In contrast, 
the APM Galaxy Survey was  designed specifically to measure $\w$ at large
scales and has several important advantages over the Lick survey
that allow more stringent controls of systematic errors:

\item{(a)} The APM Galaxy Survey gives positions, magnitudes and 
profile information for each image.  The photometry on neighbouring
plates in the APM Survey is matched by identifying individual
galaxies in the plate overlap regions. In contrast, plates in the Lick
survey are matched by comparing the galaxy counts on overlapping
plates, which are strongly affected by galaxy clustering. The plate
matching errors in the APM Galaxy Survey are consequently smaller than
those for the Lick survey.

\item{(b)} The magnitudes of individual galaxies in the APM Survey
can be compared with CCD magnitudes to provide checks of the
photometric accuracy.

\item{(c)}
The APM Survey covers a large area of sky and contains
over $2$ million galaxies.
It can therefore  be divided into subsamples
to provide empirical estimates of sampling fluctuations and
systematic errors. Although the Lick catalogue contains 1 
million galaxies, only about half of these are at 
sufficiently high Galactic latitude to be useful in studies
of the galaxy distribution.

\item{(d)}
The magnitude data in the APM catalogue allow us
to apply the scaling test derived from Limber's equation
({\it e.g.} GP77) to check whether the angular correlation function
scales with limiting magnitude as  expected if $w(\theta)$ is
measuring real structure in the
galaxy distribution.

In addition, the APM Survey contains other data such as surface brightness
and position angles, which enable a variety of other analyses
to be carried out, {\it e.g.} analyses of galaxy ellipticities,
large-scale alignments {\it etc} (Lambas, Maddox and Loveday 1992).
Higher order correlations from the APM Survey are discussed by
Gastanaga \& Frieman (1994) and Szapudi \etal (1995). Catalogues
of rich clusters of galaxies selected from the APM Survey, and
the spatial distribution of APM clusters, are described in
Papers IV and V of this series (Dalton \etal 1994a, 1995) and
by Dalton \etal (1992, 1994b).

In this paper, we present measurements 
of $w(\theta)$ from the APM Galaxy Survey and an
analysis of systematic errors and how these affect $w(\theta)$. 
In Section 2 and the Appendix, we discuss various 
estimators for $\w$ which we have tested on simulated galaxy 
distributions. We then present measurements of $\w$ from the APM Survey.
Section 3 gives a detailed discussion of systematic errors in
the survey. We compare our magnitudes to independent CCD photometry 
and address the points raised by Fong \etal (1992)
concerning large scale gradients in the APM survey.
We also present estimates of the angular correlation
functions after correcting for systematic errors.
In Section 4 we use deep redshift surveys to determine the distribution
in redshift of APM galaxies as a function of
limiting magnitude and these are used in Limber's equation
to test the scaling of $\w$ with depth.  We discuss briefly how our
results compare with those from other catalogues and
we investigate the effects of  filtering 
our galaxy maps by subtracting smooth fits to the galaxy distribution.
Some cosmological implications of our results are discussed
in Section 5. Our main conclusions are summarized in Section 6.

\vfill\eject

\noindent
{\bf  2 Measurements of $\w$ } 

\vskip 0.1 truein

\noindent
{\it 2.1 Estimators of the correlation function.}

Several methods of estimating the angular correlation function from
galaxy surveys have been discussed in the literature, and in Appendix
A we describe tests of various estimators on simulated galaxy 
catalogues.
The results of these tests led us to choose two estimators to measure
$w(\theta)$ for the APM Galaxy Survey.

The first of these is defined by the equation
$$
\eqalignno{
\w &= { \left<N_i N_j\right> \over \left<N_i\right> \left<N_j\right> } - 1,
&(5) \cr}
$$
where $N_i$ is the count of galaxies in cell $i$ and the angle brackets
denote averages over pairs of cells at separation $\theta$.  
As explained in Appendix A, this estimator is unaffected by
first-order errors caused by density variations correlated with the
survey boundaries.
In Appendix A we define an estimator of $\w$ based of the Fourier
transform of the two-dimensional power-spectrum of the survey which
has similar statistical properties to the estimator (5) (see Appendix
A, Section A3).
However, although the Fourier transform estimator offers a
considerable time saving over equation (5), it does not provide
separate inter- and intra- plate estimates of $\w$, which are useful
as a check on systematic errors in the survey ({\it cf} Groth and
Peebles 1986a and Section 3.1.2).
We have therefore used equation (5) to calculate the large-angle
correlation functions presented in this Section.

The second estimator that we have used is based on a
direct count of galaxy pairs. 
For this estimator, we count pairs of galaxies on each field taking into
account edges and holes in the survey (see Plate 2 of Paper II) by
cross correlating the galaxy distribution with a catalogue of random
points within the same field boundaries.
The estimate of $\w$ is defined by 
$$ \eqalignno{
 \w &= 2F {DD \over DR} - 1,  & (6) }$$
where $DD$ is the number of distinct galaxy-galaxy 
pairs with separations
in the range $\theta \pm \delta \theta$,  $DR$ is the number
of galaxy-random pairs, and the density of random
points is set to $F$ times the number of galaxies. 
The estimator of equation (6) is easy to evaluate and does not require
gridding of the data.
It is, however, subject to first order errors in the galaxy density
contrast (Hamilton, 1993a; Appendix A), but this is not a problem
because we use it to estimate $\w$ on small angular scales where the
amplitude is high, and we average over a large number of fields so that
first order errors cancel.
In fact, over the range $0.09 ^\circ < \theta < 0.9^\circ$ the
differences between the $\w$ estimates from equations (5) and (6) are
found to be negligible.
Generalizations of equation (6) that have similar statistical
properties to equation (5) are discussed by Hewett (1982), Hamilton
(1993a) and Landy and Szalay (1993).

We have applied equation (5) to estimate $\w$ over the angular range
$0.12^\circ < \theta < 30^\circ $ from several equal area maps of the
galaxy distribution with cell sizes varying from $0.12 ^\circ \times
0.12^\circ$ to $0.9 ^\circ\times 0.9 ^\circ$. 
On smaller angular scales, $0.09 ^\circ < \theta < 0.9 ^\circ $, we
apply equation (5) to cell counts on each Schmidt field.
For very small separations, $ 0.0009^\circ < \theta < 0.9 ^\circ $ we
estimate $\w$ from equation (6).

As discussed in the following subsection and in Appendix A, $\w$ is biased
low if the mean density for each Schmidt field is estimated from the
number of galaxies within it.  This bias can be eliminated by using all
fields to estimate the mean density, even though the pair counts are
computed separately for each field.  On scales $0.1 ^\circ < \theta < 0.5
^\circ$, the mean of these single plate estimates agrees well with $\w$
measured from the full map (see Section 2.3).

We apply a correction to each estimate of $\w$ to account for the
residual contamination of the galaxy sample by non-galaxy images.
Since the large majority of contaminating images are stars or
star-star mergers, it is a good approximation to assume that they are
uncorrelated. 
There is a $\sim 10\%$ difference between the density of contaminating
objects at the Galactic pole compared to low Galactic latitude ($
\vert b \vert \simlt 40^\circ$), but this gradient has very little
effect on $\w$ (see Section 3.4.3). 
We thus simply multiply the $\w$ estimates by $1/(1-f)^2$ where $f$ is
the contamination fraction estimated from the visual classifications
in Table 5 of Paper~I.

As discussed in Paper I, the APM Galaxy Survey has a high level
of completeness.  
However, most of the galaxies missing from our survey will be at the
extremes of high and low surface brightness, and so incompleteness may
lead to a small error in $\w$ if these galaxies are clustered
differently to normal galaxies.
Variations in the completeness from plate-to-plate will cause an error
in $\w$ similar to magnitude matching errors: this small effect is
discussed in Section~3.1.1.

\vskip 0.15 truein

\noindent
{\it 2.2 Integral constraints and single plate measurements of
$w(\theta)$}

Figure $1$ shows the average $w(\theta)$ estimated from equation (6)
using galaxies in the magnitude range $17 \le \bj \le 20.0 $ within
the central $4.64 ^\circ \times 4.64^\circ$ area on each of $185$
Schmidt plates.
In calculating $DD$ in equation (6) we count galaxy pairs only if both
galaxies are on the same plate.
The filled circles in Figure 1 show estimates of $\w$ in which the
mean surface density in each single plate estimate is determined from
the the number of galaxies within each field ({\it i.e.} rather than
using a global estimate of the mean density derived from the entire
survey).
This method of estimating $\w$ is therefore similar to that applied by
other investigators to single Schmidt fields ({\it e.g.} Shanks \etal
1980, Hewett 1982).
We use only the centre of each field because the errors in the field
corrections and galaxy classification become larger near the edges
(see Papers I and II for quantitative details). 
%In fact since there are so many plates in the APM survey that the mean
%of $\w$ is insensitive to these errors.

At angular scales $ \theta < 14^{\prime\prime}$, corresponding physical
separation $\simlt 20 {\rm kpc}$,  the amplitude of
$w(\theta)$ decreases with decreasing pair separation.
This is because close pairs of galaxies are often recorded as one
object by the APM machine (see Paper I) leading to an underestimate
of  $\w$ at small angular separations. 

Over the angular range $ 0.01 ^\circ < \theta < 0.1 ^\circ $, the
filled circles in Figure 1 approximately follow a power law.
The results for each plate have been fitted to a power law $w(\theta)
= A \theta^{1-\gamma}$ by least-squares. 
After correcting for the dilution of clustering caused by the 6\% 
residual contamination by stellar mergers and misclassifications, 
the mean and  standard deviation of $\gamma$ and $A$ averaged 
over the $185$ fields are
$$\eqalign{\gamma &= 1.72 \pm 0.01\cr
         A &= 0.023 \pm 0.001 \cr} \bigg \}
\quad {\rm for}\;\; 0.01^\circ < \theta
< 0.1^\circ. 
$$
where the amplitude $A$ is given in units in which $\theta$ is
in degrees. 
It is evident that the filled circles in Figure 1 fall below the
extrapolation of a power-law with $\gamma = 1.7$ on scales $\theta >
0.5^\circ$.
However, this is not a real feature of the galaxy distribution.
As we will now demonstrate, estimates of $w(\theta)$ from individual
Schmidt plates are biased low if the mean density is determined from
the galaxy counts within each field.
We can estimate the bias as follows (see also the Appendix).
The net galaxy  overdensity on each field is 
$$ \eqalignno{
\overline \delta &=  {{\cal N }  - \overline {\cal N} \over 
\overline {\cal N} }, & (7) }
$$
where $\overline {\cal N}$ is the true mean surface density 
and ${\cal N}$ is the surface density measured on each field.
The expectation value of $\overline \delta$ is zero and the variance
is given by an integral over the two-point
correlation function, 
$$ \eqalignno {
 \langle \overline \delta \rangle   & = 0 , &(8a)\cr
 \langle \overline \delta^2 \rangle & = 
{1 \over \overline {\cal N} \Omega } + {1\over
\Omega^2}  \int\int_\Omega w (\theta_{12}) \;d\Omega_1 d\Omega_2 ,&(8b)\cr}$$ 
(Peebles 1980 \S36),
where $\Omega$ is the solid angle of the field. If ${\cal N}$
is used as a measure of the mean density, the estimate of
$\w$ obtained by averaging over many fields, $w^e(\theta)$, will
be biased low by 
$$ \eqalignno {\langle w^e(\theta) \rangle &
\approx \w - \langle \overline \delta^2 \rangle, &(9)\cr}
$$ 
We show in Appendix A that the precise form of the bias depends
on the particular estimator that is used to measure $\w$, but
equation (9) is a good approximation for all of the
estimators considered in this paper if the angular separation $\theta$
is small in comparison to the size of a survey field.
 
An equivalent result to equation (9) can be derived by considering 
the total number of pairs that are counted on a single plate. 
If the total number of galaxies on a particular field is $N$, the
total number of pairs must be $N(N-1)$. Thus, the estimator
$w^e (\theta)$ satisfies a relation of the form
$$ \eqalignno{ 
\int\int_\Omega w^e(\theta_{12})\,d\Omega_1 d\Omega_2 &\approx 0, 
&(10)\cr}$$ (GP77) where the integral extends over the area of
a single field. The bias in $\w$ (equation 9) thus arises 
because the estimator is subject to an integral constraint.

We can estimate $\langle \overline \delta^2 \rangle$ directly from 
the variance in field number counts. For galaxies in the magnitude 
range $17 \le b_J \le 20$, we find $\overline \delta^2$
$= 1.6\times10^{-2}$ at $\bj = 20$. 
The variance caused by plate matching errors accounts for 
$\approx 1.6\times 10^{-3}$ (see Paper II and Section 3.1) 
and so the variance attributable to real galaxy clustering is 
$1.4\times 10^{-2}$.  
The mean angular correlation function, after adding this constant is
shown by the stars in Figure $1$.
The amplitude is increased and the mean slope of $\w$ becomes slightly
shallower.
The mean parameters of least squares fits to $\w$ over the
angular range $ 0.01^\circ < \theta < 0.1^\circ $ are
$$\eqalign{\gamma &= 1.679 \pm 0.01 \cr
         A &= 0.0282  \pm 0.0012 \cr} \bigg \}
\quad {\rm for}\;\; 0.01^\circ < \theta
< 0.1^\circ. \eqno(11) 
$$
As before the amplitude has been corrected to account for the residual
stellar contamination. 
Adding the constant $\overline \delta^2$ eliminates the sharp break in
$\w$ at $\theta \sim 0.2^\circ$ and brings the estimates into good
agreement with the results obtained using the global mean to normalize
the estimator of equation (6) (shown by the open circles in Figure~1).
This analysis of bias in estimators of $\w$ from single Schmidt plates
agrees with results from Monte Carlo simulations described in Appendix
A.
Hewett (1982) has also pointed out that the integral constraint will
introduce an artificially sharp cutoff in estimates of $w(\theta)$
from single Schmidt plates and inferred that there must
be significant correlations on pair separations of 
at least $10 \hmpc$ to explain the observed scatter in single
plate estimates of $w(\theta)$.
Our results disagree with those of Shanks \etal (1980) and Stevenson
\etal (1985) who analyzed clustering on single UKSTU fields limited at
$b_J =20$ and concluded that the break in $\w$ from an power law at
$\theta \approx 0.3^\circ$ is attributable to a real feature of the
galaxy distribution.

In summary, the results of this subsection show that the integral
constraint introduces an artificial cut-off in the estimates
of $\w$ determined from single Schmidt fields.
The bias in $\w$ depends on the variance in the galaxy counts
on the scale of a Schmidt plate. 
The high photometric accuracy of the APM survey allows us to correct
for the integral constraint bias bringing the single plate estimates
of $\w$ into excellent agreement with those determined by using a
global estimate of the mean galaxy surface density.

\vskip 0.15 truein

\noindent
{\it 2.3 Measurements of $\w$ from galaxy density maps.}

To estimate $\w$ on angular scales $\simgt 0.1^\circ$, we apply equation
(5) to maps of the galaxy surface density.  The maps were constructed  by
counting the number of galaxies in each $0.23^\circ \times 0.23^\circ$
cell of an equal area projection centred on the South Galactic pole. This
produces a 512 square array covering a $120^\circ$ patch of sky.  Cells
containing holes in the survey (areas near  bright stars, globular
clusters and calibration spots) and those outside the survey area were
flagged and not used in the analysis.  Figure 2 shows $\w$ for galaxies
with $17 < \bj <20$ computed from the whole survey area. We have estimated
$\w$ from maps with cell sizes different by factors of 0.5, 2, and 4, from
those in the $512 \times 512$ map and find essentially perfect agreement
between the various estimates.  

We have also estimated $\w$ separately for each of four zones containing
45, 45, 46 and 49 plates corresponding roughly to equally spaced strips in
right ascension.\footnote*{We also computed $\w$ in four different zones, 
where we divided the survey in half in both right ascension and
declination.  The mean $\w$ and the zone-to-zone scatter from these
zones were found to similar to those for the four zones discussed 
in the main body of the text. } The mean of these four  estimates is
less sensitive to any large-scale gradient in the survey and the scatter
between the estimates provides a realistic measure of the 
sampling errors in $\w$.
The solid line in Figures~2a and 2b show $\w$ measured using the full
survey area and the open circles show the mean of $\w$ in each of the four
zones, with $1\sigma$ error bars computed from the scatter between the
zones.  The full area estimate is not significantly different  from the
average over the four zones.  This demonstrates that gradients in the
survey on scales comparable to the size of a single zone  do not
significantly affect our determination of $\w$.
The solid points in Figure 2 show the mean of the single plate
estimates described in Section 2.2 where we used the global mean
surface $\overline {\cal N}$ to normalize the pair counts.
The estimates of $\w$ from the maps are in
excellent agreement with the single plate pair count estimates.

After correcting for the residual contamination by mergers, the mean
of least squares fits of a power law to $\w$ between $\theta=
0.01^\circ $ and $\theta= 1^\circ $ in each zone gives
$$\eqalign{\gamma &= 1.699 \pm 0.032,\cr
         A &= 0.0284 \pm 0.0029 \cr} \bigg \}
\quad {\rm for}\;\; 0.01^\circ < \theta
< 1^\circ. \eqno(12)
$$
The quoted errors are from the variance in the mean of the four zones. 
At larger angles the slope of $\w$ gradually steepens. 
If we fit a power law between $0.07^\circ <\theta <0.9^\circ $, i.e.
similar physical scales as used by GP77, we obtain a slope of $-0.734,$
in agreement with their value of $-0.741\pm 0.035$. 
The break at $\sim 3^\circ $ is extremely significant given the
scatter between zones. In Figure~2b we plot $\w$ on a linear
scale showing that the estimates are very close to zero 
(to within $\sim 5 \times 10^{-4}$) for $\theta \simgt 6 ^\circ $. 

As described in Section 3.1, it is possible to correct these estimates
of $\w$ for residual plate-to-plate errors in magnitude and
classification by subtracting the correlation function of the expected
errors.
The estimates in Figures 1 and 2 have not been corrected for these
errors, but as we will show in Section 3, we estimate that the errors
on $\w$ are $\simlt 2 \times 10^{-3}$ at small angles, and decreases
rapidly at angles $\simgt 1^\circ$.
Since this is much smaller than the amplitude of $\w$ even in the
region near the break at $\sim 3^\circ$, we believe that our results
are not significantly affected by errors in the survey.
This point is so important for the interpretation of our results that
we  devote Section 3 of this paper to a lengthy analysis of
systematic errors in the survey before we present 
corrected estimates of $\w$. 

\vskip 0.15 truein

\noindent
{\it 2.4 Variation of $\w$ with magnitude} 

The variation of $\w$ as a function of magnitude allows us to check
that the angular correlation functions measure intrinsic clustering
rather than systematic errors.
If photometric errors and intervening obscuration are negligible, $\w$
is determined by the projection of the three-dimensional two point
correlation function $\xi(r)$ via Limber's equation (equation 31), and
hence $\w$ should obey a scaling relation as the depth of the survey
is changed (Peebles 1980 \S 50 and Section 4.1.1 equation 33).
To test if the correlation function scales as predicted, we evaluated
$\w$ for each of six {\it disjoint} magnitude slices between $\bj=17$
and $20.5$.
The magnitude slices are defined using uncalibrated APM magnitudes,
and so are not exactly 0.5 magnitudes wide.  
The exact magnitude ranges in the calibrated $\bj$ system
are listed in Table 4 (Section 3.5) and in
the Figures.

We have limited this analysis to the central 120 plates of our survey
which lie at Galactic latitudes $b \simlt -50^\circ$.  At such high
latitudes no significant errors are introduced by uncertainties in
Galactic absorption or from the increase in the numbers of merged
stellar images at low galactic latitudes (see Sections 3.4.1 and
3.4.3).  The results are shown in Figure~3.  Qualitatively, the
behaviour is as expected: at brighter magnitudes the amplitude of $\w$
is higher and the break from a power law occurs at larger angles.  In
Section 4.1 we quantify the expected shifts by evaluating Limber's
equation using empirical estimates of the survey selection function
derived from deep redshift surveys.

\vfill\eject

\noindent
{\bf  3.  Estimates of Systematic Errors in $w(\theta)$ }

In this Section we discuss several sources of systematic error which could in
principle affect the shape of $w(\theta)$. Errors in the APM 
Survey could be caused by:

\item{[1]} Variations in the APM scanning procedure. 

\item{[2]} Variations in plate emulsion sensitivity.

\item{[3]} Atmospheric variations.

\item{[4]} Galactic obscuration

\item{[5]} Intergalactic dust.

\item{[6]} Stellar contamination.

\noindent
Sources 1, 2 and 3 introduce plate dependent variations which should
be removed by the field corrections and plate matching procedure 
described in Paper II. Sources 3 - 6 could introduce gradients 
and features in the galaxy distribution that are continuous 
across plate boundaries and so would not be removed by the plate
matching algorithm. 

In the following subsections we describe various tests which we have
applied to check the effects of these errors.  In Section 3.1 we consider
the effects of residual errors in the photometry and galaxy selection by
analyzing various models for the errors. In Section 3.2, we test the
photometric accuracy of the survey by comparing the APM photometry with
independent CCD photometry of large numbers of galaxies.
We test for errors correlated with exposure and scanning date in Section
3.3, and for gradients associated with extinction and stellar
contamination in Section 3.4.  The dependence of systematic errors with
apparent magnitude is discussed in Section 3.5.

\vskip 0.15 truein

\noindent
{\it  3.1 Models of residual photometric errors and variations in
completeness. } 

Magnitude errors and variations in star-galaxy classification 
introduce errors in the observed number of galaxies as a function of
position in the survey leading to  errors in the measured
correlation function.
In most of the following discussion we combine both sources of error
and consider a single overall error in the number of galaxies
at each position on the sky. 
There are two distinct components to each of the errors: 
1) variations across individual Schmidt plates, which we term 
`field effects'; and 2) differences between one plate and the next, 
which we call `plate matching errors'. 

Several uncontrollable factors cause the sensitivity of a Schmidt plate
to vary as a function of position, and so there are variations in
the measured magnitudes as a function of position over each plate.
The shape and amplitude of the variations are different for each plate
and, as described in Section 2 of Paper II, we correct the magnitudes
using a two-dimensional correction function determined individually
for each plate.
Although the corrections precisely remove the errors when averaged 
over all plates, any individual plate is likely to have small
residual field effects. 
This means that each edge of a plate may have systematic errors
relative to the plate centre, and relative to the other edges. 

The parameters used for star-galaxy separation also show variations
across each plate, and so are subject to similar residual field
effects.
These errors will introduce variations in the completeness of the
galaxy sample, and also in the contaminating fraction of non-galaxy
images (mainly blended star images). 
We refer to these errors as galaxy counting errors. 

The variations in emulsion, observing conditions and measurement
parameters introduce differences in  the photometry and completeness of
each plate.  For both the photometry and star-galaxy separation we have
reduced these plate-to-plate variations by applying corrections to each
plate so that the differences between measurements on neighbouring plate
overlaps are minimized.  Any errors in the overlap offsets produce
plate-to-plate matching errors which are correlated, and so they propagate
across the grid of plates in a random walk, introducing low-amplitude
large-scale correlations in the matched parameters.  These gradients are
probably the most difficult source of error to quantify in the APM Survey.
A simple analysis of this propagation of errors was presented in Section 3
of Paper~II, and in Sections 3.1.1 and 3.1.2 below, we describe simulations 
of more complicated error models.  A direct comparison with external CCD
photometry is discussed in Section 3.2, and further tests to place upper
limits on large-scale gradients are discussed in Section 4.2.3.

Without analysing specific error models, we can estimate the effect of
correlated residual errors on measurements of $\w$ by considering the
cells estimator of $\w$ defined in equation (5).  The number of galaxies
in cell $i$ at magnitude limit $m$ is given by $log_{10}(N_i) \propto am
$, where $a \sim 0.45 $ is the slope of the galaxy number counts (Maddox
{\it et.al.} 1990d).  A magnitude error $\delta m$ will thus produce a
fractional error in the number of galaxies of  $ \delta N_i / N_i = a
\delta m \ln(10) \approx \delta m$.
Any variation in completeness or contamination rate
translates directly to a fractional error in the number of galaxies,
$\delta N_i / N_i =\delta c$, where $\delta c$ is the overall error
from these two effects, {\it i.e.} the contamination and the completeness. 
The total fractional error in the galaxy counts is $e_i = 1 + 
\delta m + \delta c $.  and will vary
between plates and as a function of position within each plate; by
definition the mean value is unity, $\left<e_i\right> = 1$.

Since our matching procedure uses the magnitudes and star-galaxy
separation parameters of individual objects in the overlap regions, 
not the galaxy surface density, the errors $e_i$ are very nearly
independent of the galaxy surface density $N_i$.
The estimated angular correlation function, $w_{est}$,
is therefore  given by 
$$ \eqalignno { 
w_{est} &= { \left<e_i N_i e_j N_j\right> \over \left<e_i N_i\right> 
\left<e_j N_j\right>  } - 1 \cr
  &= { \left<e_i e_j\right> \left<N_i N_j\right> \over 
     \left<e_i\right> \left<e_j\right> \left<N_i\right> 
     \left<N_j\right>  } - 1 \cr
 &= (1 + w_{err}) (1 + w_{true}) - 1 \cr
 &\approx w_{err} + w_{true}. & (13) } 
$$
if $w \ll 1$. The effect of errors in the survey is therefore
simply to add the 
correlation function of the errors  to the true galaxy correlation 
function. In the next section we consider several models 
for the combined effects of magnitude and counting errors, and
compute the resulting error correlation function $w_{err}$.

To determine the amplitude of $w_{err}$ we estimate the size of
magnitude and counting errors from the consistency of
measurements in plate overlaps.  
If the magnitude and counting errors are uncorrelated, the
two amplitudes should be added in quadrature; if they are perfectly
correlated they should be added, and if they are anticorrelated they
should be subtracted.
In practice we find that the magnitude errors totally dominate the
errors that propagate through the plate grid. 

\noindent
{\it 3.1.1 Uncorrelated overlap errors}

In our simplest error model we assume that the error in each overlap
is uncorrelated with those in other overlaps, as in
the error model that Groth and Peebles applied in their 
analysis of the Lick map (Groth and Peebles 1986b, hereafter GP86).
In principle our analysis applies to both magnitude and counting
errors but in practice the counting errors are negligible, and so for
simplicity we refer to magnitudes, and mention counting errors only
where necessary.

We use the definitions and notation given in Paper II unless stated
otherwise, and we begin by summarizing the important quantities. 
For each plate we define a correction $C_i$ so that an image with
measured magnitude $m_i$ has a matched magnitude $m_o$ given by $m_o =
m_i - C_i$.
In the absence of measurement errors, the difference in magnitudes
between galaxies in the overlap between plates $i$ and $j$ is a
constant
$$
\eqalignno { T_{ij} &= m_j - m_i = C_i - C_j.  \cr}
$$
In practice, there will be errors on each of these quantities
and the plate matching algorithm will converge on estimates
of the plate correction factors $C_i^e$ that differ from the
true correction factors $C_i$. 
The mean magnitude difference measured in each overlap region,
$T^e_{ij}$, will also differ from the true difference $T_{ij}$.

If we assume that the errors $\delta T_{ij}$ in the $T^e_{ij}$ are
uncorrelated with each other, so that
$$
\eqalignno { \langle \delta T_{ij} \delta T_{ik} \rangle
 &= \sigma^2 \delta_{jk},  &(14) \cr}
$$
then we showed in Paper II (equation 23) that the variance in the
difference between the $T^e_{ij}$ and the plate correction factors on
each overlap is given by
$$
\eqalignno {  \epsilon ^2 = {\rm var} (T^e_{ij} + C^e_j - C^e_i) 
 & =  { N - 1 \over N + 1} \sigma^2 \approx {2 \over 3} \sigma^2. 
&(15) \cr}
$$
where  $N$ is the mean number of overlaps on each plate. 

Using the plate overlaps to calculate the correction factors $C^e_i$
introduces correlations between the correction factors of different
plates ({\it c.f.} de Lapparent \etal 1986). The 
propagation of these errors through a periodic square grid of plates
can be analysed analytically and an
expression for $\langle \delta C_i \delta C_j (\theta)
\rangle$, based on this model is given in equation (31) of Paper II.
However, the assumption of  a periodic array of plates is
not a good representation of the real plate configuration
of the APM survey. We have therefore carried out Monte Carlo simulations
using the real plate configuration as described in Paper II. In these
simulations, a random Gaussian error was generated for each overlap and
used to estimate the  overlap offsets in the edge-matching procedure.  The
resulting correction coefficients are actually the errors that would be
introduced in the plate-correction factors by the errors in the plate
overlaps.  The correlation function of errors, $w_{err}$, is thus
given by the correlation function of the correction coefficients.  As
shown in Figure 14 of Paper II, the resulting $w_{err}$ has small
oscillations similar to the analytic function at $\theta \simlt 10^\circ$,
but on larger scales the Monte Carlo results decline smoothly while
the analytic function remains oscillatory.

For the APM survey, the variance in overlap residuals measured
directly from the overlaps is $ \epsilon ^2 = 1.9 \times 10^{-3} $
(see Paper II). 
The mean number of overlaps per plate is approximately $N=5$ for the
UK Schmidt fields, so equation (15) gives
$$\eqalignno {\sigma^2 & = 3.2 \times 10^{-3}, &(16) 
\cr}$$ 
{\it ie} the {\it rms}
mean magnitude difference in a plate overlap is $0.057$ magnitudes,
which corresponds to a 6\% fractional error in the
galaxy density on a single Schmidt plate.

The corresponding overlap measurements for the $\psi$ classification
parameter described in Paper I, which we use to separate stars and
galaxies, give a dispersion of $\epsilon = 44.5$.  The number of
galaxies as a function of $\psi$ is well fit by an asymmetrical
distribution centred on $\psi = 2166$ with dispersions $\sigma_1 = 603
$ and $\sigma_2 = 347 $, (see equation 2 of Paper I for a definition
of the parameters $\sigma_1$ and $\sigma_2$).  Our galaxy sample is
defined by selecting objects with $\psi>1000$, so the observed
dispersion in $\psi$ introduces a dispersion of roughly 0.3\% in
galaxy density on a single Schmidt plate.  Magnitude errors therefore
totally dominate the propagation of errors in the galaxy counts.  Note
however, that the galactic gradient in stellar density will cause a
gradient in the contamination rate, independent of the plate-matching
procedure.

Using the magnitude {\it rms} overlap residual to normalize the error
estimates, the analytic model 
predicts that the error correlation function at zero lag is
$$
\eqalignno{
w_{err}(0) &= \beta^2_{ii} = 0.457 \sigma^2 \approx   1.8 \times 10^{-3}. &(17) \cr}
$$
In (17) we use the notation of GP86 where $\beta^2_{ii}$ denotes the
variance in the plate zero points. 
At larger scales where the Monte Carlo estimate becomes smooth, $w_{err}$ 
can be
approximated roughly by an exponential
$$
\eqalignno{
\langle \delta C_i \delta C_j (\theta) \rangle & \approx w_{err}( \theta) 
\approx 0.49\; \sigma^2 {\rm exp}\big ( - \theta /\theta_c \big ), 
\qquad \theta_c \approx 11^\circ, \qquad \theta \simlt 50^\circ. &(18) \cr}
$$
Over the range $10^\circ \simlt \theta \simlt 50^\circ$ equation (18)
is accurate to within a few percent. 

It is possible to check if the model for $w_{err}$ is consistent with the
data by computing the intra-plate and inter-plate estimates of
$\w$ (see GP77 and Maddox \etal 1990c).
The intra-plate estimate is derived by counting only galaxy pairs
that lie on the same plate while the inter-plate estimate is derived by
counting only galaxy pairs that lie on neighbouring plates. 
Since the true correlation function is independent of the choice of
plate boundaries, any difference between the intra-plate and
inter-plate estimates should be due to the difference in the error
correlation function\footnote*{Although this is true in the mean, any
single realization of a clustered distribution is likely to have some
structure correlated with plate boundaries, so some offset is
expected. This is discussed more fully in Section 3.5}.
The  error on the intra-plate correlation function is 
$w_{err}(0)$, and the error on the 
inter-plate estimate is approximately $w_{err} (5^\circ)$, since $ \theta
= 5^\circ$ corresponds to the distance between neighbouring Schmidt
plates. In the analytic error model 
$$
\eqalignno{
w_{err}(5^\circ) & \approx \beta^2_{ij} = 0.208 \sigma^2 \approx  
1.0\times 10^{-3}, &(19) \cr}
$$
where,  following the notation of GP86,
 $\beta^2_{ij}$ denotes the covariance between the zero point
errors for neighouring plates.
Hence the predicted difference between inter- and intra-plate estimates of
$w(\theta)$ is approximately $\beta^2_{ii} - \beta^2_{ij} \approx 0.8
\times 10^{-3}$. 
The offset observed from the APM measurements 
is $ 1.7 \times 10^{-3} $, which suggests that
this uncorrelated error model may be too simplistic.
We discuss this test in more detail after considering more complex
error models in the next section. 

\noindent
{\it 3.1.2 Correlated overlap errors.}

The error model of the previous section assumes that the overlap
errors are uncorrelated. 
Though this assumption enables an analytic approach to the modeling,
it may not be a good approximation to the APM survey because the overlap
errors are dominated by residual field errors. 
There are 3000 galaxies in a typical overlap with an {\it rms} error
of 0.2 magnitudes per galaxy, so the random magnitude error in a
single plate overlap is $0.2 / \sqrt{3000} \sim 0.005 \rm\,$ magnitudes. 
In the absence of residual field effects, this would be the only
source of error in the matching procedure, and so the plate-matching
errors would be of similar order. 
However, the uncertainties in the corrections for field effects are
expected to be of order 0.05 magnitudes, and so will dominate the
error in each overlap. 
This is consistent with the observed dispersion between magnitudes in
overlaps which is 0.06 magnitudes, as described in the previous
section.
The field correction errors are very likely to be systematically
correlated for each plate: 
for example if the amplitude of the applied field correction is too
small for a particular plate,
then all of the edges would be low compared to the true value. 
We must therefore consider error models which include residual field
errors. 

Before describing these models, we point out that in the Lick survey,
the assumption of uncorrelated errors is much more likely to be valid.
The Lick plates were not hypersensitized and so the field effects
should be similar for all plates. Applying the  average field correction
to all plates should therefore be accurate.
Furthermore, the Lick plate matching used galaxy counts in plate overlaps to
estimate the different depths of neighbouring plates.
Since the counts on each unmatched plate sample the galaxy
distribution to a different depth, and galaxies are clustered,
this introduces an error in the estimated plate depths.
In fact this error is much larger than the residual field errors. 
Since it depends on the galaxy distribution, the error is uncorrelated
with the plate boundaries, and so it is reasonable to assume
uncorrelated overlap errors for the Lick survey.

To simulate the effects of residual field errors in the APM Survey, we
first generate an error, $S(x,y)$, as a function of position for each
plate in the survey.  This represents the total error in galaxy density as
a function of position (caused by magnitude and classification errors), but as
discussed in the previous section, the magnitude errors are the dominant
contribution.  We consider three functional forms to represent the
residual field errors:

\noindent
[A] Linear mode: $S(x,y) = 1 + ax + by$ where $a$ and $b$ are
independent and selected at random from Gaussian distribution with 
zero mean and dispersion $s$. 

\noindent
[B] Quadratic mode: $S(x,y) = 1 + ax^2 + by^2$ with $a$ and $b$ selected
as in mode A.

\noindent
[C] Radial mode: with $S(x,y) = 1 + a(x^2 + y^2)$, and 
$a$ selected as in modes A and B.

The error function for each plate was generated independently of the 
others  and each field was also assigned a random global zero-point error 
with a distribution similar to that observed in the survey (Figure 11,
Paper II). 

For each neighbouring plate pair the difference between the functions
at the common plate edge is used to measure an overlap offset.
As for the real data, the iterative plate matching procedure (Section
3, Paper II) was applied to the model plate grid giving a set of
estimated plate corrections which minimize the overlap residuals.
The resulting corrections are added to the original field errors for
each plate to give a map of the overall residual errors which
includes both the original field errors, and any plate-to-plate
matching errors introduced by the overlap errors.
Finally we measure the final error correlation function using
equation (5) applied to this error map. 

The mean of 100 realizations for each mode are shown in
Figures (4a)--(4c). The  filled circles joined by the solid lines 
show the inter-plate estimates of $\w$ and the open
circles joined by the dotted lines show the
intra-plate estimates.
In the models for these plots we chose the dispersion $s$ for each of
the modes so that the residual overlap error in the model is $\epsilon^2
= 1.9 \times 10^{-3} $ as measured in the real data.
Table 1 gives the values for $s$ normalized so that the plate
coordinates $x$ and $y$ are in the range $-1$ to $+1$. 

\noindent
\bigskip

$$
\vbox{
\halign { \hfil # \hfil \tabskip 1em plus 2em minus 5em
& \hfil # \hfil & \hfil # \hfil & \hfil # \hfil &
& \hfil # \hfil\cr
\multispan{6}{\hfil \bf Table 1: Offsets and normalizations for
different error modes \hfil}\cr 
\noalign{\medskip\hrule\vskip0.1truecm\hrule}
\noalign{\medskip}
 & \multispan{2} mode & $s$  & inter-intra offset & $ w_{err}(0) \sim \beta_{ii}^2$   \cr
\noalign{\medskip\hrule\medskip}
&A & linear    & $0.07$ & $1.0 \times 10^{-3}$ & $ 1.6 \times 10^{-3}$ \cr %$ 1.7 \times 10^{-3}$ \cr
&B & quadratic & $0.10$ & $1.1 \times 10^{-3}$ & $ 1.7 \times 10^{-3}$ \cr %$ 2.9 \times 10^{-3}$ \cr
&C & radial    & $0.22$ & $6.9 \times 10^{-3}$ & $ 8.3 \times 10^{-3}$ \cr %$17.6 \times 10^{-3}$ \cr
\noalign{\medskip\hrule}
}}$$

\bigskip

As seen in Figure 4 the correlation functions of the errors for modes
[A] and [B] are similar to the uncorrelated error model of the 
previous subsection.
This is reasonable because the overlap errors for modes [A] and [B]
are relatively weakly correlated and so the simple model of
Paper II provides an adequate approximation.
The intra-inter-plate offsets are slightly larger than the 
predictions of the Paper II model, because the intra-plate estimates
include the structure of the error function within each plate. 
The offsets are slightly smaller than measured from the real data. 

There are also small differences between behaviour of the inter- and
intra-plate estimates of modes [A] and [B]. 
For mode [A] the linear gradient in errors across each field produces
a gradient in the intra-plate estimates:
the offset is more positive on small scales and negative at larger
scales.
For mode [B] the intra-inter-plate offset is more uniformly 
positive, with a slight increase between $6^\circ$  and $7^\circ$,
where the intra-plate estimate is based on galaxies in the very
corners of the plates and the quadratic error functions have a large
dispersion. 

For mode [C] the overlap errors are almost perfectly correlated, so if
the plate grid were uniformly rectangular, this mode would produce
perfectly consistent plate overlaps, and there would be no residual
dispersion in the overlaps.
The real plate grid has overlaps of differing size and position for
each field, and so there are small overlap residuals, but still 
this mode very poorly constrained by plate overlaps. 
For the parameters which match the observed variance in the
overlap residuals of the APM Survey, mode [C] leads to a very high
intra-plate estimate of $\w$, which is inconsistent with the data. 
In fact, comparing the observed and predicted inter-intra-plate offset
in $\w$ provides a much better 
limit to the normalization of this mode of errors. 
Since the overlap residuals are so small in this error mode, the 
errors propagate less strongly through the plate network in
comparison to the other two modes, leading to much smaller errors on
scales larger than $7^\circ$. 
Note that the small overlap residuals seen in the
$\psi$ matching do not constrain possible mode [C] field errors in the
star-galaxy separation. 
Although the counting errors do not contribute significantly to errors which
propagate through the survey, they may still be significant in terms
of radial field errors. 

The intra- and inter-plate measurements from the real data in the
magnitude range $17 \le \bj \le 20$ are plotted in Figure 5.
The open circles show the intra-plate estimate of $\w$ and the filled
circles the inter-plate estimate of $\w$.
In each case we show the mean $\w$ from the four zones as described
Section 2.3, and plot $1 \sigma $ error bars from the scatter between
the four zones.
In the range $1.5^\circ < \theta < 4^\circ$, where the errors on both
estimates are relatively small we find a mean offset $w_{intra} -
w_{inter} = 1.7 \pm 0.7 \times 10^{-3} $. 
The size of the offset is similar to that expected from  modes [A] and 
[B], but much smaller than mode [C]. 
In principle, we could reproduce the shape of the observed offset using
some linear combination of the three modes, but the large uncertainties
in the observations would make such a decomposition unreliable. 

We have estimated the uncertainty in the observed offset from the
dispersion between the four zones, as quoted above and plotted as
error bars in Figure 5.
This suggests that the $w_{intra}- w_{inter}$ is positive at only the
one sigma level.
The large scatter in the offsets measured in the different zones 
may seem surprising, but is in fact consistent with the scatter
expected from `cosmic' variance. 
We have tested this by computing inter- and intra- plate estimates of $\w$
for several 
hundred simulations of clustered distributions using the
Soneira-Peebles (1978) technique described in Appendix A1.
For each realization, we imposed an artificial plate grid to calculate
intra- and inter-plate estimates of $\w$.  
No errors were added to the simulated distribution, and the grid was
used only to decide which pairs of cells to count for the intra-plate
estimates and which for the inter-plate estimates.
The rms scatter in the inter- intra-plate angular correlations was
found to be $1 \times 10^{-3}$, in good agreement with the observed scatter
between the 4 zones of the APM survey.
Therefore the observed value $w_{intra}-w_{inter} = 1.7 \times
10^{-3}$ is only marginally significant evidence for errors in the APM
measurements.
Nevertheless we can use the measurement to limit the amplitude of mode
[C] errors to be $\simlt 2 \times 10^{-3}$ at zero lag.

%A simple calculation can show that this is reasonable: 
%The total number of galaxies in each zone is $\sim 4 \times 10^5$, and
%so the expected Poisson error in the estimated mean galaxy density is
%$1.6\times 10^{-3}$

In summary, the observed overlap residuals provide strong constraints
on errors of the form [A] or [B], so that the zero-lag amplitude must
be $ 1.7 \times 10^{-3} \simlt \beta_{ii}^2 \simlt 3\times 10^{-3}$,
corresponding to a magnitude error of between 0.041 and 0.054
magnitudes. 
The observed inter-intra-plate offset limits mode [C] errors to 
to have an amplitude  $\simlt 2 \times 10^{-3}$ at $\theta \simlt 0.5^\circ$
Including all modes [A]-[C], these constraints limit the error correlation
function to be $w_{err} \simlt 6 \times 10^{-4} $ on angular scales $\theta
\simgt 5^\circ$. 

%\vfil\eject

\noindent
{\it  3.2 Comparison of APM photometry with CCD magnitudes} 

Two methods of measuring correlated errors in the APM photometry
were analyzed in detail in Paper~II: 

\item{(a)} The covariance of residuals between the matched magnitudes
and  41 CCD sequences distributed over the survey. This directly
measures the contribution to $\w$ from gradients in the magnitude
matching and indicates that $w_{err} \simlt 10^{-3}$ (see
Figure 24 Paper~II). 

\item{(b)} 
The covariance of the estimated plate zero-points (see Figure~16 of
Paper~II).   Assuming that the {\it true} plate zero points $C_i$ are
uncorrelated,  correlations between the {\it estimated} plate zero points
$C^e_i$ provide a measure of correlated plate-matching errors. As
described in Paper II, the correlation function of the estimated plate
zero points sets a limit of $\simlt 10^{-3}$ on the error in
$w(\theta)$ arising from large-scale gradients in the survey.

Fong \etal (1992, hereafter F92) have argued that some of the signal in
our estimates of $w(\theta)$ from the APM Survey
might be caused by large-scale gradients which we have failed
to detect in these and other tests. Tests such as (b) are indirect and it
is possible, though extremely unlikely, that the true plate zero points
are anticorrelated with the estimated plate zero points to a high precision
thus masking gradients in the survey. A direct test, {\it i.e.}
in which independent photometry is compared 
with that of the APM Survey, is clearly the most convincing
way of checking the uniformity of the APM magnitudes. 
Recently we have been
able to make such a comparison using CCD magnitudes of $13570$ galaxies
measured by Shectman \etal (1992, 1995). This is described in Section 3.2.1. We
postpone a more detailed discussion of Fong \etal's criticisms until
Section 3.2.2.

\noindent
{\it 3.2.1 Comparison of Las Campanas CCD and APM magnitudes}

The CCD Photometry used in this Section is part of the
Las Campanas Deep Redshift Survey (LCDRS) described by 
Schectman \etal (1992).
The CCD photometry is in the $R$ band and was obtained from
drift scans with the Las Campanas 1-metre Swope telescope.
The scanned areas that have been kindly made available to us
consist of a number of ``bricks'' of area $1.5^\circ 
\times 3.0^\circ$ in three declination strips 
running along the entire width of the APM survey, as shown in 
Figure 6. 
We have correlated the positions of APM galaxies
with the LCDRS images and computed the mean magnitude difference
$\Delta m = b_J - m_{CCD}$ as a function of UKSTU field. The CCD magnitudes
are limited to $m_{CCD} \simlt 18$, and the mean colour difference is $
\langle \Delta m \rangle = 1.251$. Because of the large wavelength
difference of the passbands in the two surveys, the {\it rms} dispersion
of $\Delta m$ 
about the mean is large and is equal to $0.45$ magnitudes.
Fortunately, there are a large number of galaxies with CCD magnitudes
on most of the UKSTU fields and so the net scatter on the transforms
arising from colour differences is small.

%\vfill\eject

\topinsert 
$$
\vbox{
\halign { \hfil # \hfil \tabskip 1em plus 2em minus 5em
& \hfil # \hfil & \hfil # \hfil & \hfil # \hfil & 
\hfil # \hfil & \hfil # \hfil & \hfil # \hfil &
\hfil # \hfil & \hfil # \hfil & \hfil # \hfil &
\hfil # \hfil &
&\tabskip=0pt \hfil # \hfil\cr
\multispan{12}{\hfil \bf Table 2: Las Campanas CCD-APM magnitude residuals \hfil}\cr
\noalign{\medskip\hrule\vskip0.1truecm\hrule}
\noalign{\medskip}
 &     &  UKSTU & $N_g$ & \multispan3{\hfil $\alpha^\dagger$ \hfil} &
\multispan3{\hfil $\delta^\dagger$ \hfil} & $\delta m$ & $\sigma_m$\cr
\noalign{\medskip\hrule\medskip}

 &$\;\;1$&  $  236$&  $ \;\;141$&  $21$ &   $42$& $50$& $-48$& $\;\;6$& $36$&  $-0.001$&  $  0.724$ \cr
 &$\;\;2$&  $  237$&  $ \;\;379$&  $21$ &   $51$& $28$& $-48$& $\;\;9$& $35$&  $\;\;0.012$&$  0.500$ \cr
 &$\;\;3$&  $  241$&  $ \;\;284$&  $23$ &   $59$& $24$& $-43$& $51$& $36$&  $-0.050$&  $  0.488$ \cr
 &$\;\;4$&  $  242$&  $ \;\;198$&  $\;\;0$& $28$& $37$& $-45$& $\;\;1$& $12$&  $-0.017$&  $  0.555$ \cr
 &$\;\;5$&  $  245$&  $ \;\;312$&  $\;\;1$& $50$& $55$& $-44$& $37$& $11$&  $\;\; 0.043$&  $  0.399$ \cr
 &$\;\;6$&  $  246$&  $ \;\;242$&  $\;\;2$& $23$& $40$& $-44$& $28$& $47$&  $-0.173$&  $  0.446$ \cr
 &$\;\;7$&  $  247$&  $ \;\;185$&  $\;\;2$& $43$& $45$& $-43$& $31$& $47$&  $\;\; 0.022$&  $  0.352$ \cr
 &$\;\;8$&  $  248$&  $ \;\;457$&  $\;\;3$& $13$& $50$& $-44$& $47$& $24$&  $\;\; 0.004$&  $  0.340$ \cr
 &$\;\;9$&  $  249$&  $ \;\;538$&  $\;\;3$& $48$& $23$& $-44$& $50$& $24$&  $-0.135$&  $  0.431$ \cr
 &$10$&     $  251$&  $ \;\;155$&  $\;\;4$& $32$& $38$& $-45$& $10$& $48$&  $\;\;0.007$&  $  0.407$ \cr
 &$11$&     $  288$&  $ \;\;114$&  $21$   & $50$& $43$& $-47$& $22$& $48$&  $\;\;0.071$&  $  0.462$ \cr
 &$12$&     $  290$&  $ \;\;245$&  $23$   &$\;\;2$& $52$& $-44$& $57$& $ 0$&  $\;\;0.094$&  $  0.396$ \cr
 &$13$&     $  291$&  $ \;\;221$&  $23$   & $14$& $23$& $-44$& $34$& $48$&  $\;\;0.030$&  $  0.534$ \cr
 &$14$&     $  293$&  $ \;\;887$&  $23$   & $59$& $25$& $-40$& $34$& $48$&  $-0.020$&  $  0.450$ \cr
 &$15$&     $  294$&  $    1036$&  $\;\;0$& $26$& $ 3$& $-39$& $39$& $35$&  $\;\;0.012$&  $  0.444$ \cr
 &$16$&     $  295$&  $ \;\;343$&  $\;\;0$& $53$& $50$& $-40$& $49$& $48$&  $-0.065$&  $  0.442$ \cr
 &$17$&     $  296$&  $ \;\;621$&  $\;\;1$& $15$& $40$& $-39$& $54$& $ 0$&  $\;\;0.036$&  $  0.442$ \cr
 &$18$&     $  297$&  $ \;\;366$&  $\;\; 1$& $46$& $11$& $-39$& $35$& $24$&  $-0.031$&  $  0.448$ \cr
 &$19$&     $  298$&  $ \;\;273$&  $\;\;2$&$\;\;5$& $50$& $-41$& $29$& $24$&  $\;\;0.028$&  $  0.521$ \cr
 &$20$&     $  299$&  $ \;\;674$&  $\;\;2$& $36$& $14$& $-40$& $16$& $12$&  $\;\;0.027$&  $  0.447$ \cr
 &$21$&     $  300$&  $ \;\;747$&  $\;\;3$&$\;\;3$& $19$& $-39$& $47$& $59$&  $\;\;0.017$&  $  0.416$ \cr
 &$22$&     $  301$&  $ \;\;476$&  $\;\;3$& $28$& $12$& $-41$& $40$& $48$&  $-0.051$&  $  0.431$ \cr
 &$23$&     $  302$&  $ \;\;606$&  $\;\;3$& $48$& $23$& $-38$& $58$& $12$&  $-0.075$&  $  0.492$ \cr
 &$24$&     $  303$&  $ \;\;307$&  $\;\;4$& $26$& $12$& $-38$& $53$& $23$&  $\;\;0.053$&  $  0.486$ \cr
 &$25$&     $  342$&  $ \;\;386$&  $21$   &$\;\;7$& $16$& $-38$& $59$& $24$&  $-0.025$&  $  0.517$ \cr
 &$26$&     $  343$&  $ \;\;279$&  $21$   & $45$& $50$& $-39$& $\;\;2$& $59$&  $\;\;0.033$&$ 0.512$ \cr
 &$27$&     $  345$&  $ \;\;724$&  $22$   & $29$& $38$& $-38$& $58$& $12$&  $-0.049$&  $  0.458$ \cr
 &$28$&     $  346$&  $ \;\;285$&  $23$   &$\;\;5$& $43$& $-39$& $\;\;3$& $36$&  $\;\;0.093$& $0.414$ \cr
 &$29$&     $  347$&  $ \;\;345$&  $23$   & $20$& $40$& $-40$& $33$& $36$&  $\;\;0.037$&  $  0.446$ \cr
 &$30$&     $  348$&  $ \;\;285$&  $23$   & $43$& $37$& $-38$& $57$& $35$&  $\;\;0.049$&  $  0.428$ \cr
 &$31$&     $  359$&  $ \;\;199$&  $\;\;4$&$\;\;5$& $59$& $-35$& $57$& $\;\;0$&  $\;\;0.015$&  $  0.448$ \cr
\noalign{\medskip\hrule}
\cr
\multispan{3}{$\dagger$ Equinox 1990.5} \cr
\cr
\cr
\cr
\cr
}}$$
\endinsert 

\vskip 0.1 truein

%\vfill\eject

Table 2 summarizes the results of the comparison of the APM
and LCDRS photometry. The second column of the Table lists the UKSTU field
number and the third column gives the number of galaxies with CCD
magnitudes in the LCDRS survey matched to APM galaxies in the field.
Columns 4-9 list the mean right ascension and declination (equinox 1990.5)
for the matched galaxies. Column 10 gives the mean magnitude difference
$\delta m = \Delta m - \langle \Delta m \rangle$ and column 11 
gives the rms dispersion about the mean for a single galaxy $\sigma_m$.
The  error on $\delta m$ is thus $\sigma_m/\sqrt N_g$. 
We analyze only those
UKSTU fields with more than $100$ galaxies from the LCDRS CCD frames; there
are a few fields with a smaller number of galaxies, but the uncertainties
in $\delta m$ for these are dominated by the dispersion in the
galaxy colours.

From the entries in Table 2, we find
$$
\eqalignno{
\sigma (\delta m) & = 0.058, &(20a) \cr
\sigma(m)  &= 0.028, & (20b) \cr}
$$
where $\sigma(m)$ is the contribution to the variance from scatter
in the individual magnitudes on each field,
$$
\eqalignno{
\sigma^2(m) &= {1 \over N_f} \sum_i^{N_f} \sigma_m^2(i)/N_g(i), &(21)\cr}
$$
and $N_f$ is the number of fields. If errors in the CCD zero points
can be neglected, the {rms} error in the zero point of an APM field is 
$$
\eqalignno{
\sigma_{APM} &= (\sigma^2(\delta m) - \sigma^2(m))^{1/2} = 0.051. &(22)\cr}
$$
The quantity $\sigma_{APM}$ is the square root of $\beta_{ii}^2$ defined
in equation (17), which we have estimated to be between $0.041$ and
$0.054$ in Section 3.1. 

Figure 7 shows the residuals $\delta m$ plotted as a function of right
ascension.
The Figure shows that the residuals are uncorrelated with right
ascension; there is no evidence for any east-west gradient in the APM
magnitude system.
Two fields, $246$ and $249$ are outliers in Figure 7 and have
residuals of $-0.173$ and $-0.135$ respectively, showing that the
distribution of the magnitude errors is non-Gaussian.
We have checked whether these fields show any peculiarities
that might account for such large magnitude errors.
Field $249$ contains one of our calibration sequences (see Table~3), and the
residual between the APM and CCD magnitudes is only 0.037, showing no
evidence of poor matching. 
The plate correction factors $\delta C^e$ applied to these fields
($0.12$ and $0.11$ for $246$ and $249$) are not unusual when compared
to the distribution of APM plate correction factors, which is well
approximated by a Gaussian, with {\rms} $\langle (\delta C^e)^2
\rangle^{1/2} = 0.178$ ({\it cf} Figure 11 of Paper II).
The fluctuations in the galaxy surface density on each field, $\delta
n/ n$, to a magnitude limit of $b_J = 20$ are plotted against the
zero point residuals $\delta m$ in Figure 8.
The surface density fluctuations on fields $246$ and $249$ are
positive and equal to $\delta n/n = 0.148$ and $0.091$ respectively,
compared to the {rms} over all $185$ plates of $0.143$.
The surface densities on these plates are thus typical of the rest of
the survey and provide no firm evidence of large zero point errors.
Another possible cause of these outlying points might be zero point
errors in the CCD magnitudes; for example, the residual for field
$246$ would not look so improbable if the CCD zero point were in error
by $\sim 0.05$ magnitudes.

It is therefore  reasonable to exclude field $246$ from the analysis
giving
$$
\eqalignno{
\sigma (\delta_m) & = 0.050, & (23a) \cr
\sigma(m)  &= 0.028, & (23b) \cr
\sigma_{APM} &= 0.041,\qquad ({\rm minus}\;\;{\rm field}\;\;246).
&(23c)\cr}
$$
We conclude that the dispersion in the APM plate zero points lies in
the range $0.041 \simlt \sigma_{APM} \simlt 0.051$ and that the lower
value is probably more appropriate.
This estimate is in  excellent agreement with the limits $0.041
\simlt \sigma_{APM} \simlt 0.054$ determined from our analysis of of
errors in Section 3.1.

In Figure 9, we plot the correlation function for the LCDRS CCD-APM
residuals, 
$$
\eqalignno{
& \langle \delta m_i \delta m_j \rangle - \sigma^2(m) \delta (\theta)
&(24) \cr}
$$
and the error bars are computed on the assumption that the $\delta
m_i$ are uncorrelated,
$$
\eqalignno{
{\rm var} \big [ \langle \delta m_i \delta m_j \rangle 
\big ] & = {\sigma^2( \delta m) \over N_{pairs}}  &(25)\cr}
$$
where $N_{pairs}$ is the number of distinct field pairs used in the
bin centred on angle $\theta$.  
The two sets of points show the correlation functions with and without
field $246$, but apart from the small difference at zero lag, the
exclusion of this field causes no significant change to the
correlation function.

%The solid line plotted in the figure shows the correlation function
%for the Monte Carlo calculations from Figure 14 of Paper II normalized
%to be $1.76\times 10^{-3}$ at zero lag, as expected for uncorrelated
%plate overlap errors $\langle \delta T_{ij} \delta T_{ik} \rangle =
%\sigma^2 \delta_{jk}$ with $\sigma = 0.057$ (see Section 3.1).
%% this really ought to be updated to have the better models  ???

The solid line plotted in the figure shows the
correlation function for the error model [A] in Section 3.1 normalized
to be $1.7\times 10^{-3}$ at zero lag, as required to match the
observed plate overlap residuals $\epsilon ^2 = 2 \times 10^{-3}$ (see
Section 3.1).
This error mode produces the largest error correlation function on
large scales, and so is the worst case for our measurements. 
For all of our error models, the correlation function of the
magnitude errors should be $\simlt 10^{-3}$ on scales $\theta \simgt
2^\circ$, and this is compatible with the absence of any large scale
correlations in the CCD-APM residuals.
Figure~9 can be compared to similar figures plotted in Paper II
showing the covariance of the zero-point correction factors $\langle
C^e_i C^e_j \rangle$ (Figure 16 of Paper II) and the correlation
function of the residuals between our own CCD zero points and APM
magnitudes (Figure 24 of Paper II).
In all cases, these correlation functions are consistent with zero to
an accuracy of $\simlt 1 \times 10^{-3}$ at $\theta > 4^\circ$.

In summary, the comparison between the LCDRS and APM magnitudes shows
that the rms error in APM plate zero-points is $0.041 \simlt
\sigma_{APM} \simlt 0.051$, and that the resulting error in $\w$ is
less than $1 \times 10^{-3}$ at $\theta > 4^\circ$.
These estimates are in excellent agreement with the limits between $0.041
\simlt \sigma_{APM} \simlt 0.054$ and the error correlation functions 
determined in Section 3.1.

\noindent
{\it 3.2.2 Comparison of APM CCD calibrations and APM magnitudes }

In Paper II we presented a comparison of APM magnitudes with CCD
magnitudes that we determined from observations with the 1m telescope
at the South African Astronomical Observatory.

Recently, Fong {\it et al.}  (hereafter F92) have used our CCD
sequences to argue that we have underestimated the errors in the APM
magnitudes and that the magnitude errors correlate on large-scales
leading to a roughly linear gradient in the magnitude scale with right
ascension.

From the APM and CCD magnitudes, we follow F92 and compute the
statistics
$$
\eqalignno{ \sigma^2_B & =  \sum_{i=1}^k n_i {(\overline c_i - 
\overline c)^2 \over \overline n (k-1)}, & (26a) \cr
\sigma^2_W & =  \sum_{i=1}^k \sum_{j=1}^{n_i}
 {(c_{ij} - \overline c_i)^2 \over \overline n (N-k)}, & (26b) \cr}
$$ 
where
$$
\eqalignno{ c_{ij} & =  (m_{APM})_{ij} + (b_J)_{ij}/1.097, & (27a) \cr
\overline c_i &= \sum_{j=1}^{n_i} {c_{ij} \over n_i},
\qquad \overline c = {1 \over N} \sum_{i=1}^k \sum_{j=1}^{n_i} c_{ij},
&(27b) \cr
\overline n &= {N \over k}, \qquad N = \sum_{i=1}^k n_i, &(27c) \cr}
$$ 
and there  are $n_i$ galaxies in each of $k$ sequences. 
The factor of $1.097$ in equation (27a) takes into account the small
deviation from unity of the slope in the relation between APM
magnitudes and CCD $b_J$ magnitudes (see Paper I, note that the APM
magnitudes before conversion to the $b_J$ system are defined to be
negative, hence the plus sign in equation 27a).
The excess variance caused by plate matching errors is 
$$
\eqalignno{ \sigma_{APM}^2 & =  \sigma^2_B - \sigma^2_W. & (28) \cr} 
$$ 
These statistics are almost identical to the statistics used in
Section 3.2.1 to measure the excess variance from the LCDRS CCD and
APM photometry, differing only in the weighting given to sequences
containing different numbers of galaxies.
Calculating these statistics to the sequences listed in Table 3 of
Paper II, we find 
$$
\eqalignno{
\sigma_B  & = 0.1167, & (29a) \cr
\sigma_W  &= 0.0857, & (29b) \cr
\sigma_{APM} &= 0.079. &(29c)\cr}
$$
The estimate of $\sigma_{APM}$ derived here is  slightly smaller
than the value $0.084$ found by F92.  
We have combined the multiple sequences in fields $344$ and $530$ in
computing these numbers.
The residuals $\overline c_i - \overline c$ and the scatter for each
sequence are given in Table 3.

\topinsert
\vbox{
\noindent
$\;$

\vskip 9 truein

\includegraphics{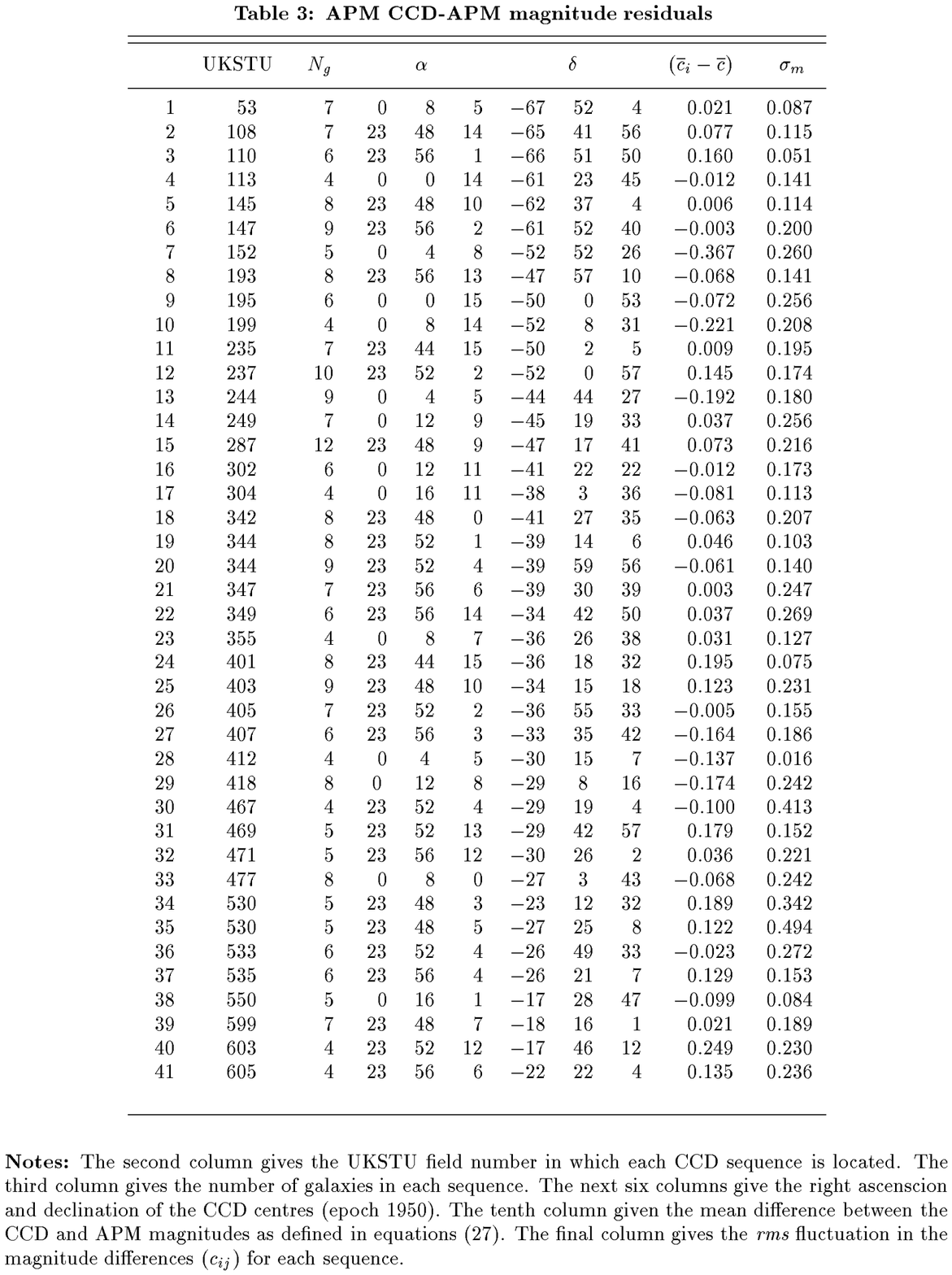}
\vskip 1 truein

}
\endinsert

As discussed in Paper II, the residual for the sequence on field $152$
is $-0.363$, differing from the mean by more than three standard
deviations.
Excluding field $152$, we find
$$
\eqalignno{
\sigma_B & = 0.1055, & (30a) \cr
\sigma_W  &= 0.0849, & (30b) \cr
\sigma_{APM} &= 0.063 \qquad ({\rm minus}\;\;{\rm field}\;\;152 ).
&(30c)\cr}
$$
F92 use numbers similar to those in equations (29) to argue that we
have underestimated the errors in the APM zero points, concluding that
the error is closer to $0.08$ magnitudes rather than the value
$0.04$--$0.05$ estimated in Paper II and Sections 3.1 and 3.2.1 of
this paper.
However, this is extremely unlikely for the following reasons:

\hangindent \parindent
[1] There can be little doubt that the CCD photometry in field $152$
is in error because the galaxy number density on the plate is {\it
underdense} by $11 \%$, whereas if the APM magnitudes in this field
were in error by $0.36$ magnitudes, we would expect the galaxy number
density in this field to be {\it high} by $\approx 40 \%$.
Consistency between the CCD zero point on this field and the observed
number counts would thus require that field $152$ is underdense by
more than $50 \%$, {\it i.e.} by about $3.5$ standard deviations,
which is implausible.
Figure 10 shows the magnitude residuals plotted against fluctuations
in the galaxy surface density to a magnitude limit $b_J=20$ in each
APM field. 
There is no trend between the magnitude residuals and the galaxy
density. 

\hangindent \parindent
[2] The errors on the zero point of each individual sequence are large
because the number of galaxies are small.  
The {\rms} deviation of the APM-CCD and APM magnitudes is $\sigma_m =
0.24$, {\it i.e.}  about half the dispersion measured in the previous
section for the LCDRS and APM magnitudes.
The smaller dispersion between the APM CCD and APM magnitudes arises
because the APM CCD sequences were taken in the $B$ and $V$ passbands.
The CCD magnitudes were transformed to the $b_J$ passband via a colour
equation (equation 33 of Paper II) and so the dispersion caused by the
mismatch between the CCD and APM passbands is much smaller that for
the R-band LCDRS photometry.
However, whereas there are typically $400$ galaxies in each LCDRS-APM
sequence, there are only $\approx 7$ in the APM CCD sequences.
The accuracy with which we can determine $\sigma_{APM}$ from the APM
calibrations is therefore relatively poor.
Figure 11 shows the distributions of $10000$ Monte Carlo computations
in which we have generated $39$ sequences with the same number of
galaxies per sequence as listed in Table 4.
Each magnitude has a Gaussian random error of $0.22$ magnitudes.
The solid and dotted histograms show the distribution of
$\sigma_{APM}$ measured from the statistics assuming that the zero
point errors are Gaussian distributed with dispersion $0.045$ and
$0.080$ magnitudes respectively.
These Monte Carlo calculations show that a measured value of
$\sigma_{APM}=0.063$ ({\it i.e.} the value determined excluding field
152) occurs with a probability of about $24\%$ if
$\sigma_{APM}=0.045$, and so is perfectly reasonable.

\hangindent \parindent
[3] If we include field F152, then the measured value of
$\sigma_{APM}=0.079$ excludes a true value of $\sigma_{APM}=0.045$ at
about the $93 \%$ confidence level.
However, we have shown that the outlying CCD sequence in field $152$
{\it must} be an error because the galaxy counts in this field are
close to the mean value for the whole survey (Figure 10).
Furthermore, high values of $\sigma_{APM}$ are inconsistent with the
results from the LCDRS CCD sequences.
For example, if the true value of $\sigma_{APM} = 0.080$, the
probability of measuring $\sigma_{APM} \le 0.051$ from the LCDRS
sequences is less than $0.64 \%$.
Fong \etal's assertion that $\sigma_{APM} \approx 0.08$ can thus be
excluded at a very high confidence level.

\hangindent \parindent
[4] Some of the scatter in the comparison with APM-CCD and APM
magnitudes might be caused by systematic errors in converting $B$ and
$V$ magnitudes to the $b_J$ system (see Paper II).
If we exclude galaxies with very blue colours, $B-V \ge 1.25$, then we
find $\sigma_{APM} =0.063$ if field $152$ is included and
$\sigma_{APM} =0.043$ if field $152$ is excluded.

F92 also point out that the APM CCD sequences indicate a large-scale
gradient correlated with right ascension.
We plot the magnitude residuals against $\alpha$ in Figure 12 and in
agreement with F92 we find some indications for a large-scale gradient in the
magnitude residuals. However, the error bars are large and so the
evidence for a linear gradient is of marginal statistical significance.
We make the following remarks concerning Figure 12:

\hangindent \parindent
[1] Even if we accept that the gradient in Figure 12 is caused by
large-scale correlated errors in the APM magnitudes, the correlation
function for these residuals (Figure 24 in Paper II) has a low
amplitude and would cause very little change to our estimates of
$w(\theta)$.

\hangindent \parindent
[2] The comparison of APM with LCDRS photometry as a function of right
ascension (Figure 7) shows no evidence for a large-scale gradient in
the APM magnitude residuals.
If there is a large-scale gradient in the APM photometry, it is
evidently {\it not} a simple gradient in right ascension.

\hangindent \parindent
[3] The lower panel in Figure 12 shows fluctuations in the mean galaxy
surface density to $b_J = 20$ plotted as a function of right ascension
for the plates containing APM CCD fields (open circles) and for all
$185$ fields (filled circles).
In each case, there is no evidence of a gradient in the galaxy counts.
In their discussion of gradients, F92 consistently ignored the
important point that our estimates of $w(\theta)$ are derived from the
{\it galaxy counts}.
It is evidently incorrect to infer that magnitude errors introduce
errors in $w(\theta)$ via large-scale gradients in the galaxy counts
if, in fact,  there are no such large-scale gradients in the counts.

\hangindent \parindent
[4] The absence of gradient in the galaxy counts could be caused by an
unlikely compensation between systematic variations in magnitude 
and star galaxy classification.
Large-scale gradients caused by classification errors are discussed in
more detail in Section 3.5.

\vskip 0.15 truein

\noindent
{\it 3.3 Tests for systematic errors correlated with scan and exposure dates}

\noindent
{\it 3.3.1 Scan dates} 

First we consider the plate scanning process.  The best check for errors
in this step is simply to scan the same plate more than once and compare
the image parameters measured from each scan.  For a few plates we 
compared scans made at different times during the survey and these showed
no evidence for significant variations in the measured image parameters.
Such a comparison is impractical for more than a few fields, but we have
tested for systematic trends in all fields by producing a scatter plot of
galaxy density for each Schmidt plate in order of APM scan.  As shown in
Figure~13, the galaxy density shows no systematic trend with scan date.

Furthermore, we can check directly that the scanning date has no
effect on the $\w$ measurements by evaluating $\w$ separately for
plate pairs with a scan interval $< 1$ year and $> 1$ year. 
The resulting estimates, together with the total $\w$ are shown in
Figure~14. 
The difference between the overall estimate (solid line) and the other
estimates (dotted and dashed lines), is $\sim 10^{-3}$.
From our lengthy discussion in Sections 3.1 and 3.2, we expect to see
random errors of order $10^{-3}$, and so this difference is not
statistically significant.  
We also evaluated $\w$ after simply discarding plates scanned during 
some periods. 
This produced virtually no change in $\w$. 
Thus it is unlikely that APM scan variations have seriously biased our
correlation functions. 

\noindent
{\it 3.3.2 Exposure dates }

Other possible sources of plate-to-plate scatter include variations in
emulsion sensitivity and observing conditions.  These would show up as
changes in the detection efficiency as a function of the exposure date
of each plate.  Figure 15 shows the galaxy surface density on each
plate plotted against its exposure date; no significant trend is seen
showing that there is no evidence for any long-term temporal dependence
in the completeness of the APM Survey.

The clustering of the galaxy distribution introduces a scatter of
$\approx 11\%$ in the number of galaxies per plate, so we we must
average over many plates to test for variations with smaller
amplitude.  An average of roughly 20 plates per year were taken for
the survey, and so it is impossible to check directly for $ \simlt
3\%$ variations over time-scales less than about a year.  However,
short term variations can be excluded as a significant source of error
on scales $5^\circ - 10^\circ$ by considering the distribution of
exposure dates across the sky as plotted in Figure 17 of Paper II.
Only 54 out of the 505 pairs of neighbouring plates were exposed
within a month of each other compared to 36 expected for a random
distribution.  The \rms of the unmatched plate zero points is only 0.2
magnitudes, so in the worst case, where the 18 excess plate pairs have
perfectly correlated errors of 0.2 magnitudes, the error in $\w$ would
be only $ (18/505) \times 0.2^2 \sim 1.4 \times 10^{-3} $.

\noindent
{\it 3.4 Plate-independent Errors} 

Obscuration by galactic and intergalactic dust, variable contamination
by stellar images and atmospheric absorption can introduce errors in
the survey which would continue smoothly across plate boundaries.
Such errors would not be corrected by the plate-matching procedure,
and would not show up in the tests carried out in the previous
sections.  In this subsection we present tests that show none of the
effects significantly bias our measurements of $\w$.

\noindent
{\it 3.4.1 Galactic Extinction }

Extinction by dust in our Galaxy introduces a gradient in galaxy
density with Galactic latitude, and the patchiness of the 
obscuration can produce
spurious clustering on smaller scales.  The extinction can be
estimated from the 100 micron emission as mapped by the IRAS
satellite.  Assuming that all the dust is at the same temperature, the
thermal emission will trace the column density of dust in any
particular direction, and hence it is possible to predict
the extinction at visual wavelengths.  This
has been studied by Rowan-Robinson et al (1991), who estimate a
constant to convert from 100 micron surface brightness to an
extinction value, $A_V = 0.06 I_{100}.$

In our equatorial survey covering the sky with $22^{hr} < \alpha <
4^{hrs} $ and $ -17.5^\circ < \delta < +2.5^\circ $, the extinction is
high enough that we can see an anti-correlation between the surface
density of galaxies and $I_{100}$ if no correction is applied.
Applying magnitude corrections from the IRAS map removes the overall
correlation between the galaxy density and $I_{100}$, but inspection of
the corrected galaxy maps reveals that there are several small patches
of high $I_{100}$ emission which are still underdense in galaxies.  It
is probable that these patches correspond to clouds of dust that are
colder than the mean dust temperature, leading to a higher extinction
per unit 100 micron flux.  In the region with $\delta < -17.5^\circ $ the
predicted extinction varies by less than 0.05 magnitudes, and there
are none of these anomalous patches, so the simple single temperature
model should
be accurate.

The points in Figure 16 show the  correlation function of
galaxies with magnitude $17<b'_{J} <20$ where $b'_J$ is the magnitudes
after correcting for obscuration using the $I_{100}$ estimate.  The
solid line plotted in Figure~16 is the same as in Figure~2a, and it
can be seen that the changes introduced by the correction are very
small.

We have estimated the covariance function of the obscuration by
making a map of the predicted extinction over the area that we use for
our correlation function estimates.  We then calculated $w_{obsc} =
<A_i A_j> $, where $A_i$ is the mean obscuration in cell $i$ of the
map.  This is plotted in Figure~17, and it corresponds to the error
introduced in the observed galaxy correlation function by the
obscuration.  It can be seen that the effect of Galactic extinction is
less than $ 10^{-3}$ on all angular scales.

\vfill\eject

\noindent
{\it 3.4.2 Atmospheric Extinction }

The UK Schmidt Telescope is situated at latitude $-31^\circ $ and the
southernmost plates in the survey are centred at $-70^\circ $.  The
survey plates were exposed as they cross the meridian, so the airmass
for the most southerly plates is 1.28.  Using $A_V=0.11 sec(z), \
A_B=0.27 sec(z) $ and $B_J = B - 0.28 (B-V) $, the predicted maximum
extinction is only $\sim 0.06$ magnitudes relative to the zenith.  In
fact all but 13 out of the 185 plates are centred at $\delta \ge
-60^\circ $ and thus have an airmass $\le 1.14$, so the atmospheric
extinction for most fields is smaller than $0.02$ magnitudes.  The
variation in galaxy density introduced by this effect is smaller than
the variations caused by real structure in the galaxy distribution.
This can be seen in Figure~18 which shows galaxy density as a function
of $\delta$ before and after correcting the magnitudes for atmospheric
extinction.

\noindent
{\it 3.4.3 Stellar Contamination Gradient }

The increase in the number of stars towards the Galactic plane leads
to a gradient in the number of blended stellar images which
contaminate the galaxy sample.  After applying the extinction
corrections described above, this Galactic gradient is seen as an
apparent increase in galaxy density towards lower latitudes.  In
Figure~19 we plot the apparent surface density of galaxies as a
function Galactic latitude, $b$, and this shows that the change in
mean density from the Galactic pole to $|b|=40^\circ$ is 10\%.  This
variation is consistent with our visual classification checks at
different Galactic latitudes, which showed the contamination is $\sim
5\%$ near the pole, and $\sim 15\% $ at $ |b|= 40^\circ $.  The change
is not a simple linear gradient: for $|b| \simgt 50 $ the density is
essentially constant, but there is a significant rise in the density
for $40<|b|<50$.  This reflects the rather sharp increase in stellar
density as we start to look through the Galactic bulge.  In the area
of sky used for our best correlation function estimates
($\vert b \vert \simgt 50^\circ$), there is
almost no gradient, even with no further correction.

The surface density of contaminating stellar blends, $N_{blend}$, is
approximately proportional to the square of the density of stars,
$N_{stars}$.  We can therefore use a map of stellar density to
estimate the number of contaminating images, and so calculate a
correction to the equivalent galaxy map.  We determine the relation
between the $N_{blend}$ and $N_{star}$ empirically by considering the
mean surface density of ``galaxies + contaminating blends'', $N_{tot}
= N_{gal}+N_{blend}$ as a function of $N_{stars}$.  Since there should
be no correlation between the number stars and the number of true
galaxies, any correlation must be due to contaminating images.  The
scatter plot of the apparent number of galaxies ($N_{tot}$) against
the number of stars is shown in Figure 20.  To produce this plot we
created $512\times 512 $ pixel maps of the number of galaxies and the
number of stars in $0.23 ^\circ \times 0.23 ^\circ$ cells, using an
equal area projection centred on the South Galactic Pole.  The large
scatter in the number of galaxies per cell is caused by the clustering of
the galaxy distribution.  The dashed line in the figure shows the mean
galaxy density as a function of stellar density, and as expected there
is an increase in $N_{tot}$ at higher values of $N_{star}$.  The solid
line in the figure is a quadratic fit to the mean of $N_{tot} $ as a
function of $N_{star}$.  We use this fit to estimate the change in
$N_{blend}$ as a function of $N_{star}$, and by subtracting this from
the raw galaxy map we produce a corrected galaxy map.  Figure 21 shows
the corrected density of galaxies as a function of Galactic latitude,
and there is no longer any significant gradient.

\noindent
{\it 3.4.4 The overall effect of extinction and gradient corrections. }

Figure 22 shows the angular correlation function of the final
map of galaxies with $ 17 < \bj < 20.0$, corrected for mean atmospheric and 
galactic extinction, and the contamination by blended stars. 
It is almost indistinguishable from the correlation function of the
uncorrected galaxy map, which is shown as the solid line in the
figure.  
This is not surprising, because the main effect of both
extinction and contaminating images, is to introduce a gradient in
number of galaxies as a function of galactic latitude. 
Since the gradients are of opposite sign and nearly equal
amplitude they almost exactly cancel. 

These effects are very small, and the uncertainty in the overall
correction is as large as the correction itself.  Therefore we feel
that the accuracy of the corrected $\w$ estimate is not significantly
better then the uncorrected estimate.  To keep our analysis as simple
as possible we decided to exclude the low latitude part of the survey
where these effects may be significant.  For the 120 high latitude
fields used in our best $\w$ estimates, the corrections are completely
negligible.

\noindent
{\it 3.5 Magnitude dependent errors } 

The discussion of errors in the previous sections are based mainly on
the galaxies to a limit of $\bj =20.0$, but since we also measure
clustering as a function of magnitude, we have investigated
magnitude-dependent errors.  Possible sources of magnitude-dependent
error include contamination by merged stars, incompleteness of the
galaxy sample, and emulsion saturation.

Table~4 shows various estimates of the plate-matching errors, measured
for each half-magnitude slice.  The variance of the overlap residuals
$\epsilon^2$ and the inter-intra-plate offset are model independent
ways to quantify the matching errors.  The values of $\sigma^2$,
$\beta_{ii}^2$ and $\beta_{ij}^2$ are derived quantities dependent on
assumptions concerning the propagation of errors across the
plate network. In Table 4 we
we have used the simple analytic model of equations (17) and (19).  The 
inter-intra-plate offset gives an estimate of errors that is 
independent of the overlap
residuals, but the relation to an overall error correlation
function depends on which error model is assumed, as described in
Section 3.1.

%\vbox{ \vskip 0.1in 
%Table xx. Magnitude dependence of error estimates. 

%\halign { \hfil # \hfil & \hfil # & \hfil # & \hfil # & \hfil # &
%\hfil # & \hfil # & \hfil # \cr

%mag range  & $\sigma^2$ from & $\beta_{ii}^2 $ & $ \beta_{ij}^2 $ &
%$\w$  & offset & 1 mag &  matching \cr
%           & overlaps        &                 &                  & 
% all & centre & all & centre \cr
%20.0 - 20.5 & 3.13 & 1.43 & 0.65 & 3.8 &  3.4 & 1.8 & 1.6 \cr
%19.5 - 20.0 & 4.14 & 1.89 & 0.86 & 1.6 &  0.9 & -0.4 & -0.5 \cr 
%18.9 - 19.5 & 3.96 & 1.80 & 0.82 & 3.1 &  4.0 & 0.4 & 0.9 \cr
%18.4 - 18.9 & 3.73 & 1.70 & 0.78 & 6.1 &  9.2 & 3.7 & 6.7 \cr
%17.9 - 18.4 & 2.34 & 1.07 & 0.46 & 6.0 & 10.7 & 5.6 & 12.4 \cr
%17.3 - 17.9 & 1.54 & 0.70 & 0.32 & 7.5 & 17.2 & 5.9 & 9.0 \cr 
%}
%}

\midinsert
$$
\vbox{
\halign { %\hfil # \hfil \tabskip 1em plus 2em minus 5em & 
\hfil # \hfil & \hfil # \hfil & \hfil # \hfil & \hfil # \hfil &
\hfil # \hfil &\tabskip=0pt \hfil # \hfil\cr
\multispan{6}{\hfil \bf Table 4: Magnitude dependence of 
photometric errors \hfil}\cr
\noalign{\medskip\hrule\vskip0.1truecm\hrule}
\noalign{\medskip}
 magnitude range  & $\epsilon^2$& $\sigma^2$ & $\beta_{ii}^2$ & $\beta_{ij}^2$
 & $\langle w_{\rm intra} - w_{\rm inter}\rangle$ \cr
\noalign{\medskip\hrule\medskip}
% &$17.1$ -- $17.7$&  $2.5 \times 10^{-3}$&  $1.14 \times 10^{-3}$
%&  $0.52 \times 10^{-3}$ &   $7.5 \times 10^{-4}$ \cr
% &$17.7$ -- $18.3$&  $2.6 \times 10^{-3}$&  $1.17 \times 10^{-3}$
%&  $0.53 \times 10^{-3}$ &   $6.0 \times 10^{-4}$ \cr
% &$18.3$ -- $18.9$&  $2.7 \times 10^{-3}$&  $1.23 \times 10^{-3}$
%&  $0.56 \times 10^{-3}$ &   $6.1 \times 10^{-4}$ \cr
% &$18.9$ -- $19.5$&  $2.4 \times 10^{-3}$&  $1.08 \times 10^{-3}$
%&  $0.49 \times 10^{-3}$ &   $3.1 \times 10^{-4}$ \cr
% &$19.5$ -- $20.0$&  $2.5 \times 10^{-3}$&  $1.16 \times 10^{-3}$
%&  $0.53 \times 10^{-3}$ &   $1.6 \times 10^{-4}$ \cr
% &$20.0$ -- $20.5$&  $2.5 \times 10^{-3}$&  $1.13 \times 10^{-3}$
%&  $0.52 \times 10^{-3}$ &   $3.8 \times 10^{-4}$ \cr
 $17.1$ -- $17.7$& $0.92 \times 10^{-3}$& $1.5 \times 10^{-3}$& $0.70\times 10^{-3}$ & $0.32 \times 10^{-3}$ & $7.5 \times 10^{-3}$ \cr
 $17.7$ -- $18.3$& $1.40 \times 10^{-3}$& $2.3 \times 10^{-3}$& $1.07\times 10^{-3}$ & $0.49 \times 10^{-3}$ & $6.0 \times 10^{-3}$ \cr
 $18.3$ -- $18.9$& $2.24 \times 10^{-3}$& $3.7 \times 10^{-3}$& $1.70\times 10^{-3}$ & $0.78 \times 10^{-3}$ & $6.1 \times 10^{-3}$ \cr
 $18.9$ -- $19.5$& $2.38 \times 10^{-3}$& $4.0 \times 10^{-3}$& $1.81\times 10^{-3}$ & $0.82 \times 10^{-3}$ & $3.1 \times 10^{-3}$ \cr
 $19.5$ -- $20.0$& $2.48 \times 10^{-3}$& $4.1 \times 10^{-3}$& $1.89 \times 10^{-3}$ & $0.86 \times 10^{-3}$ & $1.6 \times 10^{-3}$ \cr
 $20.0$ -- $20.5$& $1.88 \times 10^{-3}$& $3.1 \times 10^{-3}$& $1.43\times 10^{-3}$ & $0.65 \times 10^{-3}$ & $3.8 \times 10^{-3}$ \cr
\noalign{\medskip\hrule}
}
\vskip 0.1 truein

{ \font\frs=cmr9
\frs
\smallskip\noindent
{\bf Notes:} The second column gives $\epsilon^2$ estimated from the
plate-overlap residuals (equation 15).  
The third column gives $\sigma^2$ determined from this assuming the
overlap errors are uncorrelated (equation 14). 
The next two columns give estimates of $\beta^2_{ii}$ and
$\beta^2_{ij}$ from equations (17) and (19).
The final column gives the observed offset between the intra- and
inter-plate angular correlation functions measured for each magnitude
slice. }

}$$

\endinsert 

The residual errors in the overlaps for the magnitude matching are
essentially constant as a function of magnitude.  This is good
evidence that the magnitude errors do not vary as a function of
magnitude.  Since the overlap errors depend on the precision of the
field corrections for each plate, the constant value of the residuals
suggests that the field corrections are just as accurate at bright
magnitudes as at faint magnitudes.

The observed offsets between intra- and inter-plate estimates of $\w$
do increase as a function of magnitude.  As mentioned in Section
3.1.2, the cosmic variance in the estimated offset is large for a
realistic clustered galaxy distribution.  Nevertheless, it may be
considered worrying that the offsets for all 6 slices are positive at
about the $1 \sigma $ level.  We therefore investigated the dependence
of the offsets in the Soneira-Peebles simulations described in
Appendix A1 by selecting samples with different magnitude limits from
each realization.  The variance in offsets for brighter magnitude
limits increased roughly in proportion to the amplitude of the
correlation function.  Also, for a particular realization, the offsets
for different magnitude limits are highly correlated.  This might have been
expected, since features in the distribution that are correlated with
plate boundaries at one magnitude limit will be correlated at other
magnitude limits as well.  The observed APM offsets thus follow the
model predictions {\it ie.} they increase roughly in proportion to the
amplitude of $\w$, remaining about $1 \sigma$ positive for all
magnitudes ranges brighter than $\bj=20$.

The exception is the $20.0 < \bj < 20.5 $ slice, where the offset is $
3.8 \times 10^{-3}$; much larger than expected from the magnitude
errors and the cosmic variance.  The reason for this is that
star-galaxy classification breaks down near the magnitude limit of the
survey.  The completeness is fairly constant at 90\% for samples
brighter than $\bj = 20.3$, but rapidly drops for galaxies fainter
than this, reaching a mean of 75\% for galaxies with $ 20.4< \bj
<20.5$.  The incompleteness becomes significant first in the vignetted
corners of each field, producing errors like the mode (C) simulations
described  in Section 3.1.2, which lead to a large offset.  Nevertheless, even
for this worst case the errors are not large enough to hide the break
in $\w$.

\noindent
{\it 3.6 Best estimates for $\w$ } 

We correct for the residual matching errors by simply subtracting our
estimated error correlation function (plotted as the solid line in Figure~9) 
from the raw $\w$ measurements. 
Except for the faintest slice, the matching errors are very nearly
constant as a function of magnitude, so we use the same amplitude for
the errors. 
For the faintest slice the variations in completeness become significant,
so the errors from the magnitude matching do not represent the full
errors. 
Since the offset for this slice is larger than the cosmic variance
introduced by clustering, we can use the observed offset to estimate
the overall matching errors. 
We therefore scaled the error correlation function to match the
observed offset, and subtracted this  from the
measured $\w$. 
To account for the contaminating fraction of stellar images, each
slice has also been multiplied by a factor $1/(1-f)^2$, where $f$
is the fraction of stellar image misclassified as galaxies and
varies from 4\% for the brighter magnitudes, to 15\% at the faintest
limit. This correction has a significant effect on the 
amplitude of the correlation functions but does not change
their shape. 
As mentioned in Section 3.4.4, the corrections for extinction and
merger gradients are completely negligible for the high latitude
fields that we use. 

Figure 23 shows the final error-corrected correlation functions.  At
first sight they are not very different to those shown in Figure 3, but
careful inspection shows that the results for faintest slice no longer
cross the next brighter slice.  This is the only significant change
introduced by the exhaustive error analysis of this Section.

The various tests presented in this Section do not reveal any additive
source of error which can increase the observed $\w$ by more than $1-2
\times 10^{-3}$ at $\theta = 0$.
Some effects are likely to introduce correlated errors, but
these rapidly decrease at larger scales, with an expected amplitude 
of $\sim 5\times 10^{-4} $ on scales $\theta \sim 5^\circ$, and
$\sim 2\times 10^{-4}$ at $\theta \sim 10^\circ$.
Over most of the angular scales we consider, this means that the observed $\w$
is more than 100 times the amplitude of our estimated errors.  
Even at the largest scales, where our results have fallen below the
power law extrapolation, the amplitude of $\w$ for the different
slices varies between $ 4 - 11 \times10^{-3}$.
This means that the estimated signal to noise at the break angle (see
Section 4.1.3 for our definition of break angle), varies from about 5
for the faintest magnitude slice, to 40 for the brightest magnitude
slice.

We can be very confident that we have not significantly 
{\it underestimated} $\w$. 
It is extremely unlikely that there are errors that have avoided
detection in our tests and that have conspired to eliminate intrinsic
galaxy correlations. 
Also, for these estimates of $\w$ we have not applied any filtering to
remove clustering. 
In Section 4.2.3 we consider the effect of explicitly removing smooth
fitting functions and conclude that these filtered measurements are
much poorer estimates of the real correlation function.

%Also, an upper limit to the density contrast of $500 \hmpc$ structures
%is given by the variation in galaxy density between 4 equal zones of
%the faint magnitude map. 
%The \rms density variation is only $1.3 \%$, which is by far the
%strongest existing constraint on  $\sim 500 \hmpc$ scales. 
%Power on scales $\simgt 1000 \hmpc$ would be undetectable by our survey. 
%For such scales to produce a contribution to $\w$ of $2 \times
%10^{-3}$ would require $4.5\%$ density fluctuations, which would
%probably generate excessive microwave background anisotropies and
%large-scale motions. 

%In the next Section we compare our best estimates of $\w$ to earlier work. 

\def\simlt{\mathrel{\spose{\lower 3pt\hbox{$\mathchar"218$}}
     \raise 2.0pt\hbox{$\mathchar"13C$}}}
\def\simgt{\mathrel{\spose{\lower 3pt\hbox{$\mathchar"218$}}
     \raise 2.0pt\hbox{$\mathchar"13E$}}}

\baselineskip=13pt
\parskip=9pt

\vskip 0.15 truein

\noindent
{\bf 4. The Scaling Test and Comparisons with other Catalogues } 

\vskip 0.1 truein

In this section we first test if our best estimates of $\w$ scale with
magnitude as predicted by Limber's equation. 
This is an important test that must be satisfied if our measurements
really represent the projected clustering of the three-dimensional
galaxy distribution.
We then use the scaling relations to compare our measurements to the
Lick catalogue, and other more recent surveys.
Although there are apparent discrepancies between $\w$ from some of
these surveys, we find that the measurements are consistent to within
the errors associated with each survey.

\noindent
{\it 4.1 Limber's equation and the scaling test} 

\noindent
{\it 4.1.1 Limber's equation}

As mentioned in Section 2.4, the variation of $\w$ as a function of
magnitude allows us to check that the angular correlation functions
measure intrinsic clustering rather than systematic errors. 
If $\w$ is determined by the projection of the true spatial
correlation function, $\xi(\br)$, rather than by systematic errors
in the catalogue, the shape of $\w$ as a  function of magnitude limit 
can be calculated from the relativistic version of Limber's equation (GP77,
Phillipps \etal 1978, Peebles 1980 \S56), 
$$ \eqalignno{ \w &= 
{\int_0^\infty\int_0^\infty  r_1^2 r_2^2  p(r_1) p(r_2) 
\xi(r_{12}, z)\;dr_1 dr_2
\over \left [\int_0^\infty r^2 p(r)\;dr \right]^2  },
&(31)} $$
where $r_{12} = \vert {\bf r_1 - r_2} \vert$ and
$p(r)$ is the selection function, which gives the fraction of
galaxies at proper distance $r$ in the magnitude limited sample. 

If we assume that $\xi(r, z)$ is a power law  in $r$ and redshift $z$,
$$ \eqalignno{ 
\xi(r, z)  &=  \left({ r_0 \over r} \right )^\gamma (1 + z)^{-(3 + \epsilon)}
&(32)} $$
and we adopt the small angle approximation $r_1 \approx r_2 \gg
r_{12}$, equation (31) can be rewritten as
$$ \eqalignno{ \w &= 
\sqrt \pi
{\Gamma [(\gamma - 1)/2] \over \Gamma( \gamma/2)}
{ B \over  \theta^{\gamma-1} } r_0^\gamma, &(33a)}$$
where
$$
\eqalignno{ B &= 
{\int_0^\infty x^{5-\gamma} a^6 p^2(x) (1 + z)^{(\gamma - 3 -
\epsilon)} F(x)^{-1} \;dx
\over \left [\int_0^\infty x^2 a^3 p(x) F(x)^{-1}\;dx \right]^2  }.
&(33b)}$$
In equation (33b) $a$ is the cosmological scale factor normalized to unity at
the present epoch, $x$ is the coordinate distance at redshift $z$,
$$
\eqalignno{ x &= (2c/H_0) [(\Omega_0 - 2)(1 + \Omega_0 z)^{1/2}
+ 2 - \Omega_0 + \Omega_0 z]/[\Omega_0^2 (1 + z)], &(34a)}$$
for a Friedman-Robertson-Walker universe with zero cosmological
constant and present day density parameter $\Omega_0$
(see Peebles 1980, \S56); the factor  
$$
\eqalignno{ F(x) &= [1 - (H_0 x/c)^2 (\Omega_0 - 1)]^{1/2},
&(34b)}$$
comes from the metric and is equal to unity for a spatially
flat universe.

Equation (33b) can be written in terms of the redshift distribution
of galaxies in the sample, $dN/dz$,
$$
\eqalignno{ B &= 
{\int_0^\infty x^{1-\gamma} \left ( {dN \over dz} \right )^2
F(x) (1 + z)^{(\gamma - 3 -
\epsilon)} \left ({ dz \over dx } \right )  \;dz
\over \left [\int_0^\infty \left ( {dN \over dz } \right )\;dz  \right]^2  },
&(35a)}$$
(Efstathiou \etal 1991), since $dN/dz$ is related to the 
selection function by
$$
\eqalignno{ \left ( {dN \over dz} \right) dz &\propto
 x^2 a^3 {p(x) \over F(x)} \left ({dx \over dz} \right ) dz.
&(35b)}$$
Thus, we can evaluate equations (33b) and (35b) if we have a
model of the selection function $p(x)$ derived {\it e.g.}
from the galaxy luminosity function, or directly
from the redshift distribution of a representative subset 
of APM galaxies to a specified magnitude limit. Models
for the selection function will be described in the next subsection.

The model for the evolution of the spatial correlation function 
(equation 32) is simplistic and may well not apply 
over a wide range of redshifts and spatial scales 
(see Efstathiou \etal 1991, Efstathiou 1995b). The value
$\epsilon = 0$ corresponds to stable clustering in 
physical coordinates, which may be applicable at small
separations where $\xi \gg 1$ and most galaxy pairs are
in bound and stable systems (Peebles 1980 \S 73); the value
$\epsilon = -1.3$ corresponds to a constant amplitude of
$\xi$ in comoving coordinates for a power law
correlation function with slope $\gamma = 1.7$.
 In Section 4.1.3 we will
illustrate the sensitivity of our results to uncertainties 
in the evolution of $\xi$ and to cosmological parameters.

\vskip 0.1 truein

\noindent
{\it 4.1.2 Model for the selection function}

In Maddox \etal (1990c) we constructed  a model for the selection
function $p(x)$ derived from the galaxy luminosity function. We
reproduce the model here for reference. The selection function
for a magnitude slice $m_{min} \le b_J \le m_{max}$ is given by
$$
\eqalignno{  a^3 p(x) &\propto \int _{\ell_{min}(x)} ^{\ell_{max}(x)} 
\varphi(\ell, z) d\ell,
&(36a) }$$
where $\phi (\ell,z)\; d\ell$ is the galaxy luminosity function
at redshift $z$ and 
$$\eqalignno{ 
\ell_{min/max}(x) &= {\rm dex} 
\{ 0.4(M_\bj^*(z) - (m_{max/min} -25 -5 \log[ x(1+z) ] 
))\}, &(36b) }$$
and $M^*_\bj$ is the characteristic absolute magnitude of 
the Schechter (1976) luminosity function at redshift $z$.
The model of $\phi(\ell, z)$ adopted by Maddox \etal (1990c) is
$$
\eqalignno{ \varphi(\ell, z)\;d\ell &= \varphi^*\ell^{\alpha}{\rm
exp}(-\ell)
\;d\ell, \quad
\ell \equiv 10^{0.4(M^*_\bj - M)},  &(37) \cr
M^*_\bj(z) &= M^*_0 + M^*_1z, \quad \alpha(z) = \alpha_0 + \alpha_1z ,\cr
\varphi^* &= 1.3 \times 10^{-2} {\rm h}^{3} {\rm Mpc}^{-3}, \cr
M^*_0 &= -19.8 + 5\log(h), \;\; M^*_1 = 1, 
\quad \alpha_0 = -1, \;\; \alpha_1 = -2.\cr
}
$$
Note that equation (3) of Maddox \etal (1990c) contains a misprint in
the definition of $\ell$.  This parametric form is intended to model 
the evolution of the galaxy luminosity function 
(including the $k$-correction) and to be consistent with the 
following observations: (a) The low redshift field galaxy 
luminosity function
(Efstathiou \etal 1988, Loveday \etal 1992a);
(b) The galaxy number counts from the APM Survey (Maddox \etal
1990d) and from deeper CCD surveys ({\it e.g.} Tyson 1988, Cowie
\etal 1988, Metcalfe \etal 1995); 
(c) The redshift distribution in the deep spectroscopic
survey of Broadhurst \etal (1988). (d)
Constraints on the evolution of the luminosity function determined from
the Broadhurst \etal survey (see Maddox 1988, and Gaztanaga 1995). 
Note that the normalization of the luminosity function, $\phi^*$,
cancels in Limber's equation.

There has been a large increase in the number of faint galaxy
redshifts in the last few years and so it has become possible
to constrain the selection function of the APM Survey
by direct comparison with the redshift distribution of 
magnitude limited samples. R.S. Ellis has kindly made available
to us a list of $697$ galaxy redshifts in $21$ small fields. 
These fields have magnitude limits that lie approximately within one
of the three ranges
$17 < B < 19.7$,  $19.7 < B < 20.3$ and $20.3 < B < 21.5$.
The fields include the redshifts measured by Broadhurst \etal
(1988), Colless \etal (1990, 1993) and unpublished redshifts.
The sample is a subset of the data used by Cole \etal (1994a)
to study the spatial clustering of faint galaxies. We correlated
the galaxy positions on each field with the APM Survey and computed
a mean  magnitude offset to convert from the 
$B$ magnitudes used in the redshift survey into the APM $b_J$
system. We then used the completeness factors, sampling rates
and areas to weight each field and so computed a redshift
distribution within a specified APM magnitude slice. The
results are shown in Figure 24 for the magnitude slices
$17 \le b_J \le 19$, $17 \le b_J \le 20$ and $17 \le b_J \le 21$.
In each case, the histograms have been normalized to the 
actual ({\it i.e.} unweighted) number of galaxies $N_g$ 
in each magnitude slice, as listed in the Figure. 

The dashed lines in Figure 24 show the redshift distributions
computed from the Maddox \etal (1990c) model of equation (37).
The median redshifts predicted by this model are slightly too high
and the model predicts more high redshift galaxies than are observed.
The solid lines in Figure 24 show redshift distributions in
magnitude slices computed from the distribution
$$
\eqalignno{ \left ( { dN \over dz} \right ) dz
&\propto {10^{0.45m} \over z_c^3} z^2 
{\rm exp} \left [ - \left( {z \over z_c(b_J)}
\right)^{3/2} \right ] \;dz,  &(38a) \cr}
$$
where the median redshift is given by
$$
\eqalignno{
z_m(b_J) &= 0.016(b_J - 17)^{1.5} + 0.046 \qquad b_J \ge 17 &(38b) \cr
z_m(b_J) &= 1.412 z_c. &(38c)
}
$$
This model was introduced by Baugh \& Efstathiou (1993) and the
specific parameters were chosen to match the redshift distributions of
the $b_J = 17.15$ limited Stromlo/APM redshift survey (Loveday \etal
1992a,b) and of the Broadhurst \etal (1980) and Colless (1990, 1993)
surveys, which are a subset of the data plotted in Figure 24.
As Figure 24 shows, equation (38) provides an excellent match to the
empirically determined redshift distributions. In the next subsection,
we will evaluate Limber's equation using the models of equations (37)
and (38) to illustrate the sensitivity of the results to errors in the
model for the selection function.
However, it is worth emphasizing that the predicted amplitude of $\w$
depends mainly on the median redshift of the sample, which is well
determined from the redshift surveys rather than on the precise shape
of the redshift distribution.

\vskip 0.1 truein

\noindent
{\it 4.1.3 Application of the scaling test} 

In this subsection we evaluate Limber's equation to scale the angular
correlation functions for the six magnitude slices plotted in
Figure~23 to a common depth.
As a reference depth, we have chosen to scale the angular correlation
functions to a magnitude limit of $b_J = 18.4$.  
At this magnitude limit, the galaxy counts in the APM Survey are about
equal to those in the Lick survey, thus our scaling will transform
each of the $\w$ estimates in Figure~23 to about the same depth as the
Lick catalogue.

For each magnitude slice, we compute the shifts $\delta {\rm log}
\theta$, $\delta {\rm log}\; \w$ required to transform the angular
correlation functions to the Lick depth, assuming a two power model
for $\xi(r)$ with a slope $\gamma_1 = 1.7$ at small separations
(equation 12) and a slope $\gamma_2 = 3.0$ at large separations (see
GP77, Figure~15, for an explanation of how these shifts are computed
in the two power-law model).
The results are listed in Table 5. 
The scaling factors are relatively insensitive to the evolution
parameter $\epsilon$ in equation (32) and to the cosmological model,
as expected at the low redshifts of the APM survey.
The scaling factors are more sensitive to the model for the redshift
distribution, though even for redshift distributions as different as
the two models adopted here, the scaling factors do not differ by more
than $0.03$ in the log.
In the rest of this subsection, we use the scaling factors in the
fourth and fifth columns of Table 5 based on equation (38),  since this
provides a good fit to the observed redshift distributions (Figure~24),
and assuming $\epsilon = 0$ and $\Omega = 1$,

Figure~25 shows the six $\w$ estimates from Figure~23 scaled to the
the Lick depth.  The agreement between these curves is excellent.  The
rms scatter in the mean of the 6 slices sample using 80
logarithmically-spaced bins in $\theta$ is only 4.5\%.  The agreement
between the six magnitude slices is better than plotted in Figure~2b
of Maddox \etal (1990c) for two reasons.  Firstly, the model of the
redshift distribution (38) is more accurate than the model of equation
(37) used by Maddox \etal (1990c), and this improves the accuracy of
the scaling correction.  Secondly, the corrected $\w$ for the faintest
magnitude slice has a lower amplitude at large angles ({\it c.f}
Figures~3 and 23) and this again improves the agreement of the scaling
test.  The improvement in the accuracy of the scaling test
supports our error analysis of the previous
sections which indicates that systematic errors are significant only
close to the magnitude limit of the survey.  The scaling test thus
provides strong evidence that our angular correlation functions are
measuring real clustering in the galaxy distribution.

$$
\vbox{
\halign { \hfil # \hfil \tabskip 1em plus 2em minus 5em
& \hfil # \hfil& \hfil # \hfil & \hfil # \hfil & \hfil # \hfil 
& \hfil # \hfil & \hfil # \hfil & \hfil # \hfil 
&\tabskip=0pt \hfil # \hfil\cr
\multispan{9}{\hfil \bf Table 5: Scaling factors to Lick depth \hfil}\cr
\noalign{\medskip\hrule\vskip0.1truecm\hrule}
\noalign{\medskip}
   &  \multispan{2}{\hfil (equation 37)\hfil} &\multispan{6}{\hfil
(equation 38)\hfil} \cr
   &  \multispan{2}{\hfil ($\epsilon=0,\;\;\Omega=1$) \hfil}
&\multispan{2}{ \hfil ($\epsilon=0,\;\;\Omega=1$) \hfil}
 &\multispan{2}{\hfil ($\epsilon=-1.3,\;\;\Omega=1$) \hfil}
&\multispan{2}{\hfil ($\epsilon=0,\;\;\Omega=0$) \hfil} \cr
 mag range  & $\delta {\rm log}\theta$ & $\delta{\rm log}\;w$ &
 $\delta {\rm log}\theta$ & $\delta{\rm log}\;w$ &
 $\delta {\rm log}\theta$ & $\delta{\rm log}\;w$ &
 $\delta {\rm log}\theta$ & $\delta{\rm log}\;w$ \cr
\noalign{\medskip\hrule\medskip}
 $17.1$ -- $17.7$&  $\;\;0.038$ & $-0.117$ & $\;\;0.015$ & $-0.114$ &
$\;\;0.013$ & $-0.112$ & $\;\;0.016$ & $-0.116$ \cr
 $17.7$ -- $18.3$&  $\;\;0.125$ & $-0.021$ & $\;\;0.143$ & $-0.046$ &
$\;\;0.141$ & $-0.054$ & $\;\;0.147$ & $-0.043$ \cr
 $18.3$ -- $18.9$&  $\;\;0.205$ & $\;\;0.071$ & $\;\;0.230$ & $\;\;0.057$ &
$\;\;0.230$ & $\;\;0.037$ & $\;\;0.237$ & $\;\;0.064$ \cr
 $18.9$ -- $19.5$&  $\;\;0.280$ & $\;\;0.159$ & $\;\;0.303$ & $\;\;0.164$ &
$\;\;0.303$ & $0.132$ & $\;\;0.311$ & $\;\;0.176$ \cr
 $19.5$ -- $20.0$&  $\;\;0.341$ & $\;\;0.233$ & $\;\;0.358$ & $\;\;0.254$ &
$\;\;0.359$ & $0.210$ & $\;\;0.367$ & $\;\;0.272$ \cr
 $20.0$ -- $20.5$&  $\;\;0.393$ & $\;\;0.299$ & $\;\;0.402$ & $\;\;0.333$ &
$\;\;0.403$ & $\;\;0.276$ & $\;\;0.412$ & $\;\;0.355$ \cr
\noalign{\medskip\hrule}
}}$$

Another way of demonstrating the accuracy of the scaling relation is
shown in Figures~26a and 26b. For each of the magnitude  slices we have fitted
a power law $\w = A \theta^{1-\gamma}$ over the angular range $ 0.01
^\circ \le \theta \le 1^\circ$, with the slope $\gamma = 1.7$. 
The resulting amplitudes are plotted against the corresponding
magnitude limit in Figure~26a.
The line in Figure~26a shows the predicted variation from the model
of equation (38) with $\epsilon = 0$ and $\Omega=1$ assuming a slope
$\gamma=1.7$ and normalizing the amplitude of $\xi$ 
to minimize the {\it rms} residuals about the line.
The amplitude of the faintest slice is about 8\% lower than predicted
by the model. 
As explained in Section 3.6, the overall normalization of $\w$ for the
faintest slice is uncertain because of residual contamination by
stars, but this is unlikely to introduce errors larger than 5\%.
Thus the discrepancy in the scaling may be marginally significant and
may indicate the start of a decline away from the model predictions
that has been found from deeper photographic and CCD surveys
(Efstathiou \etal 1991; Neuschaefer, Windhorst and Dressler 1991,
Pritchet and Infante 1992, Roche \etal 1993,  Neuschaefer and Windhorst 1995).

In Figure~26b we plot the characteristic break angle $\theta_{break}$
as a function of limiting magnitude against the predictions of the two
power-law model.  We define the break angle as the angle at which the
measured $\w$ has fallen to half the value expected by extrapolating
the power law fitted at smaller angular scales as described above.
The break angles decreases at fainter magnitudes in 
good agreement with the model predictions.  This provides strong
evidence that the break from a power-law is a real feature of the
galaxy distribution.  Figure~26b contradicts the claim by F92 that the
scaling of $\w$ measured from the APM survey is well approximated by a
shift only in amplitude.  It shows that a combination of shifts in
amplitude and angle is seen, as expected if the break in $\w$ is
associated with a physical feature in the shape of $\xi(r)$.

It is also interesting to compute the amplitude of $\xi(r)$
implied by Limber's equation. Using the redshift distribution
of equation (38) and assuming a power law $\gamma = 1.7$,
the parameters of the angular correlation function for
the magnitude slice $17 \le b_J \le 20$ of equation (12)
give a correlation length $r_0$ (equation 32) of
$$
\eqalignno{  r_0 & = 4.81 \pm 0.28 \; \hmpc \qquad (\epsilon = 0,\; 
\Omega_0=1) &(39a) \cr
r_0 & = 4.43 \pm 0.26 \; \hmpc \qquad (\epsilon = -1.3,\; 
\Omega_0=1) &(39b) \cr
r_0 & = 5.05 \pm 0.29 \; \hmpc \qquad (\epsilon = 0,\; 
\Omega_0=0) &(39c) \cr
}
$$
These numbers are in good agreement with the correlation length
of $r_0 \approx 4.7 \hmpc$ determined by GP77 from the Lick 
survey, with the correlation lengths of $\sim 5 \hmpc$ determined
from the CfA-2 (Park \etal 1994) and Stromlo-APM (Loveday \etal 
1995) redshift surveys (though redshift space distortions, Kaiser
1987, complicate the comparison with redshift surveys as described
in Section 5 below). 
Notice that the statistical uncertainty in the amplitude of $\xi(r)$
implied by equation (12) corresponds to only $6$\% in $r_0$ and is 
smaller than the 
error of about $15$\% arising from uncertainties in the evolution of
$\xi$ and the cosmological model, which are probably slightly larger
than the residual errors arising from uncertainties in the shape of
the redshift distribution.

\vskip 0.15 truein

\noindent
{\it 4.2 Comparison with $\w$ from other  surveys } 

\noindent
{\it 4.2.1 Small area surveys}

Several groups have used small area surveys to estimate $\w$.
Measurements of small numbers of Schmidt and 4m plates have 
produced galaxy surveys covering a few hundred square degrees to a
limit $\bj\sim 20$ and a few square degrees to a limit of $\bj\sim 23$
(Sebok 1986, Pritchet and Infante 1986, Koo and Szalay 1984, Stevenson
\etal 1985, Hewett 1982, Shanks \etal 1980). 
The results from these surveys show power-law behaviour in $\w$
on small angular scales, with a sharp break at larger angles. 
The power-law slope of the estimated correlations functions are
fairly consistent, varying between $\sim -0.7$ and $-0.8$. 
The corresponding amplitudes of $\w$, however, show a large scatter. 
Cosmic variance alone will introduce a large scatter in the amplitude
of $\w$ determined from a small number of Schmidt plates.
This can be seen from the simulations described in Appendix A and from
the scatter in the single-plate $\w$ estimates from the APM Survey
described in Section 2.2.

The break in $\w$ from a power law seen in all these surveys is caused
by the integral constraint in the estimator of $\w$, as discussed in
Section 2.2.
This effect also explains why the slopes of power law fits to $\w$
from small samples are slightly steeper than the fits to our $\w$
measurements described in Section 2.2.
One of the most important results from our analysis of the APM Survey
is the demonstration that large areas of sky are required to determine
$\w$ reliably on angular scales of a few degrees. 
This point is discussed further in the next subsection.

\noindent
{\it 4.2.2 Comparison with the Edinburgh-Durham catalogue}

As mentioned in the introduction, the Edinburgh-Durham Southern Galaxy
Catalogue (EDSGC) is the largest machine constructed catalogue of
comparable size to the APM Galaxy Survey.  It is based on the same
photographic plate material as the APM Survey scanned with the COSMOS
measuring machine, but comprises only $60$ plates centred on the South
Galactic Pole, {\it i.e.} about one-third of the area of the APM
Galaxy Survey.  The EDSGC is described by Heydon-Dumbleton \etal
(1989).  Estimates of $\w$ from the EDSGC have been presented by
Collins \etal (1989, 1992), and an analysis of errors in 
the $\w$ estimates from the EDSGC is described by Nichol \& Collins (1993).

In Figure~27, we compare the EDSGC estimates of $\w$ at a magnitude
limit of $b_J = 19.5$ with the APM estimates for the magnitude slice
$17 < b_J < 20$.  We have corrected for the small difference in
limiting magnitudes by scaling the EDSGC results to the APM depth
using the model of equation (38) with $\epsilon=0$ and $\Omega_0=1$.
The open circles in Figures~27 show the uncorrected estimates of $\w$
for the EDSGC from Figure~2 of Collins \etal (1992).  The solid points
in Figure~27a are the means of the APM $\w$ estimates from 4
approximately equal area zones in the survey.  Notice that the open
circles lie above the APM estimates at $\theta
\simgt 1^\circ$. 
The discrepancy between these estimates is probably caused by the
larger calibration errors of the EDSGC.

Collins \etal (1992) show that the dispersion in the plate overlap
residuals in the EDSGC is about $\epsilon = 0.08$ magnitudes.
The EDSGC matching procedure is not based wholly on plate
overlaps, and so our analysis of Section 3.1 cannot strictly be used
to relate the overlap residuals to a plate zero-point error. 
A reasonable approximation for the EDSGC is that $\beta_{ii}^2 \sim
0.5 \epsilon^2 = 3.2\times 10^{-3}$, {\it i.e.} about twice the value
measured for the APM Survey.
Larger calibration errors in the EDSGC are plausible because they use
a less accurate technique for correcting for plate field effects, and
because they use CCD sequences with rather few galaxies to determine the
the plate correction factors for about half the survey, rather than
using the plate overlaps.

Nichol and Collins (1993) have analyzed the effects of plate matching
errors in the EDSGC survey using simulations.
Their simulations do not include the effects of the propogation of
errors, which we discussed in detail in Section 3,
since they make extensive use of calibrating photometry and this should
limit the large-scale propogation of errors. 
As seen in Figure~3 of Nichol and Collins (1993), the resulting error
correlation function has a high amplitude on scales less than the size
of a plate, $\theta \simlt 5^\circ$, but the errors do not propagate
to larger scales.  
The dotted lines in Figure~27 show the EDSGC $\w$ after correcting
for plate errors and for the effects of variable Galactic obscuration
(although plate matching errors dominate the correction).
The dotted line is in much closer agreement with the APM points, though
it lies slightly higher an angular scales greater than $\sim
2^\circ$.

%-------------
%There is something funny going on with their error corrections. 
%If they have really done what they say, the error correction should
%have virtually no effect at $10^\circ$, but the difference seems to
%be about $5\times 10^{-3}$. 
%-------------

Figure~27b shows the individual estimates of $\w$ for each of the $4$
roughly equal area 
zones of the APM Survey plotted together with the uncorrected and
corrected EDSGC $\w$.
Each of the $4$ APM zones has about the same area as the EDSGC, hence
Figure~27b gives a good indication of the sampling errors for a
catalogue of the size of the EDSGC.
The $\w$ estimates for the $4$ zones show a large scatter, especially
at $\theta \simgt 4^\circ$, showing that a large area of sky is
required to determine $\w$ accurately at amplitudes $\simlt
10^{-2}$. 
The corrected EDSGC estimates of $\w$ lie within the scatter of the $4$
APM zones, hence we conclude that the corrected EDSGC estimates of
$\w$ are consistent with ours within the sampling fluctuations.
Since the EDSGC catalogue is based on the same photographic material
as the APM Survey, it is possible to make a detailed image-by-image
comparison of the two surveys.
Such an analysis is in progress and will be described in a future
paper.

\noindent
{\it 4.2.3 Comparison with the Lick survey and the effects of
filtering the APM catalogue}

The Lick survey samples a large enough area of sky that sampling
fluctuations and biases in the estimates of $\w$ caused by the
integral constraint should be relatively small.  Thus $\w$ from the
Lick survey should be comparable to the estimates of $\w$ derived from
the APM Survey. Figure~28 shows the GP77 estimates of $\w$ for the
Lick catalogue compared to the APM $\w$ estimates for the magnitude
slice $17 \le b_J \le 20$. The APM $\w$ has been scaled to the Lick
depth as described in Section 4.1.2. Figure~28a shows the GP77
estimates derived from four separate zones of the Lick catalogue which
were filtered to remove large scale gradients in the galaxy
counts (see GP77, Section III).  The Lick $\w$ is in excellent
agreement with the APM results on angular scales less than $\theta
\approx 3^\circ $.  On larger angular scales our results disagree.
GP77 find that $\w$ breaks sharply from a power law on scales of $\sim
3^\circ $, whereas we find a more gentle decline.

The most likely reason for this discrepancy is that the GP77 filtering
of the Lick map has removed a component of intrinsic clustering. In
Figure~28b we compare the APM $\w$ to the unfiltered $\w$ from the
Lick survey from Figure~2 of GP77. The results from the unfiltered
Lick map lie slightly higher than the APM $\w$ on large angular
scales. Evidently, filtering has a large effect on the Lick angular
correlation function and could 
account for the discrepancy with the APM $\w$.

Seldner \etal (1977) find that the mean square error in the fractional
galaxy count in a single plate overlap after applying the Lick plate
correction factors is $\sigma^2 = 0.025$, {\it i.e} about eight times
larger than for the APM Survey (equation 16).  As described by GP86
and in Section 3.1, these plate matching errors propagate through the
network of plates and introduce large-scale gradients in the corrected
Lick counts. GP86 estimate that the Lick plate matching errors lead to a
variance in the counts on a single plate of $\beta_{ii}^2 = 0.0127$
({\it c.f.} equation 17).  
Subtracting the contribution $4\sigma^2/25$ caused by plate-to-plate
errors GP86 conclude that large-scale gradients caused by plate
matching errors introduce a variance of $0.0087$ in the Lick counts.
However, GP86 fit a polynomial map to the Lick counts which has a
variance of $0.0194$, more than twice the predicted value, hence it is
plausible that the filtering has removed a component of real
clustering in the galaxy distribution in addition to large-scale
gradients in the counts caused by plate matching errors. 

To summarize, the expected errors in the Lick $\w$ arising from errors
in the plate matching are $w_{err} \sim 10^{-2}$, about an order of
magnitude larger than the errors estimated for the APM survey.
This estimate of the Lick errors is of the same order as the
differences seen between the APM and Lick angular correlation
functions and between the angular correlation functions estimated from
the filtered and unfiltered Lick maps.
We conclude, therefore, that it is difficult to disentangle real
galaxy clustering on scales of a few degrees from artificial
large-scale gradients in the Lick catalogue.

We have tested the effect of filtering the the APM Survey, by
evaluating the correlation functions for our six magnitude slices after
first subtracting a bicubic spline function fitted independently to
the counts in each slice (see also Appendix A2).
The results are shown in Figure~29. 
The estimates of $\w$ for the faintest slices change only slightly,
but for the brighter slices the break steepens considerably.
Figure~20b shows the filtered estimates of $\w$ scaled to the Lick
depth as in Figure~25.
The agreement between the scaled slices is significantly worse after
spline fitting.
The {\it rms} fractional variation of the spline fits is $1.5, 0.8,
2.5, 6.5, 13, 20 \times 10^{-3}$ for the 6 slices in faint-to-bright
order.
As described in Section 3.5, it is extremely unlikely that our
systematic errors are increasing this rapidly at brighter magnitudes,
thus we conclude that the spline fits to the bright slices are
dominated by true large-scale galaxy clustering and that the
unfiltered APM results are close to the correct values.

GP77 argued that their $\w$ shows anticorrelation of only $ \approx
-0.004$ at $\theta \sim 10^\circ $, therefore they could not have
over-filtered $\w$ by much more than this amount.
However, the results from spline fitting our maps show that the
correlation function of the spline fits, $w_S(\theta)$, typically
crosses zero at $\sim 20^\circ $, and that $w_S(10^\circ ) \sim
{1\over 2} w_S(0^\circ )$.
It is therefore quite likely that GP77 have in fact underestimated
$\w$ by as much as $0.008$.  It is clear from the above analysis that
the GP77 smoothing functions are the main source of the apparent
discrepancy between our results and theirs.  For example, if we fit a
spline function to our brightest magnitude slice (at approximately the
Lick depth) or to the Soneira-Peebles simulations described in
Appendix A, we obtain a very similar result to GP77.

\vfill\eject

\noindent
{\it 4.2.4 A firm upper limit to errors in $\w$} 

A firm upper limit on all sources of error in our $\w$
measurements can be set from the angular correlation function of the
faintest galaxies in the APM Survey.  Any error that is approximately
independent of magnitude will add the same error correlation function
to each of the magnitude slices, so the observed $\w$ for
the faintest slice provides an upper limit to these errors.
Realistically we believe that the errors are more than an order of
magnitude smaller than this, so this is gross overestimate of the most
likely errors.  Nevertheless, by assuming that {\it all} of the
correlations in the faintest slice are caused by errors we can
`correct' the brighter slices by subtracting $\w$ of the faintest
slice from the $\w$ of each of the brighter slices.  Conceptually this
subtraction is equivalent to finding a filter which removes all of the
structure in the distribution of faint galaxies, and then applying the
same filter to the distribution of the brighter galaxies.

After this subtraction the brighter galaxies still show strong
clustering to large scales, as seen in Figure~30.
The brightest three slices have such a high amplitude, that any errors
and gradients of the same amplitude as those in the faintest slice are
not significant.
This rules out all sources of error which affect bright and faint
galaxies equally. 
Given that there are no significant errors in the brighter
slices, then the agreement between our best estimates of $\w$ after
scaling suggests that errors are unlikely to be important even 
at faint magnitudes.
Thus we are confident that the excess power relative to
GP77 is real, especially for the brighter slices where the signal is
much more than ten times larger than the estimated systematic errors.

To summarize, we believe that filtering of the APM Survey
is unjustified for the following reasons:

\item{(i)} The angular correlation functions of the faint slices show
more power than the GP77 result even after filtering.

\item{(ii)} The angular correlation functions for the filtered
maps do not obey the scaling relation  because the filtering
affects mainly the $\w$ estimates from the brighter magnitude slices.

\item{(iii)} Filtering simulated galaxy distributions that have the
same $\w$ as measured from the APM Survey, but no plate-matching
errors, removes large-scale power from the $\w$ estimates resulting in
an artificial break in $\w$ similar to that found by GP77 for the
filtered Lick maps (see Appendix A2).

\item{(iv)} Subtracting $\w$ for the faintest slice from $w$
for the brightest three slices does not change them significantly.  
Hence filtering is justified only if the systematic errors are 
unique to the brightest slices.

\vfill\eject

\noindent
{\bf 5. Cosmological Implications}

\vskip 0.15 truein

In this Section we consider some implications of the angular
correlation functions estimated in previous Sections. In Section
5.1, we compare our estimates of $\w$ with the angular correlation
functions calculated from a family of cold dark matter (CDM) power 
spectra. In Section 5.2, we discuss the three-dimensional power
spectrum derived by inverting $\w$ following the technique
described by Baugh and Efstathiou (1993) and we compare the
results with the power spectra determined from N-body simulations
of several CDM-like models. We compare the APM power spectrum
with the power spectra estimated from several redshift surveys
in Section 5.3.

\vskip 0.15 truein

\noindent
{\it 5.1 Angular correlations}

The spatial two-point correlation function is the Fourier transform of
the three-dimensional power spectrum $P(k)$ (equation 1). Thus, 
given a theoretical prediction for $P(k)$, we can compute the spatial
correlation function $\xi$ and hence $\w$ by integrating Limber's
equation (31).  In this subsection, we compare the APM $\w$ with the
predictions of a family of CDM-like power spectra. According to linear
perturbation theory, the present day power-spectrum of mass
fluctuations in a CDM dominated universe arising from scale-invariant
adiabatic fluctuations in the early universe is well approximated by
the formula 
$$
\eqalignno{  P(k) &=  {B k \over 
(1 + [ak + (bk)^{3/2} + (ck)^2]^{\nu})^{2/\nu}}, &(40) \cr a =
(6.4/\Gamma) \hmpc&, \quad b = (3.0/\Gamma) \hmpc, \quad
c=(1.7/\Gamma)\hmpc, \quad \nu=1.13. \cr } $$ 
Equation (40) is accurate to a few percent for spatially flat 
universes in which baryons
make a negligible contribution to the total mass density (Bond \&
Efstathiou 1984). The parameter $\Gamma = \Omega_0 h$ defines a useful
sequence of models with varying amounts of large-scale power
(Efstathiou, Bond and White 1992). In the standard version of the CDM
model ({\it e.g.} Blumenthal \etal 1984, Davis
\etal 1985) $\Omega_0 = 1$ and $h \approx 0.5$, thus $\Gamma \approx 0.5$.

Figure~31 shows $\w$ computed from the power spectra (40)
compared to the APM measurements for the magnitude slice
$17 \le b_J \le 20$. We have adopted our standard model for 
the APM selection function (equation 38) and have assumed
$\epsilon = 0$ and $\Omega_0 =1$ in evaluation Limber's 
formula.  Each curve in Figure~31 has been normalized
so that the variance of the mass fluctuations in a
sphere of radius $x_f = 8 \hmpc$,
$$
\eqalignno{  \sigma^2(x_f) &=  {V \over 2 \pi^2}
\int_0^\infty P(k) W^2(kx_f) k^2\; dk,   &(41)  \cr
 W(kx_f)  & = {3 \over (kx_f)^3}
[{\rm sin} kx_f - kx_f\; {\rm cos}\; kx_f ].
  &  \cr
}
$$
is equal to $(0.9)^2$. This normalization has been chosen so
that each of the curves in Figure~31 passes through the
APM data points at $\theta \approx 0.4^\circ$. The
angular scale corresponding to a physical scale
of $5 \hmpc$ at the median depth $d^* \approx 315 \hmpc$
of the sample is shown by the arrow. Larger angular 
scales probe spatial separations at which $\xi (r) \simlt 1$ (equation 39) 
and so  typical
mass fluctuations should smaller than unity 
unless galaxies are strongly antibiased
relative to the mass distribution.

The shape of the APM $\w$ at large angles is well approximated 
by equation (40) with $\Gamma$ in the range $0.2$ -- $0.3$,
rather than the value $\Gamma = 0.5$ appropriate for
the standard CDM. This is consistent
with the results of previous papers (Maddox
\etal 1990c. Efstathiou \etal 1990, Efstathiou 1993,
Peacock and Dodds 1994). We make several points
concerning this Figure:

\item{(i)} The best fitting value of $\Gamma$ in Figure~31
is slightly larger than the value $\Gamma \approx 0.2$ 
favoured in some of our earlier papers. This is because we
compared the CDM models to the scaled results from the
six magnitude slices plotted in Figure~3. The faintest slice
is noticeably affected by systematic errors as described in
earlier sections and this biases $\Gamma$ to slightly lower
values. In contrast, the effects of  systematic errors 
on the estimates plotted in Figure~31 should be no larger
than $\approx 10^{-3}$ in $\w$.

\item{(ii)} The normalization of the CDM curves plotted in
Figure~31 is arbitrary. In principle, we could raise the
amplitude of the $\Gamma = 0.5$ curve in Figure~31 to fit
the angular correlation function at $\theta \sim 5^\circ$
(this would require a normalization of 
 $\sigma_8 \approx 1.6$) but the predicted $\w$ would
then lie well above the data points on angular scale
$\simlt 2^\circ$.

\item{(iii)} The points at $\w$ on scales $\simgt 1^\circ$
correspond to physical separations at which the mass fluctuations
are likely to be in the weakly non-linear regime. Thus, it is
possible that non-linearities may distort the shape of the
power spectrum over much of the angular range plotted in Figure~31.

We discuss the last two points in more detail in the next
subsection.

\vskip 0.15 truein

\noindent
{\it 5.2 The power spectrum determined from the APM Survey}

Baugh and Efstathiou (1993, hereafter BE)
show that the three-dimensional power spectrum $P(k,z)$ is related
to $\w$ by an integral equation 
$$
\eqalignno{ 
w (\varpi) &= \int_0^{\infty} P (k) k g (k \varpi)\; d k, \qquad 
 \varpi = 2 \;{\rm sin} \left( \theta/2 \right),  & (42a) \cr}
$$
where the kernel $g(k \varpi)$ is equal to 
$$
\eqalignno{ 
g (k \varpi) &= 
{{1\over 2 \pi} \int_0^{\infty} {F (z) \over (1 + z)^{\alpha}} 
\left({dN \over dz} \right)^{2} \left({dz \over dx} \right)
\;J_0 (k \varpi x) dz \over \left [ \int_0^\infty
 \left ( {dN \over dz} \right)\;dz \right]^2}, & (42b) \cr}
$$
and the time evolution of the power spectrum has been
parameterized  as 
$$
\eqalignno{  
P (k,t) &= {P (k) \over (1 + z)^{\alpha}}. & (42c) \cr}
$$
In these equations $k$ is defined as a comoving wavenumber, hence
for a power law spatial correlation function of the form given
in equation (32), the exponent $\alpha$ giving the time dependence
of $P(k,t)$ is equal to $\alpha = 3 + \epsilon - \gamma$. In 
the analysis below, we set $\alpha = 1.3$, {\it i.e.} we assume
that the small scale power-spectrum is constant in physical
coordinates, consistent with the assumption $\epsilon = 0$
that we adopted to compute the scaling of $\w$ in Figure~25.
 Equation (42a) can be inverted
using Lucy's (1974) deconvolution algorithm to yield
$P(k)$ from observations of $\w$. The technique is described
in detail by BE. A similar technique,
in which observations of the two-dimensional power spectrum
are inverted to recover $P(k)$ is discussed by Baugh and 
Efstathiou (1994), together with an application to the 
two-dimensional power spectra measured from the APM Galaxy 
Survey.

Figure~32a shows the BE technique applied to the $\w$ estimates for
the magnitude slice $17 \le b_J \le 20$.  The kernel (42b) has been
evaluated assuming $\Omega=1$; BE discuss the sensitivity of the
inversion to assumptions concerning the cosmological model and the
evolution of $P(k)$, showing that these have little effect on the
shape of the power spectrum but introduce an uncertainty of $\sim 20
\%$ in the amplitude of $P(k)$ even at the moderate depths of the APM
Survey ({\it cf} the estimates of $r_0$ in equations 39 derived from
Limber's equation). The dotted lines in Figure~32a show the inversion
of the $\w$ estimates for each of the four nearly equal area zones
({\it cf} Figure~27b). The points show the mean of these curves
together with the standard deviation of the mean. The thick solid line
shows the inversion of the $\w$ estimate for the full APM area
(Figure~2).  Figure~32b shows a check of the inversion procedure; here
we have integrated equation (42a) by linearly interpolating the
inverted power spectrum for the full APM area and compared the
resulting $\w$ (plotted as the solid line) against the APM data
points. The agreement is excellent, showing that the inversion is
extremely accurate (see BE for further tests of this method).

Figure~32a shows that we recover the power spectrum accurately for
wavenumber $k \simgt 0.05 h\;{\rm Mpc}^{-1}$. At smaller wavenumbers
there is considerable scatter between the estimates from different
zones and at these wavenumbers, systematic errors of $\sim 10^{-3}$ in
$\w$ can bias the estimates of $P(k)$ ({\it cf} Figure~7 of BE). Thus,
although we see tentative evidence for a peak in the power spectrum at
$k \sim 0.05 h\;{\rm Mpc}^{-1}$, we caution readers that the scatter
between zones is likely to underestimate the errors in $P(k)$, which
are probably dominated by systematic errors at such small wavenumbers.

We have integrated equation (41) using the power spectra
for each of the four zones plotted in Figure~32a to
computed $\sigma_8$. The result is
$$
\eqalignno{
     \sigma_8 & = 0.96 \pm 0.04 & (43) \cr}
$$
where the error is determined from the scatter from the four
zones. Uncertainties in the cosmological model and the exponent
$\alpha$ in equation (42c) introduce an additional uncertainty
of about $10\%$ in this estimate. This is in agreement with
the value $\sigma_8 \sim 1$ determined by Davis and Peebles
(1983) from an analysis of $\xi(r)$ from the CfA redshift survey,
but note that the estimate (43) is based on the power spectrum
in real space, rather than in redshift space and so is 
unaffected by the distortion of the clustering
pattern caused by galaxy peculiar motions.

Figure~33 shows the inverted estimates of $P(k)$ plotted
against the family of CDM power spectra, in analogy with
Figure~31. Each of the theoretical power spectra has been
normalized to $\sigma_8 =1$.  The
standard CDM power spectrum, $\Gamma = 0.5$, provides a
very poor match to the shape of the APM power spectrum.
A value  $\Gamma \approx 0.2$ -- $0.3$ provides a much better fit to
the data, in agreement with our discussion of $\w$.

In Figure~34 we compare the APM power spectrum to the power
spectra of the mass distributions from N-body simulations of
three CDM-like models: standard CDM (SCDM) {\it i.e.} $\Omega=1$, $h=0.5$;
a low density spatially flat model (LCDM) with $\Omega=0.2$ and $h=1$; a
mixed dark matter (MDM) model with $h = 0.5$ in which CDM contributes
$\Omega_{CDM}=0.6$, light neutrinos contribute $\Omega_\nu =0.3$ and
baryons contribute $\Omega_b=1$. The MDM and SCDM simulations are as
described by Dalton \etal (1994); for each of these models we ran an
ensemble of $10$ simulations each containing $10^{6}$ particles within a
cubical computational box of length $300 \hmpc$ using the P$^3$M N-body
code described by Efstathiou \etal (1985).  The LCDM curves in Figure~34
were derived from an ensemble of $10$ simulations each containing $4
\times 10^6$ particles within a box of length $600 \hmpc$. The models were
evolved to approximately match the amplitude of microwave background
anisotropies determined from the first year COBE maps (Smoot \etal 1992); thus
 $\sigma_8$ for the mass fluctuations is equal to
unity for the LCDM and SCDM models, and is $0.67$ for the MDM model
(ignoring gravitational wave contributions to the anisotropies, {\it e.g.}
Crittenden \etal 1994).

Each of the curves in Figure~34 shows the mean of the power
spectra of the mass distributions, $P_m(k)$, estimated from the simulations.
For the MDM model, we have arbitrarily multiplied $P_m(k)$ by a 
constant $b^2$, with $b=1.3$ to match the observations
on large scales, {\it i.e.} we have introduced a linear biasing
factor in relating mass fluctuations to fluctuations in the
galaxy distribution. Figure~34 illustrates the following points:

\item{(i)} The SCDM curve has a similar shape to the linear
theory power spectrum plotted in Figure~33 over almost
the entire range of wavenumbers plotted in the Figure
($k \simlt 0.3 h\;{\rm Mpc}^{-1}$). To reconcile the
SCDM curve with the APM data would require substantial
scale-dependent biasing, with galaxies strongly biased
relative to the mass distribution at wavenumbers
$k \simlt 0.1 h\;{\rm Mpc}^{-1}$ and strongly antibiased
at larger wavenumbers.  An argument against
scale dependent biasing as a resolution of this problem
has been given by Dalton \etal (1994)
who show that the spatial two-point correlation function
of rich APM clusters of galaxies also fails to match the
predictions of the SCDM model.

\item{(ii)} The LCDM model provides a good match to the
observations at wavenumbers $k \simlt 0.1 h\;{\rm Mpc}^{-1}$
but lies above the data points at larger wavenumbers. Evidently
antibiasing is required on  spatial scales $\simlt 5 \hmpc$
if this model is to match the clustering of the galaxy distribution
({\it cf} Efstathiou \etal 1990).

\item{(iii)} The MDM model provides a satisfactory fit to the
observations over the entire range of wavenumbers plotted in
the Figure.

In summary, the APM angular correlation functions and the
three dimensional power spectrum derived by inverting $\w$
are incompatible with the power spectrum expected in the
standard CDM model with $\Gamma=0.5$. Non-linear evolution
of the mass fluctuations cannot resolve this discrepancy,
as shown by the numerical simulations plotted in Figure~34.
The observations are in better agreement with a CDM power
spectrum with $\Gamma$ in the range $0.2$-$0.3$.

\noindent
{\it 5.3 Comparison with power spectra determined from redshift surveys}

In this subsection we compare the APM power spectrum with the power
spectra estimated from redshift surveys. A number of authors have
estimated power spectra of redshift surveys: Vogeley \etal (1992) and
Park \etal (1994) have analyzed the CfA-2 survey (see Baumgart and Fry
1991, for an analysis of the CfA-1 survey); Park, Gott \& da Costa
(1992) and da Costa \etal (1994)
have analyzed the optically selected Southern Sky Redshift
Survey; Fisher \etal (1993) have computed the power spectrum of a
redshift survey of IRAS galaxies with $60 \mu$ fluxes $f_{60} > 1.2$Jy
(see Fisher \etal 1995 for details of the $1.2$Jy survey); Feldman,
Kaiser \& Peacock (1994) have estimated $P(k)$ for the sparse sampled
$1$ in $6$ QDOT survey described by Lawrence \etal (1995).

Figure~35 shows the APM power spectrum (as in Figure~34) together
with estimates of $P(k)$ for IRAS and optically selected galaxies.
The open symbols show $P(k)$ from a reanalysis of the $1.2$Jy and
QDOT IRAS surveys by Tadros \& Efstathiou (1995a). The open
triangles show $P(k)$ for the CfA-2 survey from Park \etal
(1994); these are derived from a volume limited subset of
the CfA-2 survey consisting of $1509$ galaxies to a coordinate
distance of $101 \hmpc$. The three lines show estimates of
$P(k)$ in {\it redshift space} derived from the three 
ensembles of N-body simulations described in the previous
subsection. We make the following remarks concerning this figure:

\item{(i)} The IRAS power spectrum lies below the real-space
estimates of $P(k)$ derived from the APM survey
by a factor of $\sim 1.5$. Since redshift space
distortions are likely to boost the amplitude of the power spectrum in
redshift space compared to that measured in real space (Kaiser 1987,
compare also the redshift space and real space estimates of $P(k)$ for the
simulations in Figures~34 and 35), we conclude that IRAS galaxies are
more weakly clustered than optically selected galaxies on large
scales. The effect is relatively small, and is consistent with an
analysis of the cell-count variances of IRAS and the Stromlo-APM
redshift surveys described by Efstathiou (1995c).

\item{(ii)} The CfA-2 power spectrum has a slightly higher amplitude
than the APM estimate in the wavenumber range $0.1 \simlt k \simlt 0.3
h\;{\rm Mpc}^{-1}$. At $k \simlt 0.10 h\;{\rm Mpc}^{-1}$, the CfA-2
estimates have a slightly lower amplitude, but the errors on $P(k)$ at
such low wavenumbers are large. In principle, a direct comparison of
the APM estimates of $P(k)$ with an estimate derived from an optically
selected redshift survey would allow a determination of the effects of
redshift-space distortion. However, it is clear from Figure~35 that
the observational errors (particularly in the redshift-space estimates
of $P(k)$) are sufficiently large that it is not yet possible to set
tight limits on the effects of redshift-space distortions in the
linear regime $k \simlt 0.2 h{\rm Mpc}^{-1}$.  This point is discussed
in much greater detail in another paper (Tadros \& Efstathiou 1995b),
where we compare the APM $P(k)$ with redshift space estimates of
$P(k)$ derived from the Stromlo-APM redshift survey (Loveday \etal
1992a,b). Analyses of redshift space distortions in redshift surveys
are described by Hamilton (1993b) Fisher \etal (1994), Cole \etal
(1994b), Peacock \& Dodds (1994) and Loveday \etal (1995).

\item{(iii)} The redshift-space power spectra derived for
the MDM and LCDM models are compatible with the CfA-2 estimates with
the large observational errors. The SCDM curve lies slightly lower
than the CfA-2 estimates and slightly higher than the IRAS estimates,
though the general shape is compatible with the both estimates ({\it
cf} Bahcall \etal 1994, who make a similar point). This illustrates a
point that we have discussed in detail in other papers ({\it e.g.}
Loveday \etal 1992b, Efstathiou 1995a,c), namely that the
uncertainties on estimate of $P(k)$ derived from redshift surveys are
sufficiently large that they cannot yet distinguish between
different CDM-like models.

\vskip 0.15 truein

\noindent
{\bf 6. Conclusions}

\vskip 0.15 truein

The main conclusions of this paper are as follows:

\item{(i)} We have discussed various estimators of $\w$ and
illustrated the biases in estimates caused by integral constraints
and gradients in the galaxy surface density that are correlated
with the survey boundaries. Estimates of $\w$ derived from single
Schmidt plates are biased significantly by the integral constraint
and this leads to an artificial break in $\w$ from a power
law.

\item{(ii)} We have investigated how errors in the selection of
galaxies in the APM survey will affect our estimates of $\w$. 
We have analyzed simulations of errors including correlated plate
matching errors and field effects, and show that the observed plate
overlap residuals and the good agreement between the inter-plate and
intra-plate estimates of $\w$ 
limit the error in $\w$ to be $\simlt 10^{-3}$.

\item{(iii)} 
We have used a large body of CCD photometry from the
Las Campanas Deep Redshift Survey to estimate the errors in the APM
magnitudes. The results are consistent with the error estimates of
Paper II and show that the zero-point magnitude error for a typical
APM field is in the range $0.04 \simlt \sigma_{APM} \simlt 0.05$
magnitudes. We estimate that photometric errors lead to 
systematic errors of $\simlt 10^{-3}$ in $\w$.

\item{(iv)} We have investigated  the criticisms of the APM
photometry by Fong \etal (1992) and have shown that they are unjustified.
In particular, large photometric errors as claimed by Fong \etal are
incompatible with the good agreement that we have found between the
APM and LCDRS photometry,  with the uniformity of the
galaxy counts in the APM survey, and with various other tests of the
accuracy of the APM magnitudes described in Paper II.

\item{(v)} There is no evidence for any correlation between
errors in the APM Survey and the exposure or scanning
dates of the Schmidt plates.

\item{(vi)} The effects of Galactic obscuration have been 
investigated using IRAS maps of $100 \mu$m emission.  We estimate that
the angular correlation functions measured from the APM Survey are
biased by at most $8 \times 10^{-4}$ by variable Galactic obscuration.

\item{(vii)} The changes in stellar contamination of the APM galaxy sample
caused by the variation in stellar surface density in our galaxy are
negligible over the area we use to estimate the galaxy correlations.

\item{(viii)} We have used the magnitude residuals on plate overlaps
and inter- and intra-plate residuals in $\w$ to estimate the errors
in the APM survey as a function of magnitude. The systematic
errors in $\w$ are found to be independent of magnitude
except close to the magnitude limit of the survey
$b_J > 20$, where plate-to-plate variability in star-galaxy
classifications introduces fluctuations in the APM counts.

\item{(ix)} We have presented estimates of $\w$ as a 
function of limiting magnitude corrected for various
sources of systematic error. The corrections are
almost negligible except for the faintest magnitude
slice $20 \le b_J \le 20.5$.

\item{(x)} We have used faint redshift surveys to determine
the selection function of the APM survey as a function of
limiting magnitude. By evaluating Limber's equation, we show
that the APM estimates of $\w$ scale accurately with depth, 
as expected if we are measuring  real structure in the 
galaxy distribution.

\item{(xi)} The APM $\w$ estimates show slightly less power
at large angular scales than the estimates from the Edinburgh-Durham 
Southern Galaxy Catalogue (EDSGC). The discrepancy may be explained by the
larger photometric errors and smaller area of the EDSGC.

\item{(xii)} The APM $\w$ estimates are higher at large angular scales
that the GP77 estimates of $\w$ from the filtered Lick counts.
We have argued that GP77 may have removed real clustering by 
filtering the Lick catalogue.

\item{(xiii)} The APM $\w$ estimates are incompatible with the
linear theory predictions of the standard CDM model with
$\Gamma = \Omega_0 h = 0.5$. Our angular correlation functions
are much better described by a CDM model with $\Gamma$ in the
range $0.2$--$0.3$.

\item{(xiv)} We reach similar conclusions by comparing
the three-dimensional power-spectrum $P(k)$ determined by inverting
our estimates of $\w$. By calculating $P(k)$ from N-body
simulations we show that non-linear evolution cannot reconcile the
standard cold dark matter model with the shape of $P(k)$ determined
from the APM survey.

%\item{(xv)} The errors on the APM $P(k)$ estimates 
%are smaller than the errors on $P(k)$ derived from redshift surveys.
%Furthermore the APM $P(k)$ estimates are in real-space and so are
%unaffected by redshift-space distortions. The APM results thus
%provide strong constraints on models of 

\vfill\eject

\noindent
{\bf Acknowledgements} 

We wish to thank the staff of the UK Schmidt Telescope for providing
the excellent plate material on which this survey is based; We are
indebted to the APM group at Cambridge (Mike Irwin, Pete Bunclark \&
Mick Bridgeland) for much advice and assistance over the past seven
years and to Gavin Dalton and Jon Loveday for their contributions to
the APM Galaxy Survey.  We thank Douglas Tucker and Richard Ellis,
and their collaborators, for
supplying data in advance of publication.  GPE and WJS
would like to thank PPARC for the award of Senior and Advanced
Fellowships respectively.

\vfill\eject

\def\ref {\par \hangindent=1cm \hangafter=1 \noindent}

\centerline{\bf  REFERENCES}

\vskip 0.15 truein

\ref Bahcall N.A., Cen R., Gramann M., 1994, ApJ, {\bf 408}, L77.

\ref Baugh C.M., Efstathiou G., 1993, MNRAS, {\bf 265}, 145.

\ref Baugh C.M., Efstathiou G., 1994, MNRAS, {\bf 267}, 323.

\ref Baumgart D.J., Fry J.N., 1991, ApJ, {\bf 375}, 25.

\ref Blumenthal G.R., Faber S.M., Primack J.R., Rees M.J.,
1994, Nature, {\bf 311}, 517.

\ref Bond J.R., Efstathiou G., 1984, ApJ, {\bf 285}, L45.

\ref Broadhurst T.J., Ellis R.S.,  Shanks T., 1988, 
 MNRAS, {\bf 235} 827. 

\ref Cole S., Ellis R.S., Broadhurst T., Colless M., 1994a, MNRAS,
{\bf 267}, 541.

\ref Cole S., Fisher K.B., Weinberg D.H., 1994b, MNRAS, {\bf 267}, 785.

\ref Colless M.M., Ellis R.S., Taylor K., Hook R.N., 1990, MNRAS,
{\bf 244}, 408.

\ref Colless M.M., Ellis R.S., Broadhurst T.J., Taylor K.,
Peterson B.A., 1993,  MNRAS, {\bf 261}, 19.

\ref Collins C.A., Heydon-Dumbleton N.H., MacGillivray, H.T.,
1989, MNRAS, {\bf 236}, 7p.

\ref Collins C.A., Nichol R.C., Lumsden S.L., 1992, MNRAS,
{\bf 254}, 295.

\ref Cowie L.L., Lilly S.J., Gardner J.P., McLean I.S., 1988, ApJ,
L29.

\ref Crittenden R., Bond J.R., Davis R.L., Steinhardt P.L.,
1993, PRL, {\bf 71}, 324.

\ref da Costa L.N., Vogeley, M.S., Geller M.J., Huchra J.P.,
 Park C., 1994, ApJ, {\bf 437}, L1.

\ref Dalton G.B., Efstathiou G., Maddox S.J., Sutherland W.J.,
1992, ApJ, {\bf 390}, L1.

\ref Dalton G.B., Efstathiou G., Maddox S.J., Sutherland W.J.,
1994a, MNRAS, {\bf 269}, 151. (Paper IV)

\ref Dalton G.B., Croft R.A.C.,  Efstathiou G., Sutherland W.J.,
Maddox S.J., Davis, M., 
1994b, MNRAS, {\bf 271}, L47.

\ref Dalton G.B.,  Efstathiou G., Sutherland W.J.,
Maddox S.J., 1995, in preparation. (Paper V).

\ref Davis M., Peebles P.J.E., 1983, ApJ, {\bf 267}, 465.

\ref Davis M., Efstathiou G., Frenk C.S., White, S.D.M.,
1985, ApJ, {\bf 292}, 371.

\ref de Lapparent V., Kurtz M.J., 
Geller, M.J., 1986, ApJ, {\bf 304}, 585.

\ref Efstathiou G., 1993, {\it Proc Natl Acad Sci. USA}, 
{\bf 90}, 4859.

\ref Efstathiou G, 1995a, {\it Les Houches Lectures}, ed. R. Schaefer,
Elsevier Science Publishers, Netherlands, in press.

\ref Efstathiou G., 1995b, MNRAS, {\bf 272}, L25.

\ref Efstathiou G., 1995c, MNRAS, {\bf 276}, 1425.

\ref Efstathiou G., Davis M., Frenk C.S., White S.D.M.,
1985, ApJ, {\bf 292}, 371.

\ref Efstathiou, G., Ellis, R.S., Peterson, B.A., 1988,  
MNRAS, {\bf 232}, 431.

\ref Efstathiou G., Sutherland W.J., Maddox S.J., 1990a, Nature,
{\bf 348}, 705.

\ref Efstathiou G., Bernstein G., Katz N., Tyson J.A.,  Guhathakurta,
P., 1991, ApJ, {\bf 380}, L47.

\ref Efstathiou G., Bond J.R., White S.D.M., 1992a, MNRAS, {\bf 258}, 1p.

\ref Feldman H.A., Kaiser N., Peacock J.A., 1994, ApJ,
{\bf 426}, 23.

\ref Fisher K.B., Davis M., Strauss M.A., Yahil A., Huchra J.P.,
1993, ApJ, {\bf 402}, 42.

\ref Fisher K.B., Davis M., Strauss M.A., Yahil A., Huchra J.P.,
1994, MNRAS, {\bf 267}, 927.

\ref Fisher K.B., Strauss M.A., Davis M., Yahil A.,  Huchra J.P.,
1995, ApJS, in press.

\ref Fong R., Hale-Sutton D.,  Shanks T., 1992, MNRAS, {\bf 257}, 650. (F92)

\ref Gaztanaga E., 1995, MNRAS, in press.

\ref Gaztanaga E., Frieman J.A., 1994, ApJ, {\bf 437}, L13.

\ref Geller M.J., de Lapparent V., Kurtz, M.J., 1984, ApJ, {\bf 287}, L55.

\ref Groth E.J., Peebles P.J.E., 1977,  ApJ, {\bf 217}, 38. (GP77) 

\ref Groth E.J., Peebles P.J.E., 1986a, ApJ, {\bf 310}, 499. 

\ref Groth, E.J., Peebles P.J.E., 1986b,  ApJ, {\bf 310}, 507. (GP86) 

\ref Hamilton A.J.S., 1993a, ApJ, {\bf 417}, 19.

\ref Hamilton A.J.S., 1993b, ApJ, {\bf 406}, L47.

\ref Heydon-Dumbleton N.H., Collins C.A., MacGillivray H.T.,
1989, MNRAS, {\bf 238}, 379.

\ref Hewett P.C., 1982,  MNRAS, {\bf 201}, 867.

\ref Kaiser, N., 1987, MNRAS, {\bf 227}, 1.

\ref Kolb E.W., Turner M.S., 1990, {\it The Early Universe},
Addison-Wesley Publishing Company.

\ref Koo D.C., Szalay A.S., 1984,  ApJ, {\bf 282}, 390. 

\ref Lambas D. Maddox S.J., Loveday J., 1992, MNRAS, {\bf 258}, 404.

\ref Landy S.D., Szalay A.S., 1993, ApJ, {\bf 412}, 64.

\ref Lawrence A. \etal, 1995, MNRAS, in press.

\ref Limber, D.N., 1954, ApJ, {\bf 119}, 655.

\ref Loveday J., Peterson, B.A, Efstathiou, G.,
Maddox S.J., 1992a, ApJ, {\bf 390}, 338.

\ref Loveday J., Efstathiou, G., Peterson B.A.,
Maddox S.J., 1992b, ApJ, {\bf 400}, L43.

\ref Loveday J., Maddox S.J., Efstathiou, G., Peterson B.A.,
 1995, ApJ, in press.

\ref Lucy L.B., 1974, AJ, {\bf 79}, 745.

\ref Maddox S.J., 1988, PhD Thesis, Cambridge. 

\ref Maddox S.J., Sutherland W.J., Efstathiou G., Loveday J., 1990a,
MNRAS, {\bf 243}, 692. (Paper I).

\ref Maddox S.J., Efstathiou G., Sutherland W.J., 1990b,
 MNRAS, {\bf 246}, 433. (Paper II).

\ref Maddox S.J., Efstathiou G., Sutherland W.J., Loveday J., 1990c,
MNRAS, {\bf 242}, 43p. 

\ref Maddox S.J., Sutherland W.J., Efstathiou G., Loveday J.,
Peterson B.A., 1990d, MNRAS, {\bf 247}, 1p.

\ref Metcalfe N., Shanks T., Fong R., Roche N., 1995, MNRAS,
{\bf 273}, 257.

\ref Neuschaefer L.W., Windhorst R.A., Dressler A., 1991, ApJ,
{\bf 382}, 32.

\ref Neuschaefer L.W,,  Windhorst R.A.,  1995, ApJ,
{\bf 439}, 14.

\ref Nichol R.C., Collins C.A., 1993, MNRAS, {\bf 265}, 867.

\ref Park C., Gott J.R., da Costa, L.N., 1992, ApJ, {\bf 392}, L51.

\ref Park C., Vogeley M.S., Geller M.J., Huchra J.P.,
1994, ApJ, {\bf 431}, 569.

\ref Peacock J.A., Dodds S.J., 1994, MNRAS, {\bf 267}, 1020.

\ref Peebles, P.J.E., 1974,  ApJ, {\bf 189}, L51.

\ref Peebles P.J.E., 1980, {\it The Large-Scale Structure of the
Universe}, Princeton University Press, Princeton.

\ref Phillipps S., Fong R., Ellis R.S., Fall S.M.,
 MacGillivray, H.T., 1978, MNRAS, {\bf 182}, 673.

\ref Pritchet C., Infante L., 1986, AJ, {\bf 91}, 1. 

\ref Pritchet C., Infante L., 1992, ApJ, {\bf 399}, L35.

\ref Roche N., Shanks T., Metcalfe N., Fong R., 1993, MNRAS, {\bf 263}, 
360.

\ref Rowan-Robinson M., Hughes J., Jones M., Leech, K.,
Vedi, K., Walker D.W., 1991, MNRAS, {\bf 249}, 729.

\ref Schechter P.L., 1976, ApJ, {\bf 203}, 297

\ref Sebok, W.L., 1986, ApJS, {\bf 62}, 301.

\ref Shane C.D., Wirtanen C.A., 1967, Publ Lick Observatory,
{\bf 22}, part 1.

\ref Shectman S., Schechter P.L., Oemler A., Tucker D., Kirshner R.P.
Lin H., 1992, in {\it Clusters and Superclusters of Galaxies}, ed
A.C. Fabian, Dordrecht, Kluwer, p351.

\ref Shectman S., Landy S., Oemler A., Tucker D., Kirshner R.,
Lin H. \& Schechter  P., 1995, in {\it Wide Field Spectroscopy
and the Distant Universe}, 35th Herstmonceux Conference, eds
S.J. Maddox and A. Aragon-Salamanca, World Scientific Press, 
Singapore, p98.

\ref Seldner M., Siebers B., Groth E.J.,  Peebles P.J.E., 1977, 
 AJ, {\bf 82}, 249. 

\ref Shanks T., Fong R., Ellis R.S., MacGillivray H.T., 1980,
 MNRAS, {\bf 192}, 209.

\ref Smoot G.F. \etal 1992, ApJ, {\bf 396}, L1.

\ref Soneira R.M., Peebles P.J.E., 1978, AJ, {\bf 83}, 845.

\ref Stevenson P.R.F., Shanks T., Fong R., MacGillivray, H.T.,
1985, MNRAS, {\bf 213}, 953.

\ref Szapudi I., Dalton G.B., Efstathiou G., Szalay, A.S.,
1995, ApJ, {\bf 444}, 520.

\ref Tadros H., Efstathiou G., 1995a, MNRAS, {\bf 276}, L45.

\ref Tadros H., Efstathiou G., 1995b, submitted.

\ref Tyson, J.A., 1988, AJ, {\bf 96}, 1. 

\ref Vogeley M.S., Park C., Geller M.J., Huchra J.P., 1992
ApJ, {\bf 391}, L5.

\vfill\eject

\noindent
{\bf  Appendix A: Estimators of the Angular Correlation Function}

\vskip 0.2cm

\noindent
{\bf  A.1 Soneira-Peebles Simulations}

There has been some disagreement between authors concerning biases in 
estimators of angular correlation functions ({\it e.g.} Shanks
{\it et al} 1980, Hewett 1982). We have therefore checked a number of
estimators using simulations of clustered galaxy distributions. We have
also used these simulations to assess the effect on $\w$ of filtering
the data.

The simulations that we have used follow the prescription of
Soneira and  Peebles (1978) which uses
a simple algorithm designed to produce galaxy distributions with a
similar visual appearance and low-order correlation function as the
Lick survey. Although N-body simulations can provide a better match to
the observed ``frothy'' nature of the galaxy distribution, extremely
large simulations would be necessary to model the large volume
($\sim 10^8{\rm Mpc}^3$) sampled by the APM Survey and so would be
time-consuming to generate.

The Soneira-Peebles simulations are constructed from a large number of
`clumps': for each clump, an imaginary rod of length $L$ is randomly
oriented in space. Two shorter rods of length $L/a$ are randomly
oriented, with their centres at the ends of the first rod; two rods of
length $L/a^2$ are randomly oriented with their centres at the ends of
each of the previous two, and the process is continued for a number of
`levels', $N$. A `galaxy' is placed at the end of each rod on the Nth
level, i.e. $2^N$ galaxies in a clump of level $N$.  If all clumps
have the same $N$, the amplitudes of the three and four point
correlation functions are smaller than observed, thus a dispersion is
introduced in the number of levels (see Table~A.1) to produce
approximately the observed values.

The clumps are distributed randomly in space within a cone of
half-angle $30^\circ $ and length $1000 \hmpc$, and then ``observed''
from a point on the axis of the cone, $50 \hmpc$ inside the apex (to
avoid boundary effects): galaxies are accepted or rejected with a
probability calculated from a Schechter (1976)
luminosity function, with an apparent
magnitude limit of $b_J= 20.$.  This produces a simulated 0.85 steradian
of `sky' for which we construct an equal area maps as for the real survey
data, and use this to measure the angular correlation function. 

A few short-cuts, as described by Soneira and Peebles, were used to
speed up the computations.  The spatial volume was truncated at $1000
\hmpc$ where the selection probability is $\sim 1\%$, each clump was
reused a number of times at different positions in space (but
different galaxies visible each time), and if the selection
probability $p$ at the distance of a given clump was $\le 0.05$ we
simply chose $pN$ galaxies randomly from the clump rather than
generating a different random number for every galaxy.  None of these
simplifications will have any significant effect on the correlation
function.  The random numbers were generated by the NAG routine G05CAF
in `batches' of 1000, placed into an array and shuffled with another
1000 random numbers to ensure independence between successive random
numbers.

One set of simulations (labelled SP0) has identical parameters to those
given in Soneira and Peebles, shown in Table~A.1. We generated simulations
with additional large-scale power by scaling up the clumps: the minimum
value of $N$ was increased by 1, the length of the rods was scaled up by a
factor $1.76$ and the density of clumps was halved giving the set of
simulations labelled SP1.  Repeated scaling up gives the simulations
labelled SP2, SP3 and SP4.  This process does not change the small-scale
correlation functions but simply moves the break in $\w$ to larger angles.

\vskip 0.2cm

\vbox{ 
\smallskip
{ {\bf Table A.1} 
Parameters used in construction of clumps in the SP0 simulations}
\smallskip 
\halign{\qquad # \hfil & $ # $ \hfil & \quad #\hfil \cr 
Definition  & {\rm Parameter} & Value \cr
\noalign{\smallskip}
\noalign{\hrule\smallskip}
Min, max clump level & N_{min}, N_{max} & 7, 12 \cr
\% of clumps of level N & p_7 \ldots p_{12} & 68, 23, 6, 1.5, 0.5, 0.11 \cr
Length of longest rods & L_0 & 11.5 $\hmpc$ \cr
Mean spacing of clumps & n^{-1/3} & $20 \hmpc$ \cr
}}

\vskip 0.2cm

We also generated a simulation with power on very long wavelengths
({\it cf\/} the isocurvature baryon model, Peebles 1986) by simply
imposing a large-scale gradient of amplitude $ \pm 15\%$ across the
observed `sky' from an SP0 simulation. These simulations are labelled
SPS. 
Grey-scale pictures of some
typical simulations are shown in Figure~A.1.

\vskip 0.2cm

\noindent
{\bf  A.2 Estimating the Correlation Function}

For each of the simulations we evaluated the correlation function $\w$
using the `direct' estimator
$$\eqalignno{
\w &= { \left<N_i N_j\right> \over \overline{N}^2 } - 1, &(A1) \cr}
$$
and the `ensemble' estimator
$$\eqalignno{\w & = { \left<N_i N_j\right> \over \left<N_i\right> 
\left<N_j\right> } - 1. &(A2) \cr}
$$
In these equations $N_i$ is the galaxy count in cell $i$, $\overline
N$ is the mean galaxy count and the averages are over all cells
with separations within the separation bin centred on angle $\theta$.
We also used an estimator based on the Fourier transform of the
two-dimensional power spectrum.  To evaluate the FT estimator, we
constructed a 2-D array of galaxy counts in cells from an equal area
projection of the catalogue. The mean from all data cells was
subtracted from the data cells; cells outside the survey area
were set to zero.  We then evaluated the Fast Fourier Transform of
this array, multiplied this by its complex conjugate and performed the
inverse transform.  We then defined the `window function' by setting
cells within the observational window equal to unity and setting those
outside equal to zero: the same algorithm was then applied to this
array.  If $\delta({\bf x})$ denotes the density field with the mean
subtracted (we ignore errors in the mean density, which will
be discussed in Section A3), and
$W({\bf x})$ is the window function of the sample, then the observed
density field is 
$$
\eqalignno{
\delta_{obs}({\bf x}) & = W({\bf x}) \delta({\bf x}). &}
$$
Fourier transforming
$$
\eqalignno{
\hat{\delta}_{obs}({\bf k}) & = (\hat{W} * \hat{\delta})_{\bf k} &}
$$
and so the power-spectrum in two dimensions is
$$
\eqalignno{
P_{obs}({\bf k}) & = (\hat{W} * \hat{\delta})_{\bf k}
(\hat{W} * \hat{\delta})_{- {\bf k}}.}
$$
The inverse FT of the power spectrum, $f({\bf x})$, is
is 
$$
\eqalignno{
f({\bf x}) & = {\cal F}^{-1} (P({\bf k})) \cr
       & = {\cal F}^{-1} ((\hat {W} * \hat {\delta})_{\bf k}) * 
{\cal F}^{-1} (\hat {W} * \hat {\delta})_{-{\bf k}})  \cr
      & = ( W \delta * \overline{W} \overline{\delta}) ({\bf x})  \cr
}
$$
where $\overline{W}({\bf x}) \equiv W(-{\bf x})$ and ${\cal F}^{-1}$ 
denotes the inverse Fourier transform. 
If the window is uncorrelated with the true galaxy distribution,
$$
\eqalignno{
f({\bf x}) & = W*\overline{W} ({\bf x})\; \delta*\overline{\delta}({\bf x}) \cr
       & = \left<W({\bf a}) W({\bf a}+{\bf x}) \right>\,\left<\delta({\bf a}) 
\delta({\bf a} + {\bf x}) \right> &(A3)\cr 
}
$$
and so $f(x)$ is equal to the product of the correlation functions 
of the window function and the galaxy density. 
The FT estimate of the
correlation function $w_{FT}$ is therefore given by dividing 
the inverse FT of the power spectrum
of the data by that of the window function, then radially averaging
in bins of $\vert {\bf x} \vert$ to give $\w$.

We first evaluated the correlation functions with the ensemble and
direct estimators for a number of simulated Schmidt plates of $5^\circ
$ square using the SP0 simulation parameters. This is of interest
because several authors have published estimates of $\w$ from
individual Schmidt plates (see Section 4.2.1) and there has been some
controversy concerning the reliability of the results, for example
whether the results are biased low because of the small area used to
define the mean surface density.  Some typical results from our
simulations are shown in Figure~A.2: there is substantial scatter
between plates using the ensemble estimator, but very much large
scatter using the direct estimator.  In both cases the mean $\w$ of
the 9 plates shown is biased significantly low relative to $\w$ for
the whole area.
The large scatter for the direct estimator can be explained by
edge effects as follows: in a finite sample, cells 
have different numbers of neighbours  separated by angle $\theta$
depending on their proximity to the boundary, hence cells
have different weights. For $\theta \simlt 0.5 \times$ 
the sample size, cells at the edge have half the weight of those 
at the centre. For $\theta \simgt 0.5 \times$ the sample size, 
central cells have zero weight whereas edge cells have high weights.
If there are substantial density variations on scales comparable to 
the sample size, then the mean density for central cells may differ from that
of cells at the edges. Thus using a global estimate of the 
mean cell count $\overline{N}$ is then inappropriate and introduces
an error that is first order in $\delta \rho / \rho$ 
whereas $\w$ is second order. Thus the direct estimator becomes
unreliable if structures exist on scales comparable
to the sample area. This is discussed in further detail in
Section A.3.

After averaging over the simulations, the estimates
for both estimates  are biased low.
For the direct estimator, with a total of $N_T$ galaxies in
$n_c$ cells it is clear that an integral constraint applies, as
described in Section 2.2, since
$$
\eqalignno{
\overline{N}^2 \sum_{i,j} w(\theta_{ij}) & = \sum_{i,j} 
( N_i N_j - \overline{N}^2 )
 =  N_T^2 - n_c^2 \left({ N_T \over n_c}\right)^2 
= 0. \cr
}
$$
The bias is second order in $\delta \rho/ \rho$.
The ensemble estimator is also subject to an integral constraint
which leads to a bias that is second order in $\delta \rho / \rho$,
but in this case, the expression for the bias is more complex
(see Section A3). 

We also evaluated each estimator of $\w$ for the full $60^\circ $ cone
for each of the simulations.  Results for simulations SP0, SP2 and SP4
are shown in Figure~A.3.  In each case the ensemble and FT estimators
are in excellent
agreement with each other and with the expected behaviour
built in to the simulation.  For the SP0 model the break occurs at
$\sim 1.5^\circ $, half the scale of the Lick survey, since the sample
depth is about twice that of Lick. For SP2 and SP4 the break moves
outward as expected, occurring very close to the noise for SP4.

The direct estimator agrees well for the SP0 simulation but is
increasingly discrepant for SP2 and SP4: this is because the added
large-scale power results in fewer independent structures in
each simulation and so the edge effects become more important. This is
illustrated in Figure~A.4 which shows results for three different
realizations of the SP4 prescription; the direct estimator shows very
large scatter between different realizations.

For the SPS simulation with an added large-scale gradient, both the
ensemble and direct estimators show a positive tail $\w \sim 5 \times
10^{-3}$ as expected.  This is plotted in Figure~A5, and shows
that any `filtering' of large-scale power arising from the ensemble
estimator is negligible (see Section A3).

We have also investigated the effect on $\w$ of fitting large-scale smoothing
functions to the galaxy counts (as in GP77): we fitted a two-dimensional cubic
polynomial (i.e. a bicubic spline with no knots) to the data map,
divided the map by the polynomial and multiplied by the previous mean
density, then evaluated $\w$ using the ensemble estimator.  Some
typical results for an SP4 simulation are shown in Figure~A.6: the
result is very close to simply subtracting the correlation function of
the spline from that of the data.  This moves the break to 
smaller angles and substantially steepens the slope beyond the break,
so the $\w$ for a spline-filtered SP4 simulation is quite similar to
that for an SP0 simulation.  Slight anticorrelation is seen beyond the
break as in GP77.  This suggests that the GP77 smoothing functions may
have removed some true signal from their correlation function results.

The conclusions of this section can be summarized as follows:

\item{(i)}  The direct estimator displays large fluctuations which
are of first order in the fluctuations in the galaxy distribution
and so dominate the estimates of $\w$ at large angles. The
fluctuations in ensemble estimator are of second order in $\delta
\rho / \rho$.

\item{(ii)} Both  the ensemble and direct estimators are biased low
by integral constraints on scales comparable to the sample size.

\item{(iii)} The ensemble and FT estimators agree very well.

\item{(iv)} The  ensemble estimator is very nearly unbiased and 
does not filter out large scale power in the galaxy distribution
as has  sometimes been  claimed.

\item{(iii)} Fitting large-scale smoothing functions to the observed galaxy
distribution (cf GP77) can remove 
 power on {\it intermediate} scales, and
move the break in $\w$ to smaller angles.

\vskip 0.2cm

\noindent
{\bf  A.3 Mathematical Analysis of Estimators}

Hamilton (1993a) has recently made  a comprehensive mathematical
analysis of estimators of correlation functions and it is useful
to summarize how his results relate to those of the previous section.
Denoting the window function of the survey as $W$, the observed galaxy
density is 
$$
\eqalignno{ N & = \overline {\cal N} W (1 + \delta), &(A4) }
$$
where $\overline {\cal N}$ is the mean surface density. The 
mean overdensity of the catalogue $\overline \delta$, the
galaxy-catalogue correlation function $\psi(\theta)$, and the
windowed galaxy-galaxy correlation function $w_c$ are defined
by 
$$
\eqalignno{ \overline \delta & = {\langle W \delta \rangle
\over \langle W \rangle}, \quad \psi = {\langle W_1 \delta_1 W_2 \rangle
\over \langle W_1W_2 \rangle}, \quad w_c = {\langle W_1 \delta_1
\delta_2  W_2 \rangle
\over \langle W_1W_2 \rangle}. &(A5) }
$$
The direct estimator is related to these quantities by 
$$
\eqalignno{ w_D(\theta) & = {w_c(\theta) + 2 \psi(\theta) - 2 \overline
\delta - \overline \delta^2 \over (1 + \overline \delta)^2}, &(A6) }
$$
and the ensemble estimator gives
$$
\eqalignno{ w_E(\theta) & = {w_c(\theta) - \psi^2(\theta) 
\over (1 + \psi(\theta))^2}, &(A7) }
$$
(Hamilton 1993a). In our application, the window function is either
zero or unity, hence $\psi(0) = \overline \delta$, and so the two
first order terms in equation A6 do not quite cancel leading to a
first-order error in $\delta \rho /  \rho$ that depends on the
distribution of galaxies relative to the boundaries of the sample.
This is why the direct estimator leads to such a large scatter when
applied to the simulations of the previous section. 

It is easy to apply this analysis to the Fourier transform estimator
of $\w$ defined in the previous section. This gives
$$
\eqalignno{ w_{FT}(\theta) & = {w_c(\theta) - 2\psi(\theta) \overline \delta 
 - \overline \delta^2}, &(A8) }
$$
and so differs from the true catalogue correlation function by second
order terms. This is why the FT and ensemble estimators agree so closely.

If we average over a large number of catalogues, the first order terms
in equation (A6) average to zero, but the expectation value
of the second order term is the expected variance of the overdensity
within the field area
$$
\eqalignno{ \langle \overline
\delta^2 \rangle & = {\int W_1 W_2\;w(\theta_{12})\; d \Omega_1 d \Omega_2
\over [ \int W_1 d\Omega_1 ]^2} = 
{1 \over \Omega^2} \int w(\theta_{12})\; d \Omega_1 d \Omega_2 &(A9) }
$$
{\it i.e.} the mean of the direct estimate estimate differs from the 
true catalogue correlation function by the integral constraint
(equation 10)
as discussed in Section 2.2.

Each of the estimators (A5) - (A7) is subject to an integral-type
constraint leading to a bias that is second order in ${\delta \rho/
\rho}$, for example, the ensemble estimator is biased low by  
$$
\eqalignno{ \langle \psi^2(r) 
 \rangle & = {\int W_1 W_2 W_3 W_4 \;w(\theta_{13})\; d \Omega_1 d \Omega_2
d \Omega_3 d \Omega_4 \over [ \int W_1 W_2 \;d\Omega_1 d \Omega]^2}, &(A10) }
$$
where $W_1$ and $W_2$, and $W_3$ and $W_4$ are separated by $r$.
If the separation $r$ is small compared to the size of the 
sample, we can make the approximation $W_1 \approx W_2$, $W_3
\approx W_4$ and so $\langle \psi^2(r) \rangle \approx \overline 
\delta^2$, {\it i.e.} the offset is the same as for the direct 
estimator. This result is true for all of the estimators considered
here, including the estimator of equation (5) used for single
plate estimates in section 2.2. 
In summary, first order errors in the estimates of $w(\theta)$ 
will average out over a large number of realizations leaving second
order biases. At separations small compared to the survey size, the 
bias is approximately equal to the variance of
the galaxy overdensity within the field, but at large separations
the precise form of the correction depends on the particular estimator
used to measure $\w$.

\vfill\eject

\noindent
$\;$

\vskip 6 truein

\includegraphics{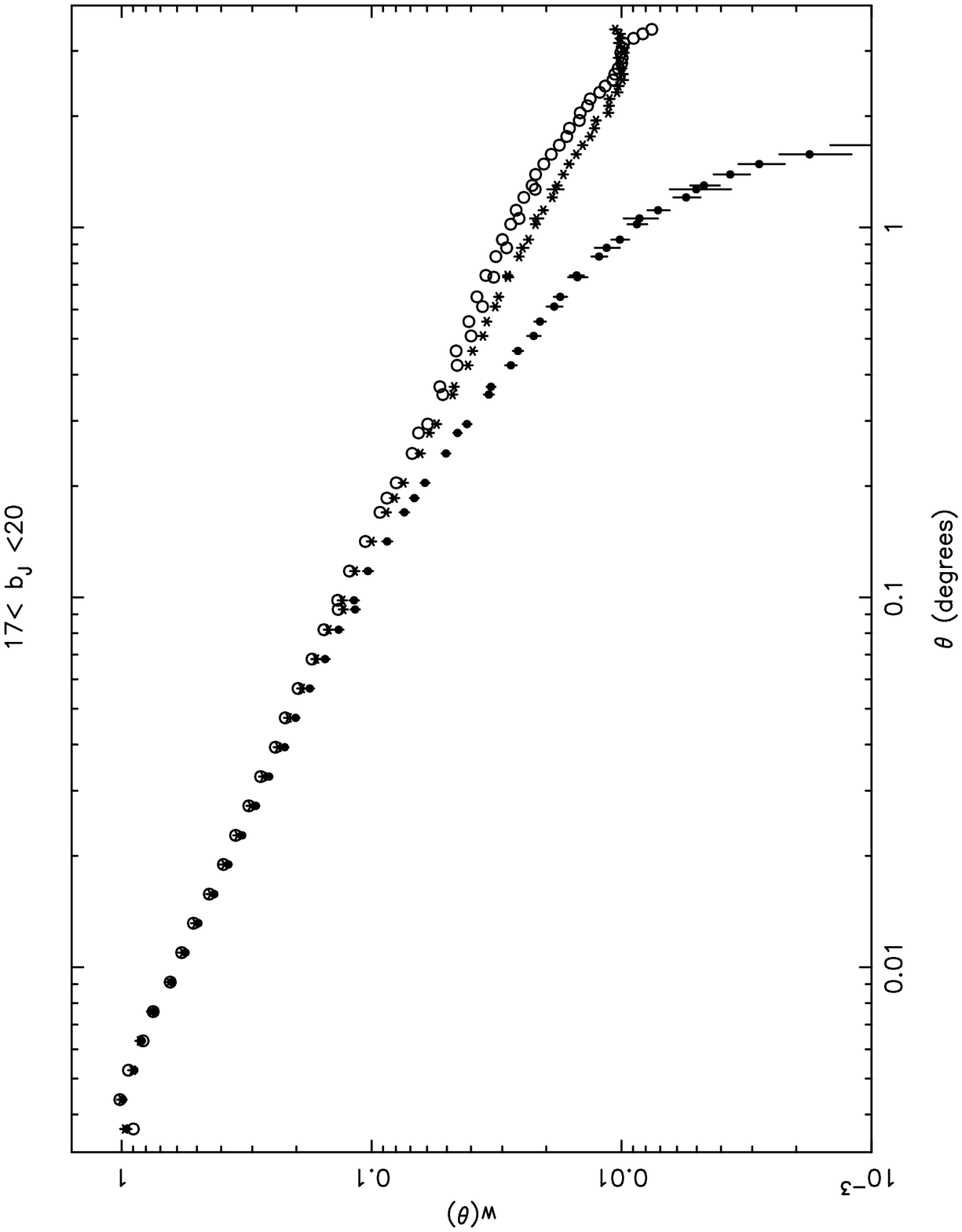}

\noindent
{\bf Figure 1}. The filled circles show the mean of $w(\theta)$ for
galaxies in the magnitude range $17 < b_J < 20$ determined by applying
the direct pair count estimator (equation 6) to $185$ Schmidt fields
using the galaxy counts on each individual plate to normalize $w$.
The open circles show the mean value of $w(\theta)$ using a global
estimate of the mean surface density to normalize $w$.
The stars show how $w(\theta)$ changes if a constant, $\overline
\delta^2 = 1.4 \times 10^{-2}$ is added to the filled circles to
correct for the bias introduced by the integral constraint (equation
10).

\vfill\eject

\noindent
$\;$

\vskip 7.2 truein

\includegraphics{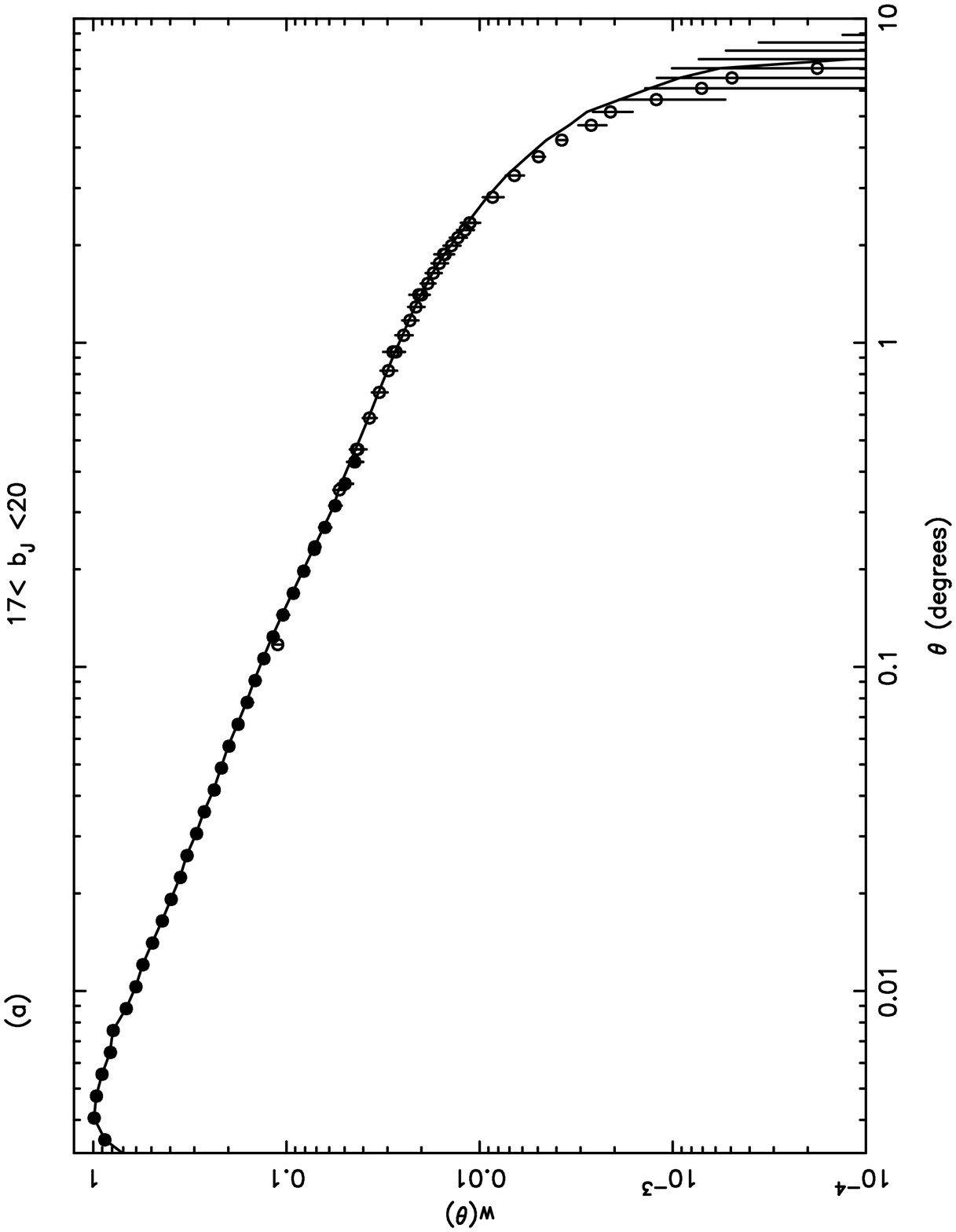}

\includegraphics{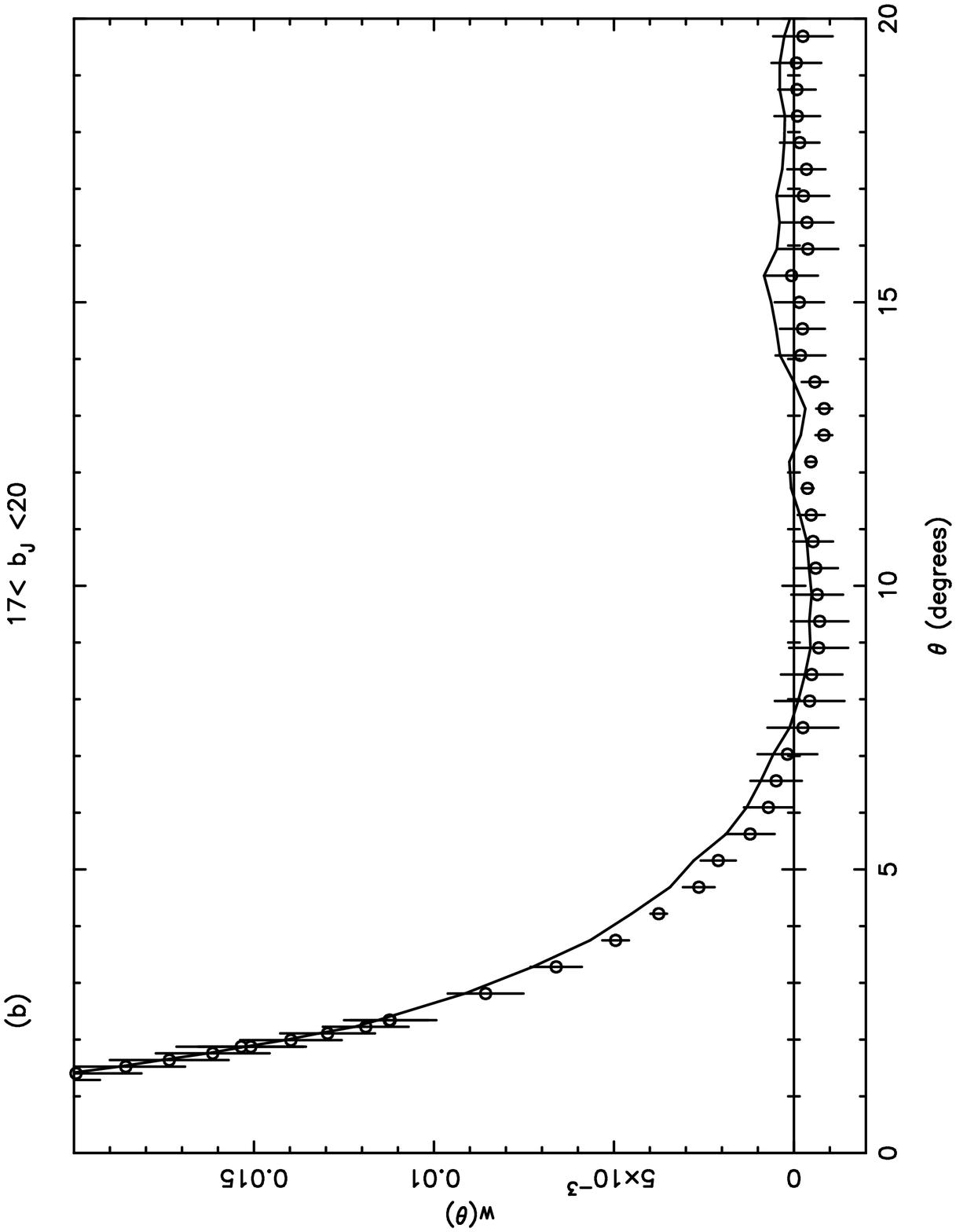}

\noindent
{\bf Figure 2}.  Estimates of $w(\theta)$ for galaxies with
magnitudes in the range $17 < b_J < 20$ plotted with logarithmic axes
in Figure (2a) and linear axes in Figure (2b).
The line shows the estimate for the full area of the survey.
The open circles at $\theta > 0.1^\circ$ show the mean of $w(\theta)$
derived by applying the ensemble estimator (equation 5) to four
separate zones of approximately equal area.
The error bars show one standard deviation on the mean determined from
the zone-to-zone scatter.
The filled circles at $\theta < 0.5^\circ$ show the mean of
$w(\theta)$ and $1\sigma$ errors using the direct pair count estimator
applied to $185$ Schmidt plates (plotted as the stars in Figure 1).

\vfill\eject

\noindent
$\;$

\vskip 6 truein

\includegraphics{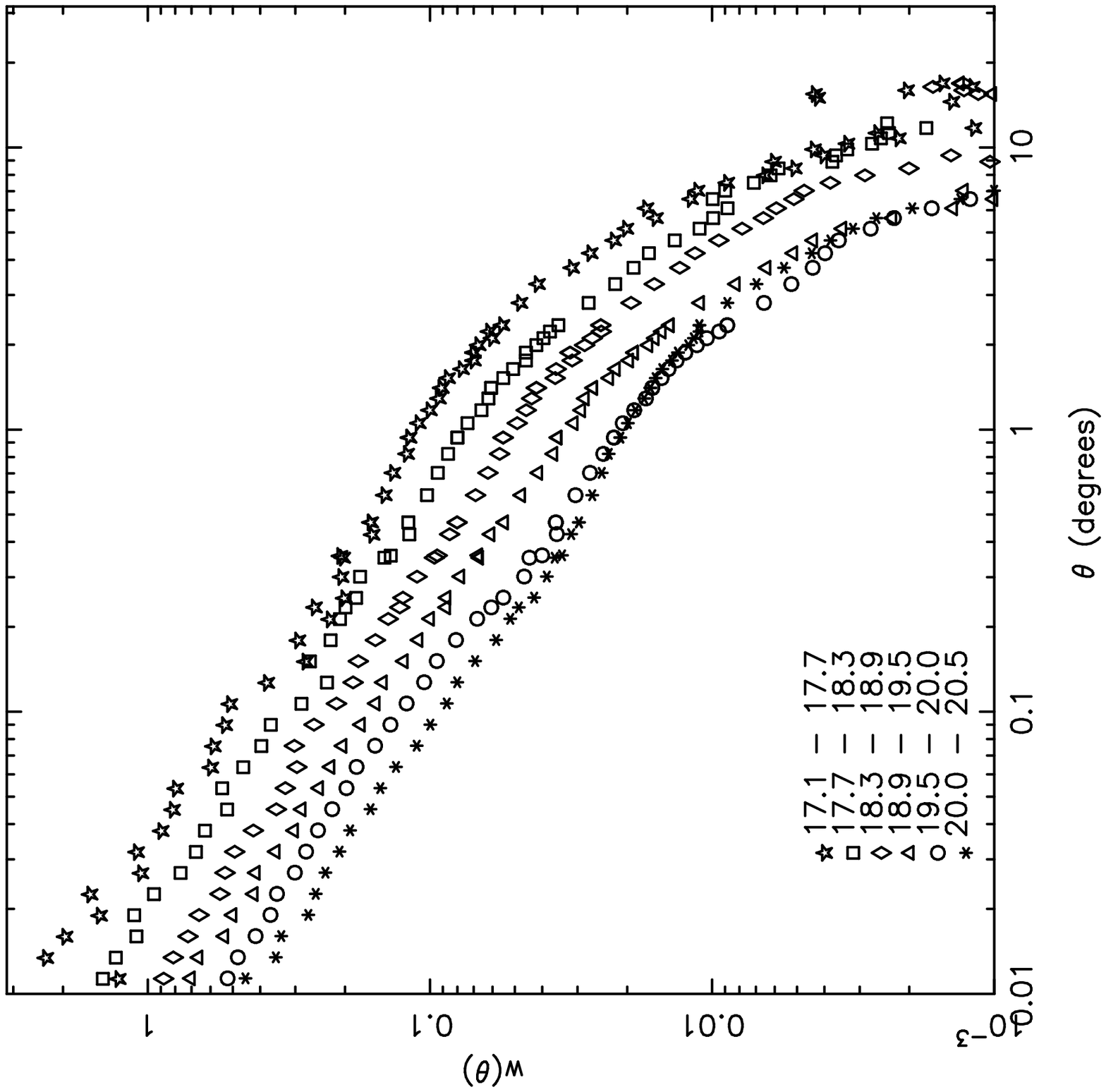}

\noindent
{\bf Figure 3}. Estimates of $w(\theta)$ for six disjoint magnitude
slices in the range $ 17.0 <b_J < 20.5$ derived from the central $120$
plates of the survey at Galactic latitudes $b < -50^\circ$.  For each
magnitude slice the angular correlation functions at $\theta >
0.4^\circ $ are derived by applying the ensemble estimator (equation
5) to the central part of the equal area maps.  At $\theta <
0.4^\circ$ $w(\theta)$ is derived by applying the direct pair count
estimator to the $120$ plates using a global estimate of the mean
surface density to normalize $w$.

\vfill\eject

\noindent
$\;$

\vskip 7.5 truein

\includegraphics{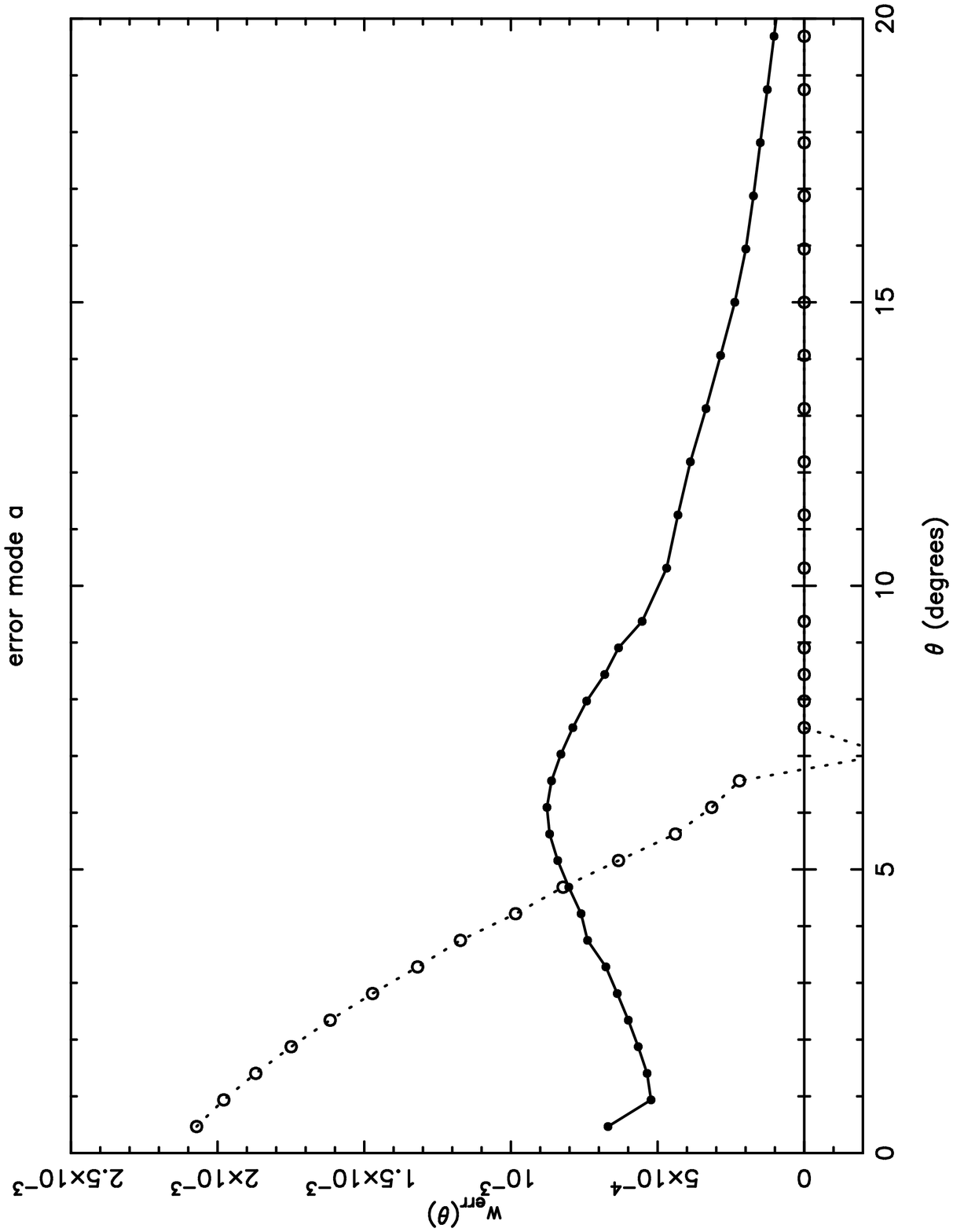}
\includegraphics{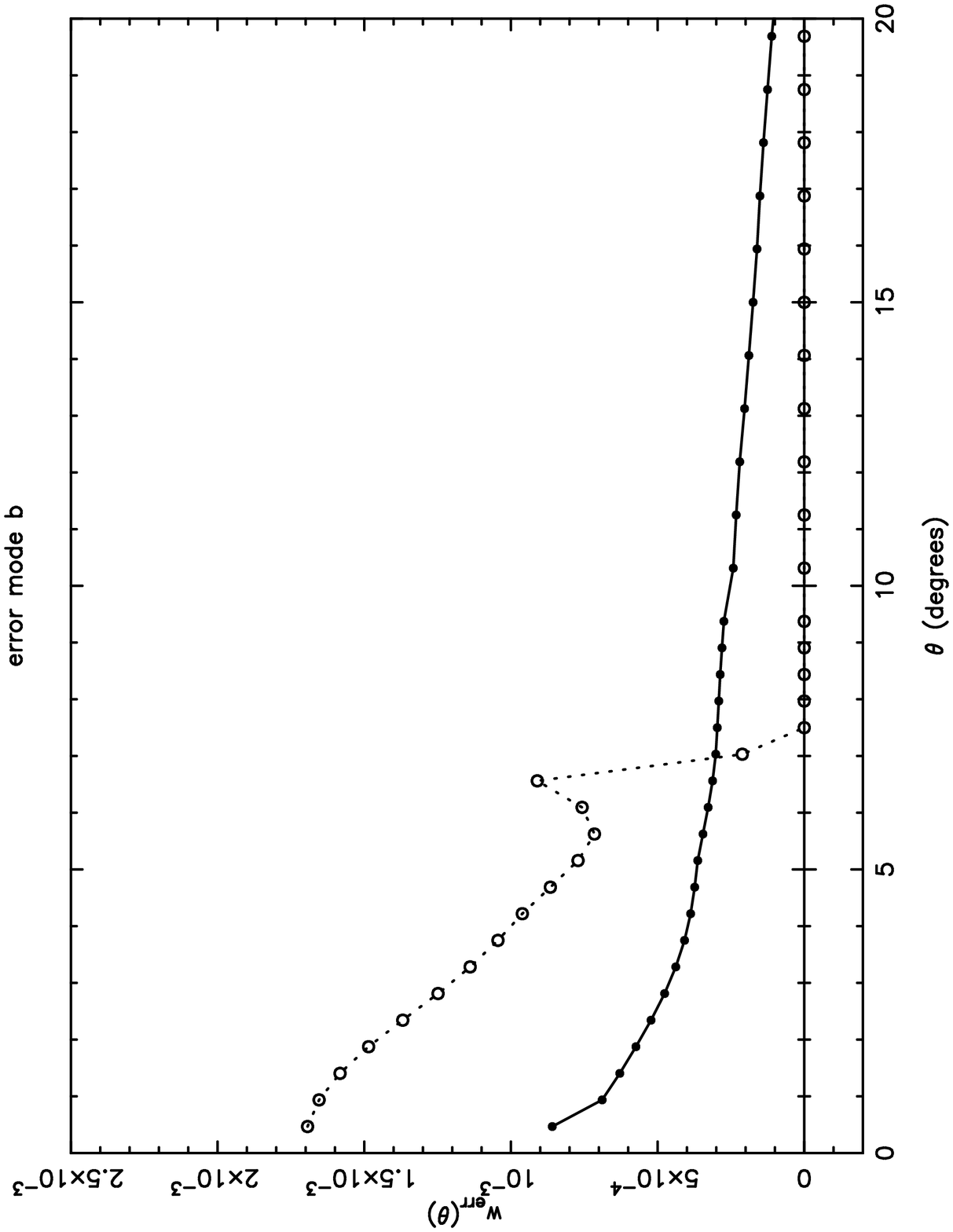}
\includegraphics{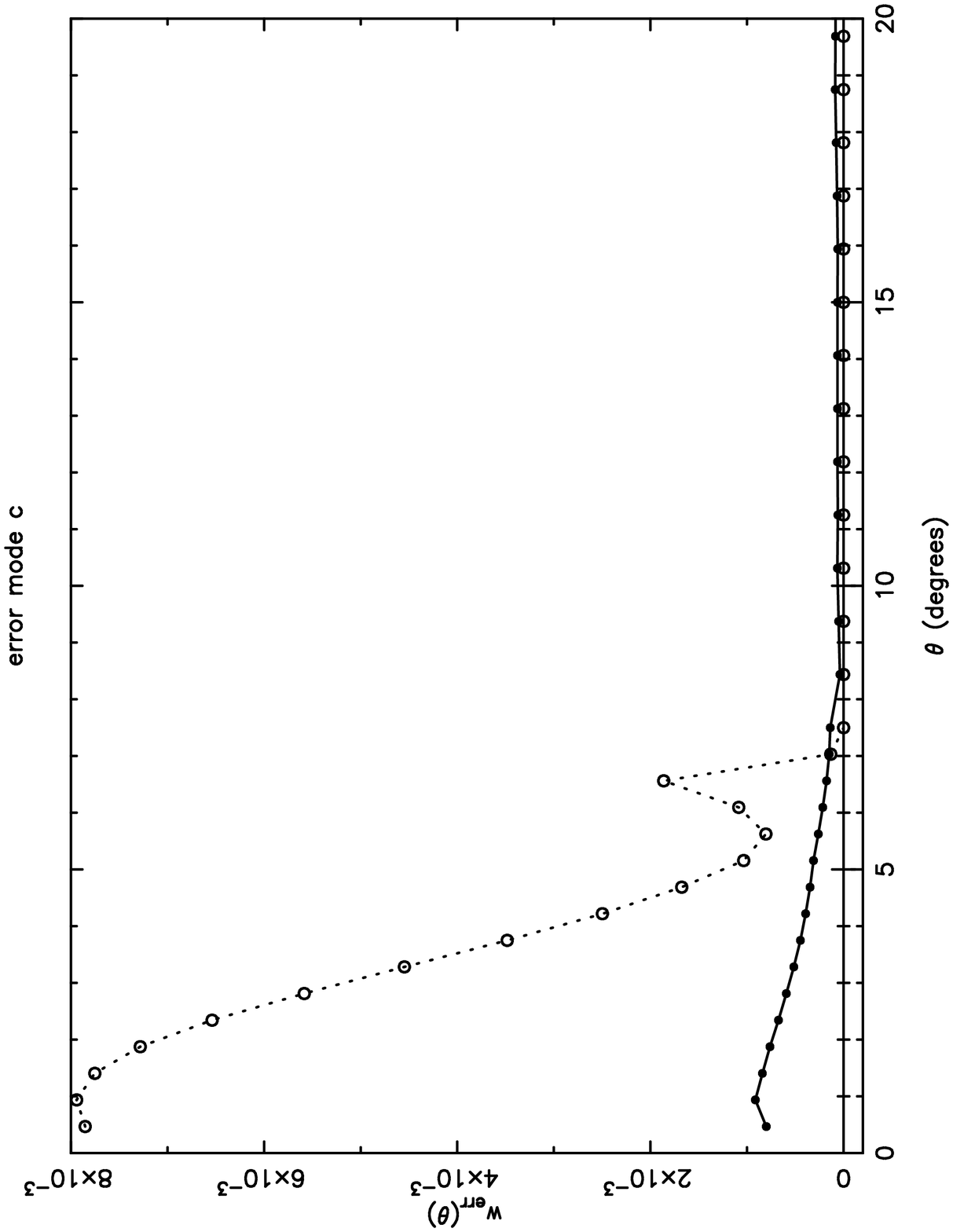}

\noindent
{\bf Figure 4}. Angular correlation functions $w_{err}$ estimated
from simulated maps which include correlated overlap errors 
according to the modes (A)-(C) described in the text. The filled
circles show the inter-plate estimates of $w$ and open circles
show the intra-plate estimates.

\vfill\eject

\noindent
$\;$

\vskip 6 truein

\includegraphics{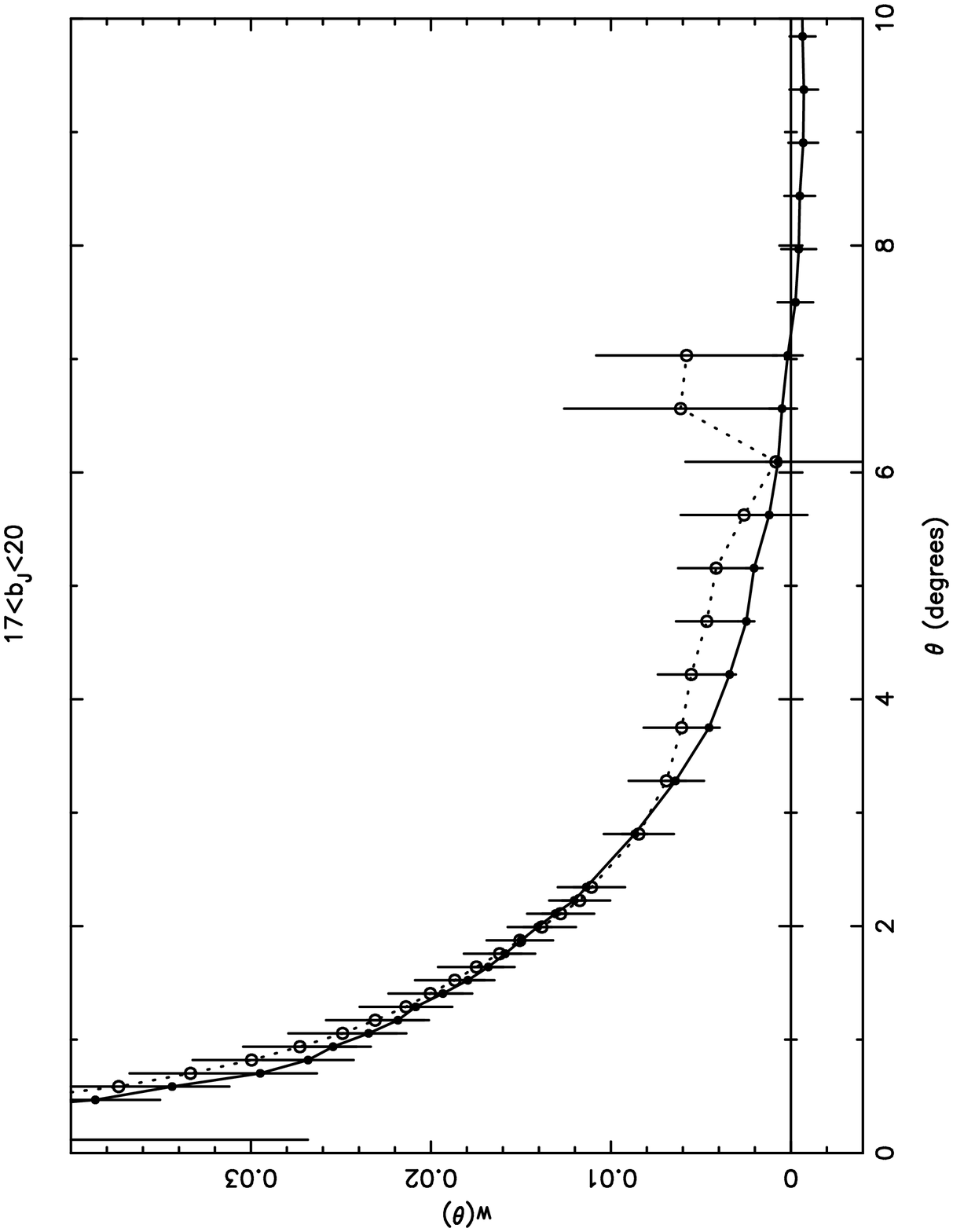}

\noindent
{\bf Figure 5.} Inter- and intra-plate estimates of $w(\theta)$
for galaxies in the APM survey in the magnitude range
$17 < b_J < 20$. The filled circles show the inter-plate
estimates and open circles show the intra-plate estimates.
The error bars are derived from the scatter of four nearly
equal area zones.

\vfill\eject

\noindent
$\;$

\vskip 6 truein

\includegraphics{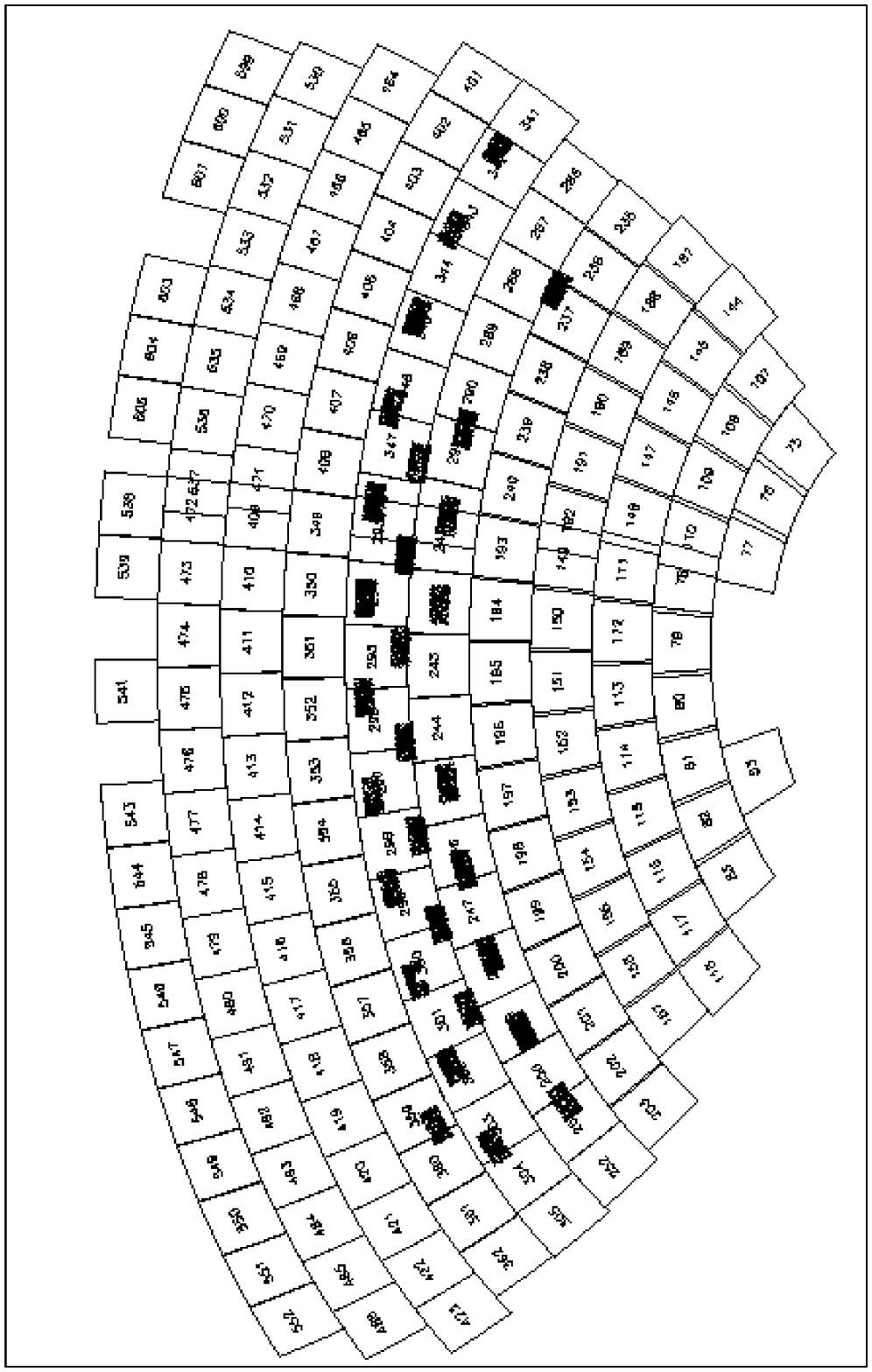}

\noindent
{\bf Figure 6.} The position of the LCDRS fields relative to the UK
Schmidt survey fields in the APM survey area. 
The map is in an equal area projection centred on the South Galactic
Pole. 
A $5^\circ$ square is plotted for each Schmidt field, and a dot is
plotted for each LCDRS galaxy. 
Each LCDRS field has an area of  $1.5^\circ \times 3.0^\circ$. 

\vfill\eject

\noindent
$\;$

\vskip 6 truein

\includegraphics{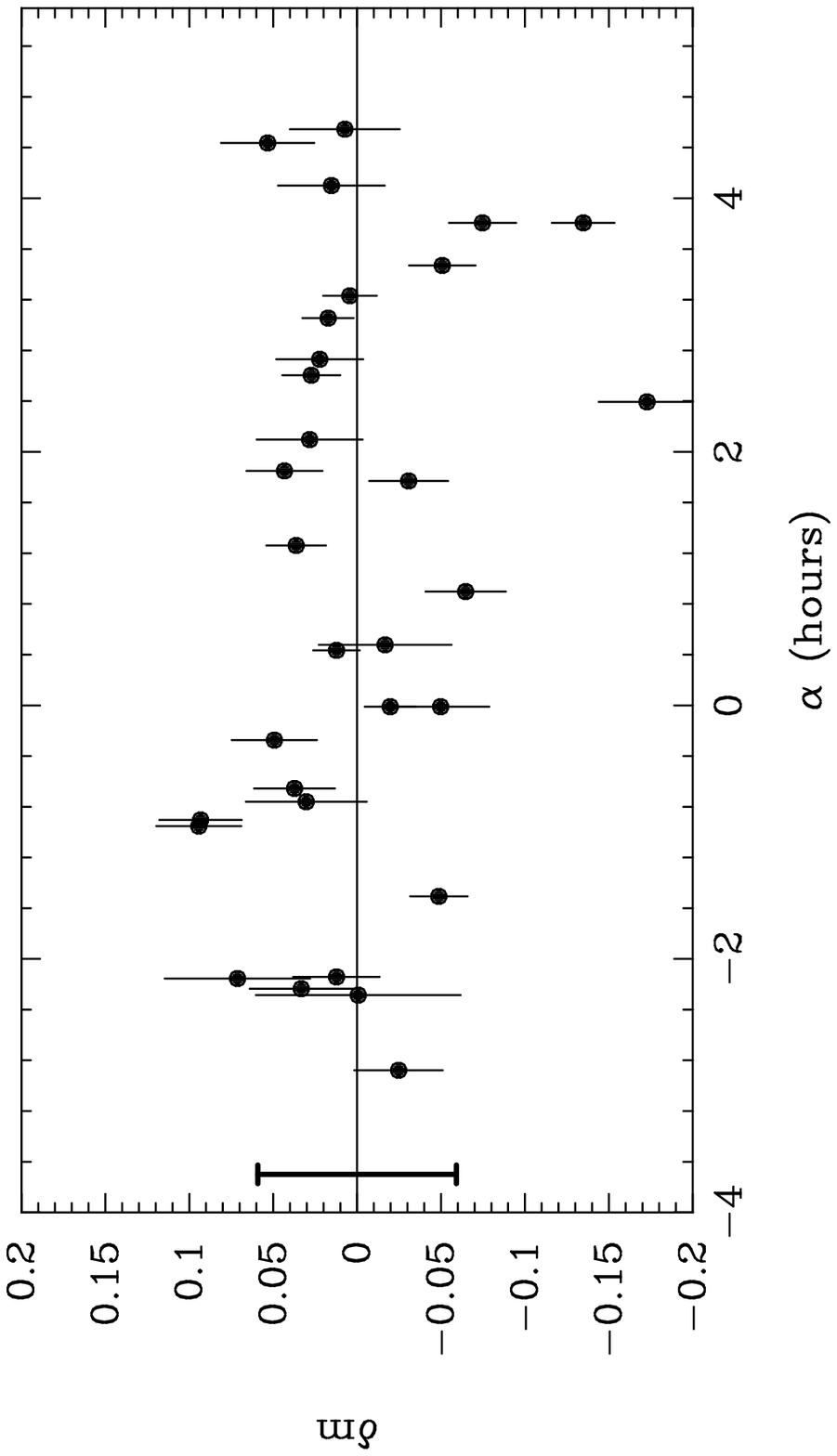}

\noindent
{\bf Figure 7}. Magnitude residuals of the Las-Campanas-APM 
photometry plotted against right ascension. The thick error
bar plotted to the left shows the {\it rms} scatter of the
points about zero (0.051 magnitudes). The error bars on the
points arise mainly from the scatter in galaxy colours.

\vfill\eject

\noindent
$\;$

\vskip 6 truein

\includegraphics{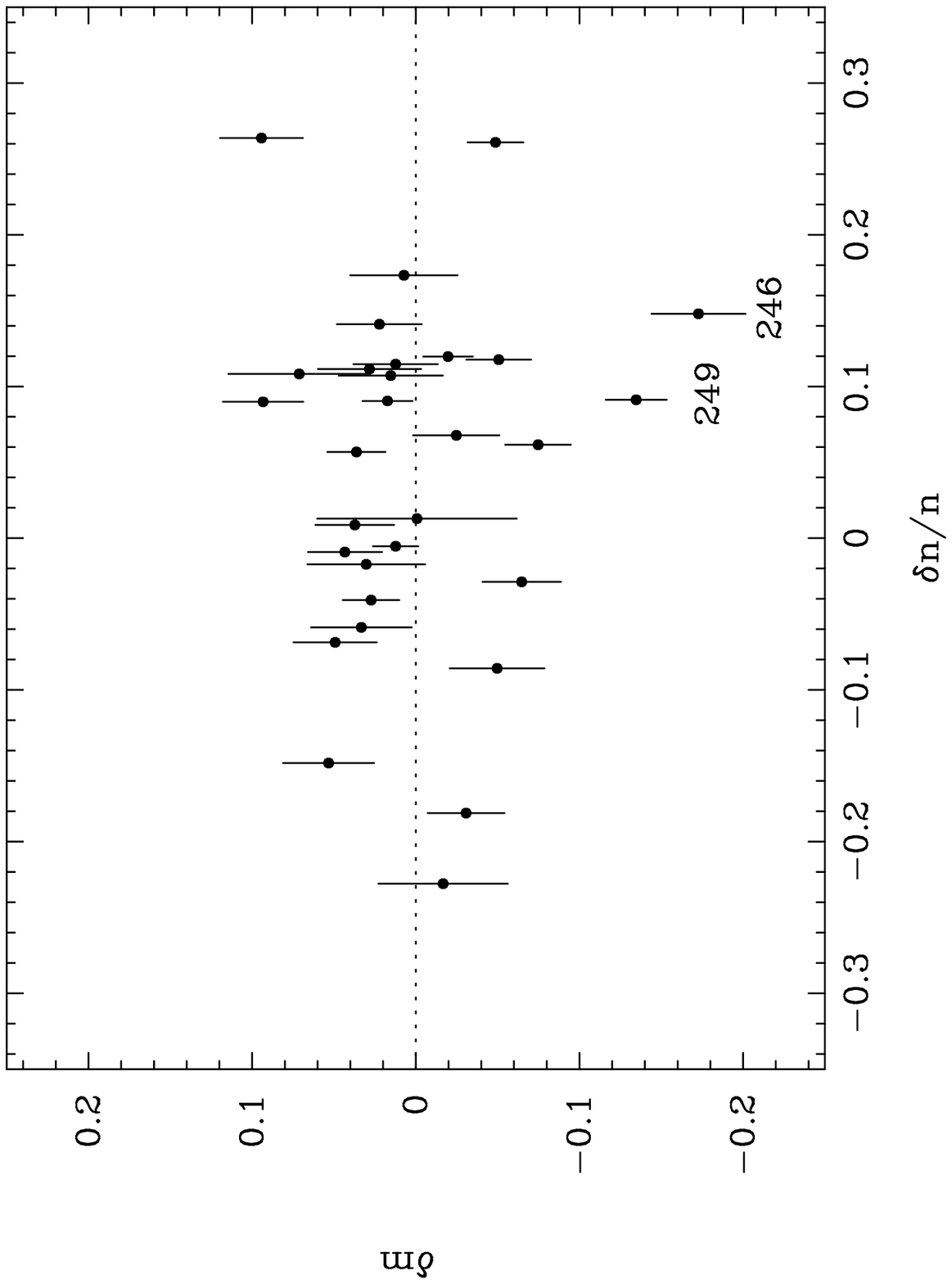}

\noindent
{\bf Figure 8}. The Las Campanas-APM magnitude residuals
plotted against fluctuations in the galaxy surface density 
in each APM field to a magnitude limit $b_J = 20$. The two
outlying points in Figure 7, APM fields 246 and 249, are
indicated.

\vfill\eject

\noindent
$\;$

\vskip 6 truein

\includegraphics{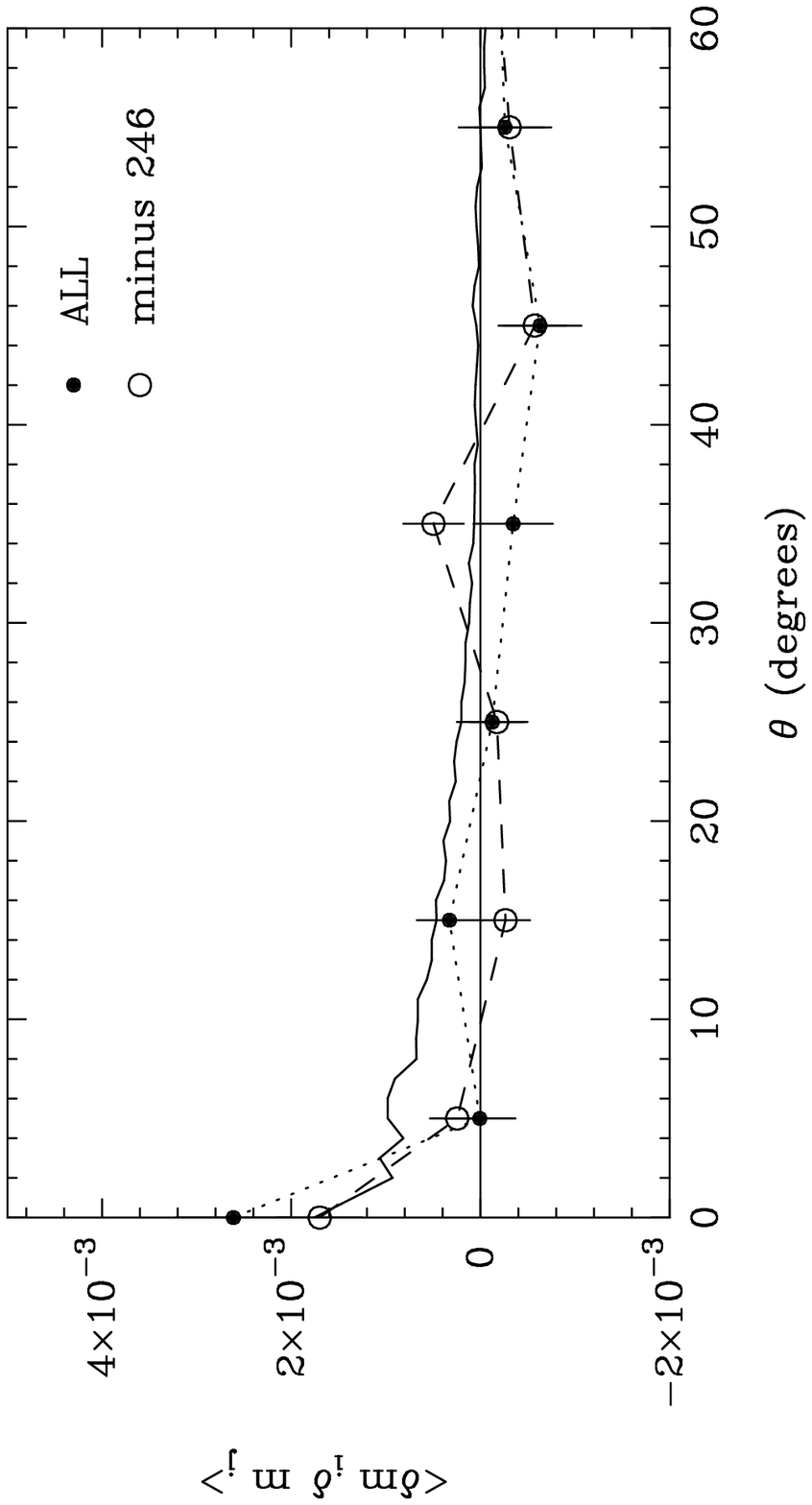}

\noindent
{\bf Figure 9}. The correlation function of Las Campanas-APM
magnitude residuals (equation 24) plotted as a function of angle. The
filled points show the correlation function of all fields and the open
circles show how the correlation function changes when the outlying
field F246 is removed.  The error bars are computed from equation
(25).  The solid line shows the Monte-Carlo model for mode [A] 
magnitude errors as described in Section 3.1.3.

\vfill\eject

\noindent
$\;$

\vskip 6 truein

\includegraphics{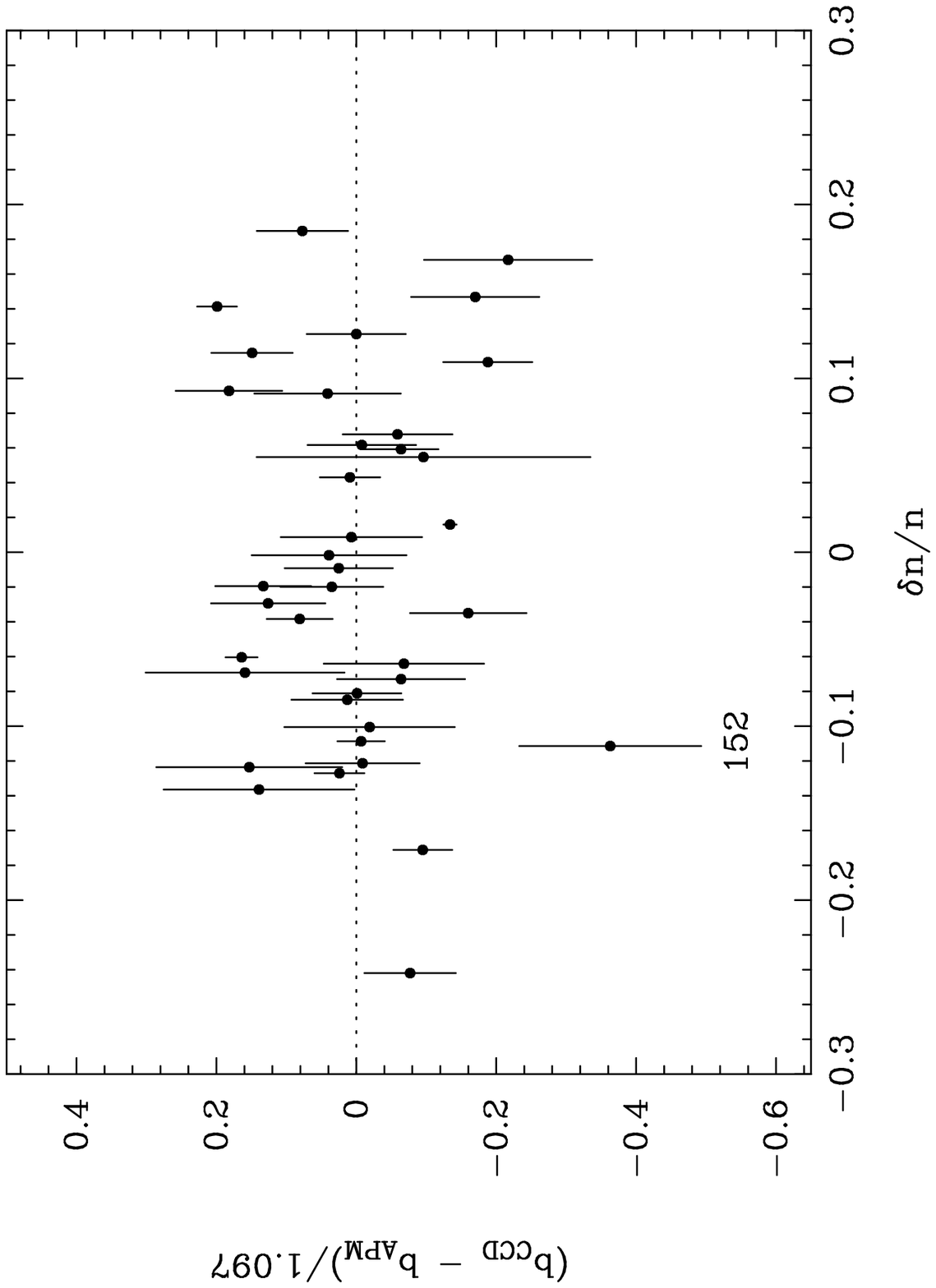}

\noindent
{\bf Figure 10}. The Maddox et al CCD - APM magnitude residuals
($\overline c_i - \overline c$ in equation 27) plotted
against fluctuations in the galaxy surface density 
in each APM field to a magnitude limit $b_J = 20$.
The error bars show the {\it rms} fluctuation of the
magnitude differences in each sequence listed in the
final column of Table 3. The outlying APM field $152$
is indicated.

\vfill\eject

\noindent
$\;$

\vskip 6 truein

\includegraphics{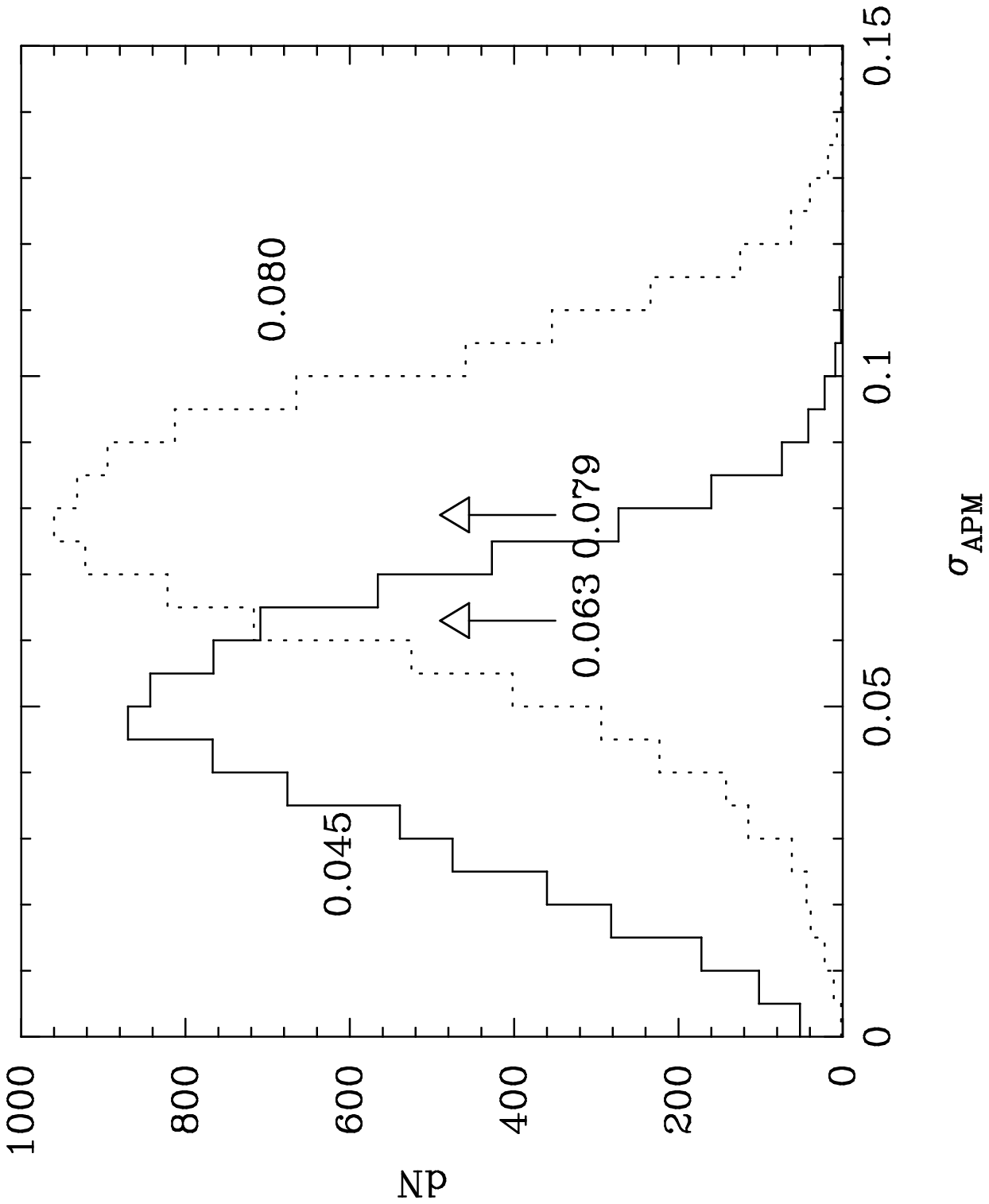}

\noindent
{\bf Figure 11}. Histograms illustrating the accuracy with
which $\sigma_{APM}$ can be determined from the Maddox et al CCD
sequences. The diagrams show the distribution of $\sigma_{APM}$
from $10000$ Monte-Carlo simulations of the Maddox et al CCD sequences
as described in the text. The solid histogram shows the distribution
if the true value of $\sigma_{APM}=0.045$ and the dashed histogram
shows the distribution if the true value is $\sigma_{APM} = 0.080$
as advocated by Fong \etal (1992). The arrows show the values
determined from the real data, $0.079$ if all fields are included
and $0.063$ if the outlying sequence on field 152 is excluded.

\vfill\eject

\noindent
$\;$

\vskip 6 truein

\includegraphics{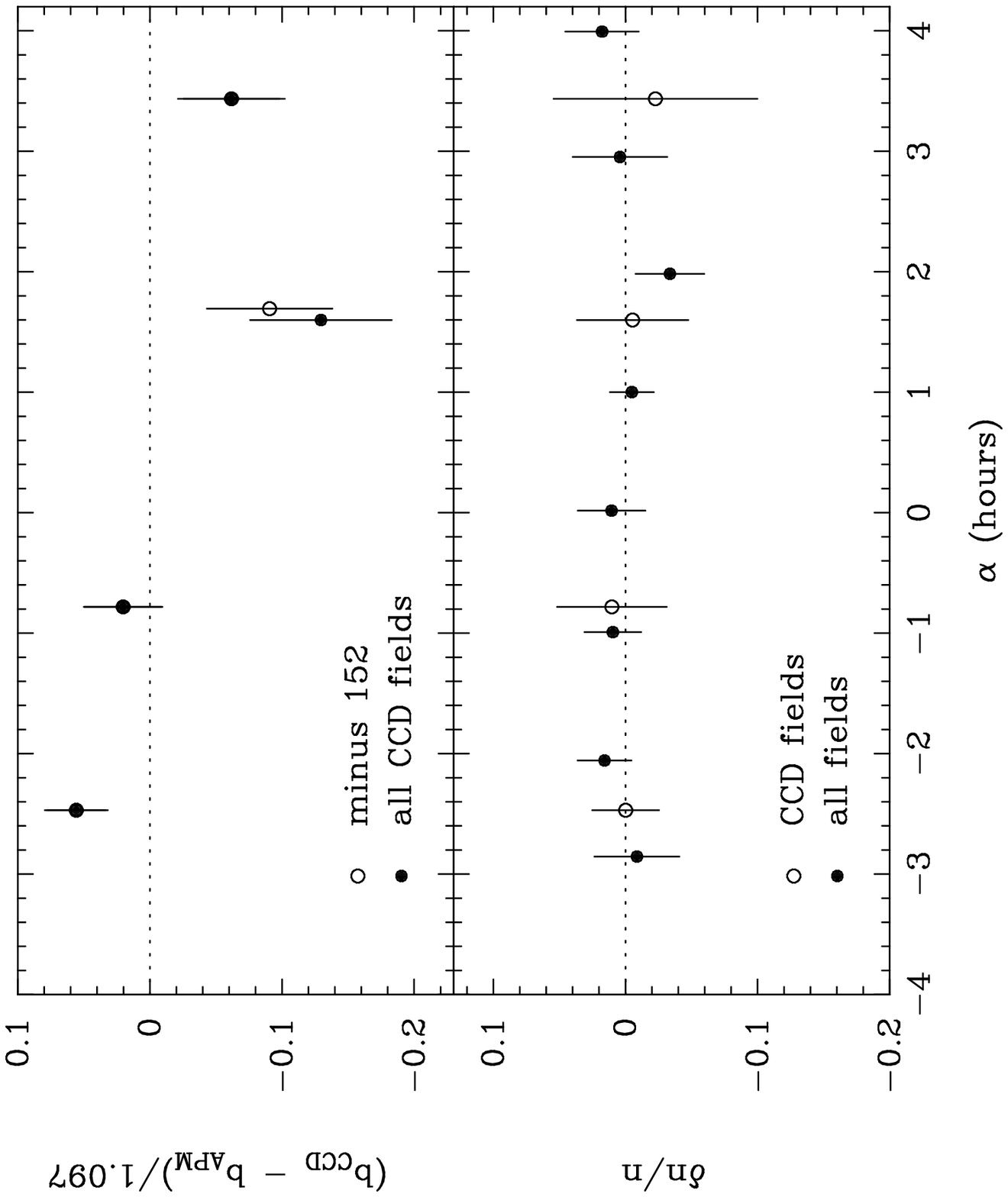}

\noindent
{\bf Figure 12}. The upper panel shows Maddox et al CCD - APM magnitude
residuals plotted against right ascension. The lower
panel shows the fluctuations in the galaxy surface density
to $b_J=20$ in the APM fields plotted against right ascension.
The filled circles in the lower panel show all fields in the 
APM survey and the open circles show only those fields 
in which there is a Maddox et al CCD sequence.

\vfill\eject

\noindent
$\;$

\vskip 6 truein

\includegraphics{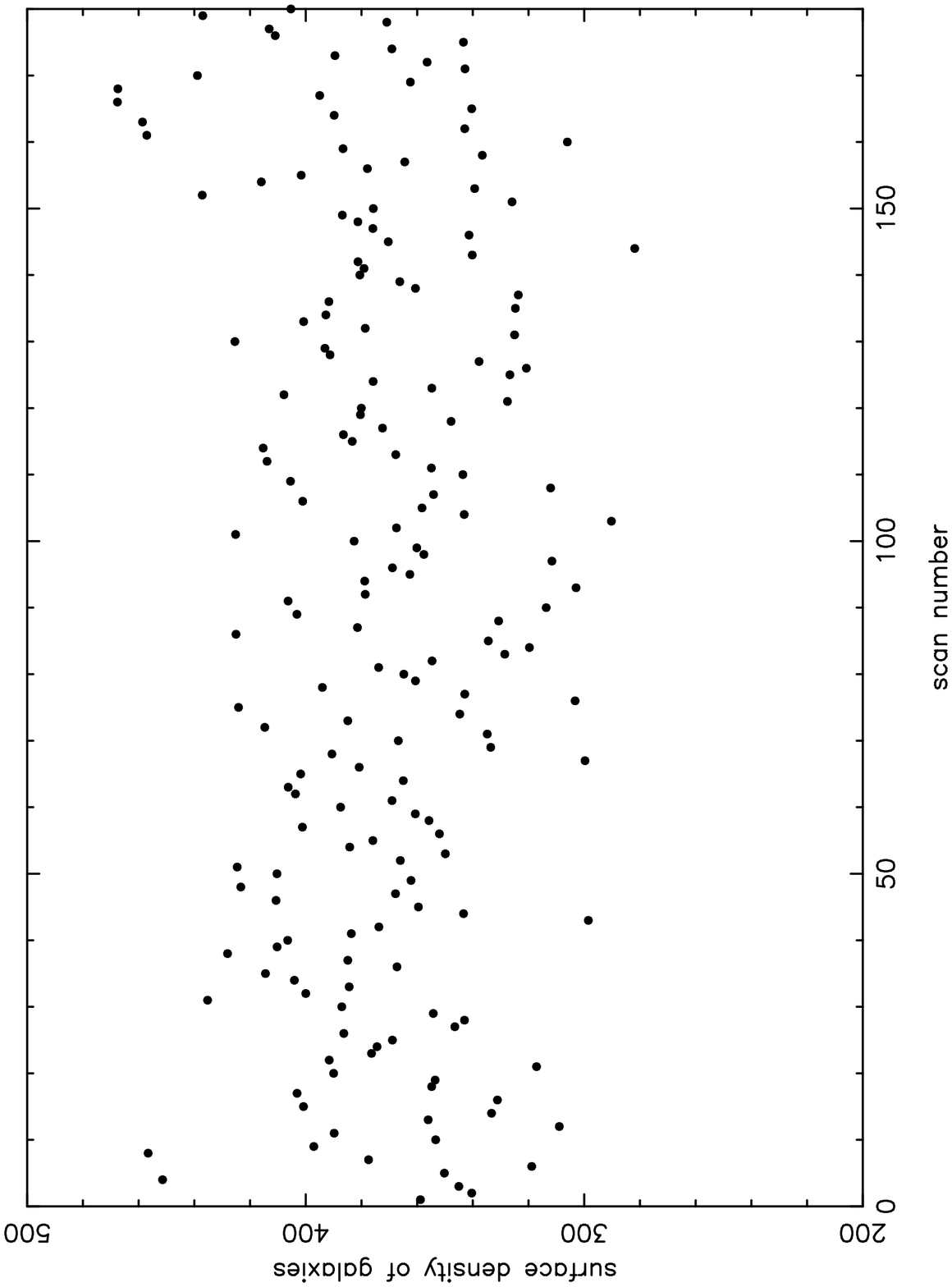}

\noindent 
{\bf Figure 13}. A scatter plot of the surface density of galaxies
brighter than $\bj = 20.5 $ on each plate in number per square degree,
plotted against the sequential number of each APM scan.

\vfill\eject

\noindent
$\;$

\vskip 6 truein

\includegraphics{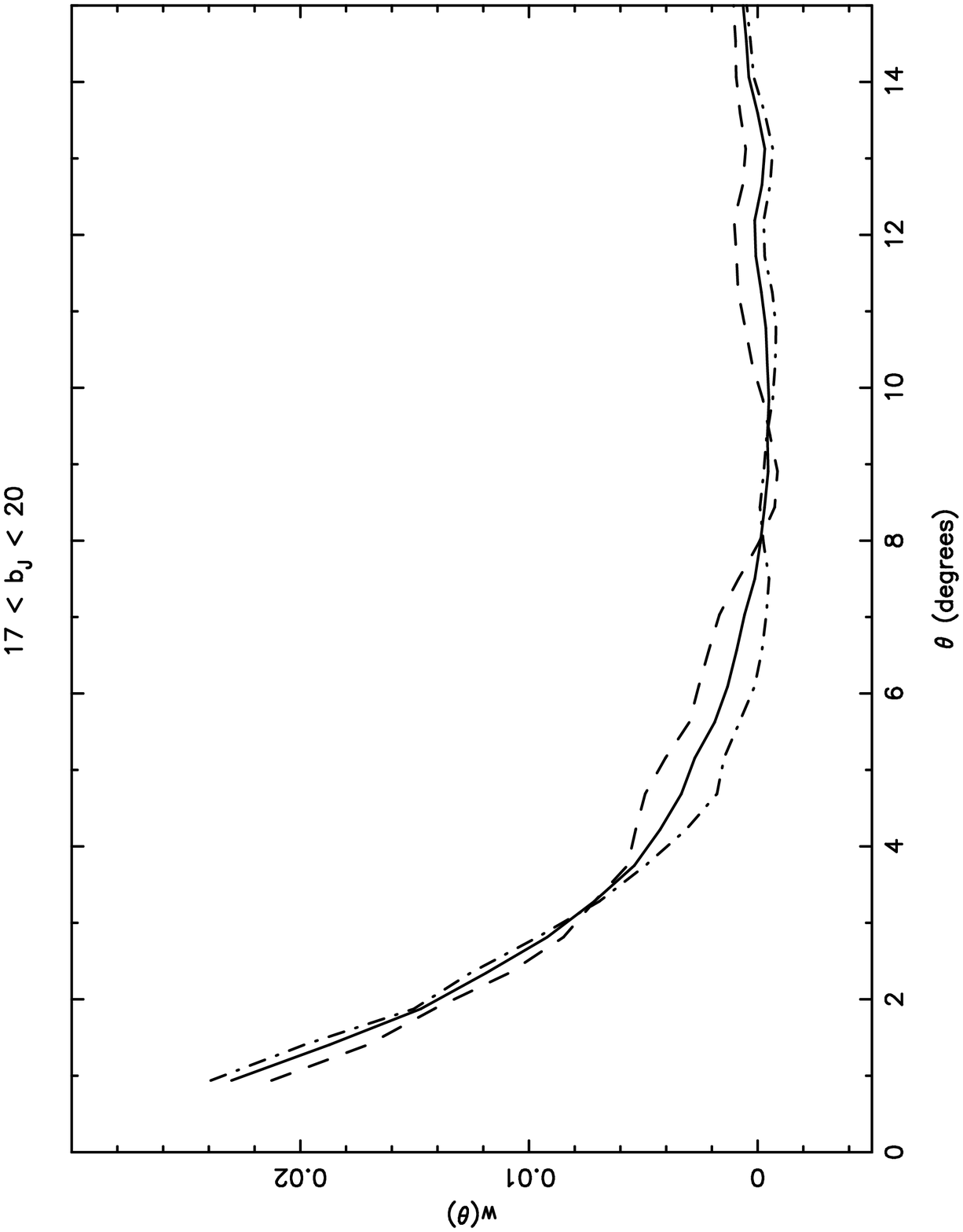}

\noindent
{\bf Figure 14}. Inter-plate angular correlation functions for plate pairs 
scanned within $1$ year of each other (dashed line) and scanned more than
one year appart (dot-dashed line). The solid line shows
the correlation function estimated from all galaxy pairs.

\vfill\eject

\noindent
$\;$

\vskip 8.
 truein

\includegraphics{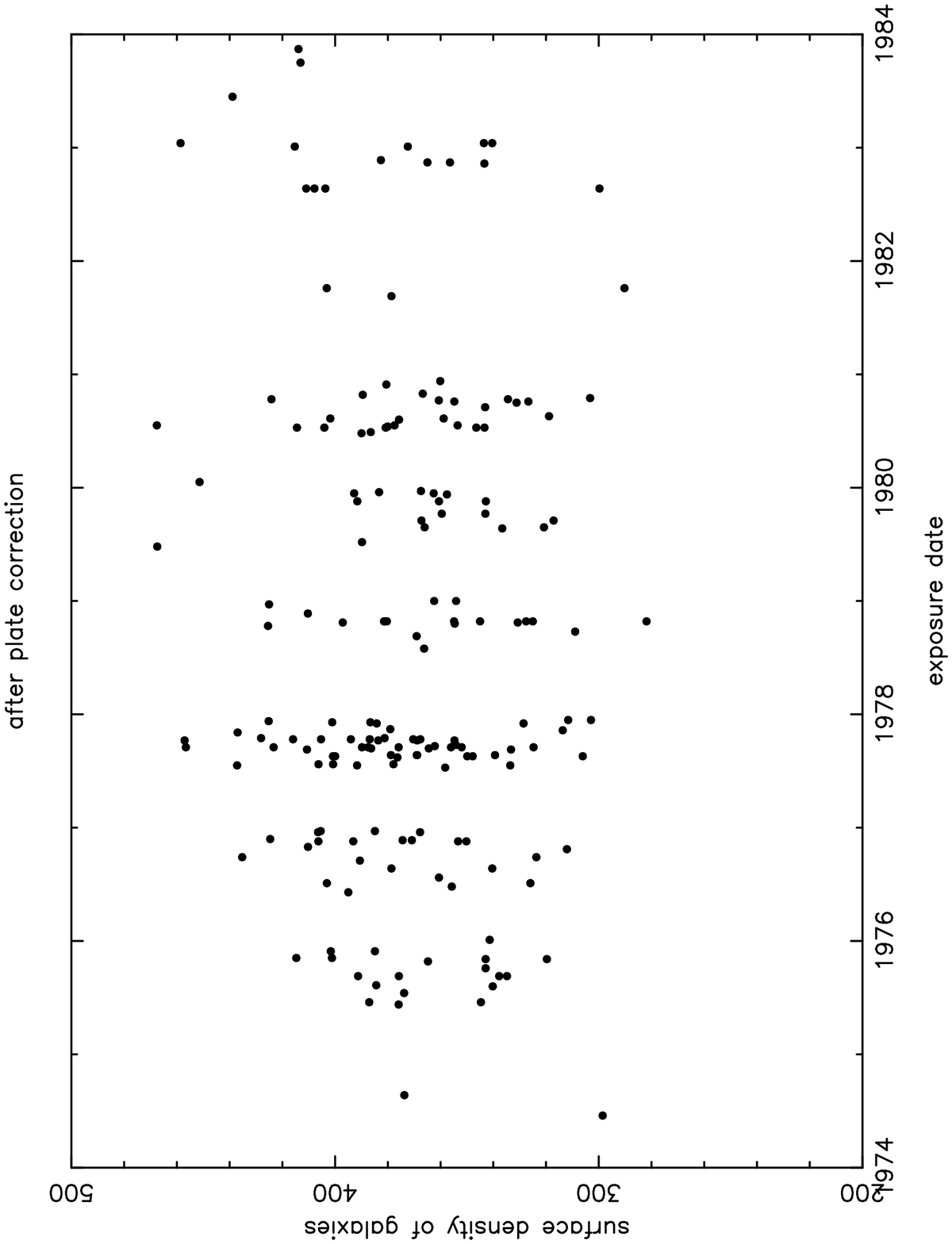}

\noindent
{\bf Figure 15}. A scatter plot of the surface density of galaxies
brighter than $\bj = 20.5 $ on each plate in number per square degree
plotted against the date of exposure of the Schmidt plate.  No
significant trends in the number of galaxies are seen.  

\vfill\eject

\noindent
$\;$

\vskip 6 truein

\includegraphics{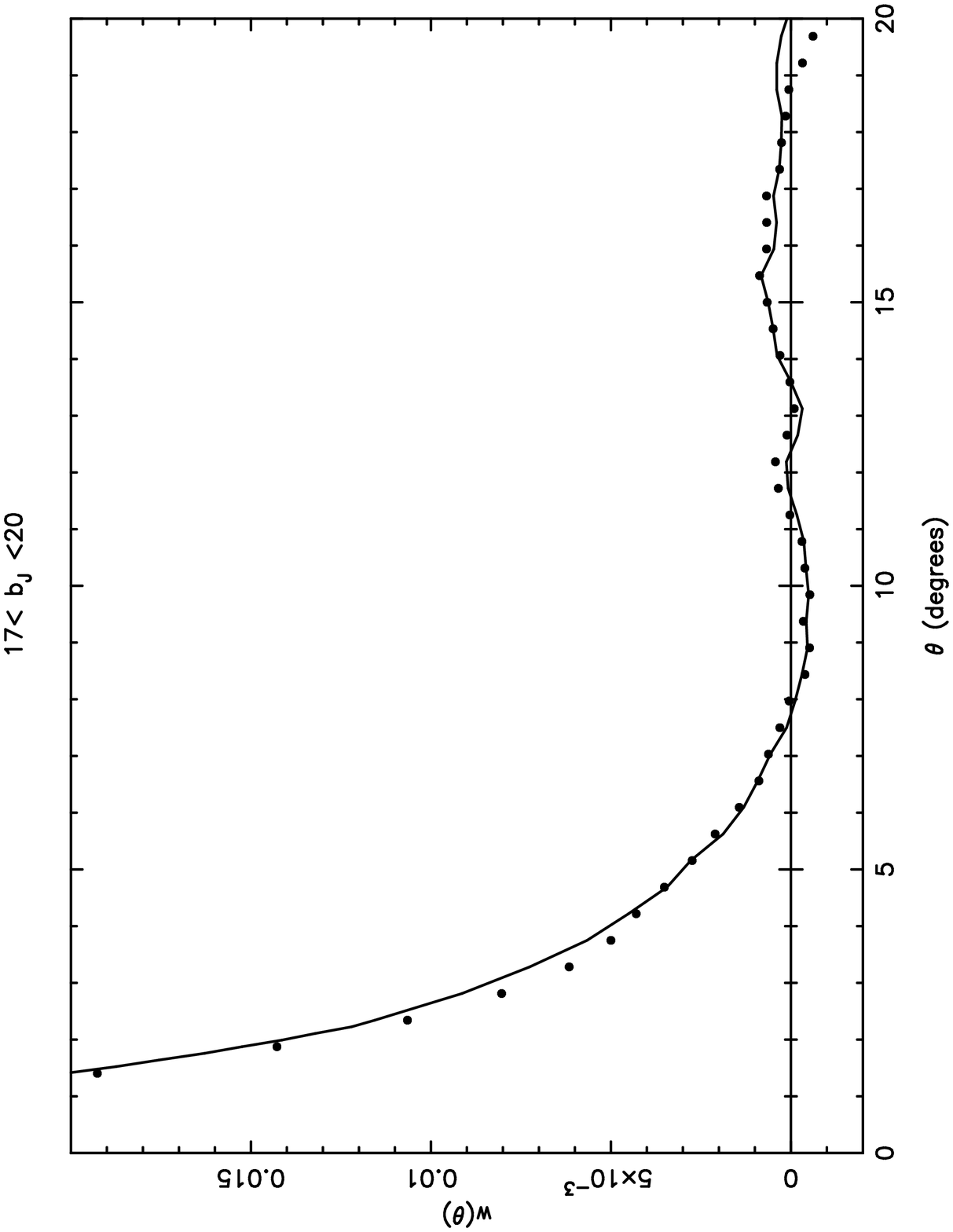}

\noindent
{\bf Figure 16}. The points show the galaxy autocorrelation function for
galaxies $17.5< b_J < 20.0 $ after correcting their magnitudes 
for the extinction estimated from the IRAS $100 \mu m$ flux. 
The solid line shows the correlation function from the uncorrected map
as plotted in Figure~2. 

\vfill\eject

\noindent
$\;$

\vskip 6 truein

\includegraphics{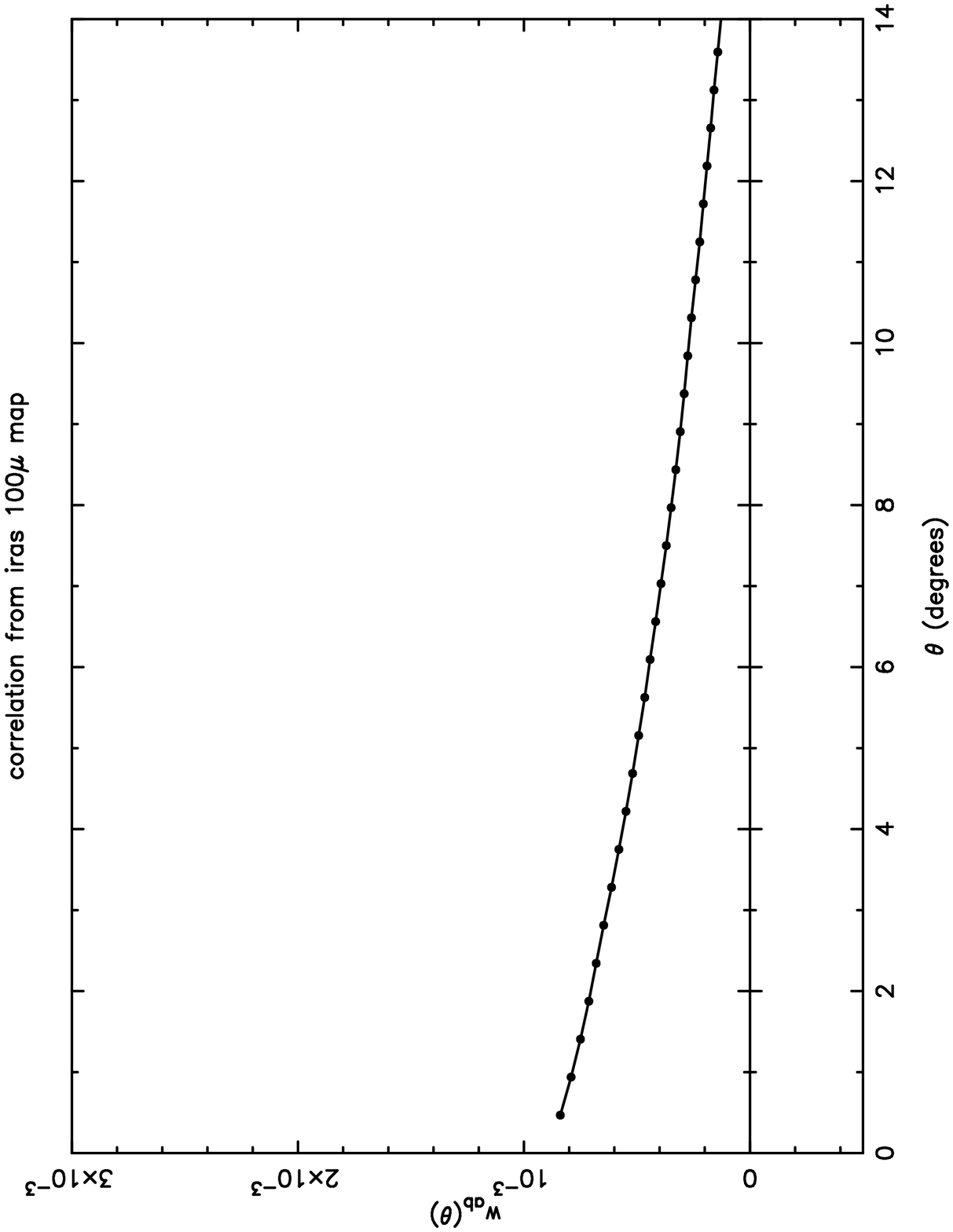}

\noindent
{\bf Figure 17}. The covariance function of Galactic
obscuration over the APM survey area as determined from IRAS
maps of $100\mu$m emission (see text for further details).  

\vfill\eject

\noindent
$\;$

\vskip 6 truein

\includegraphics{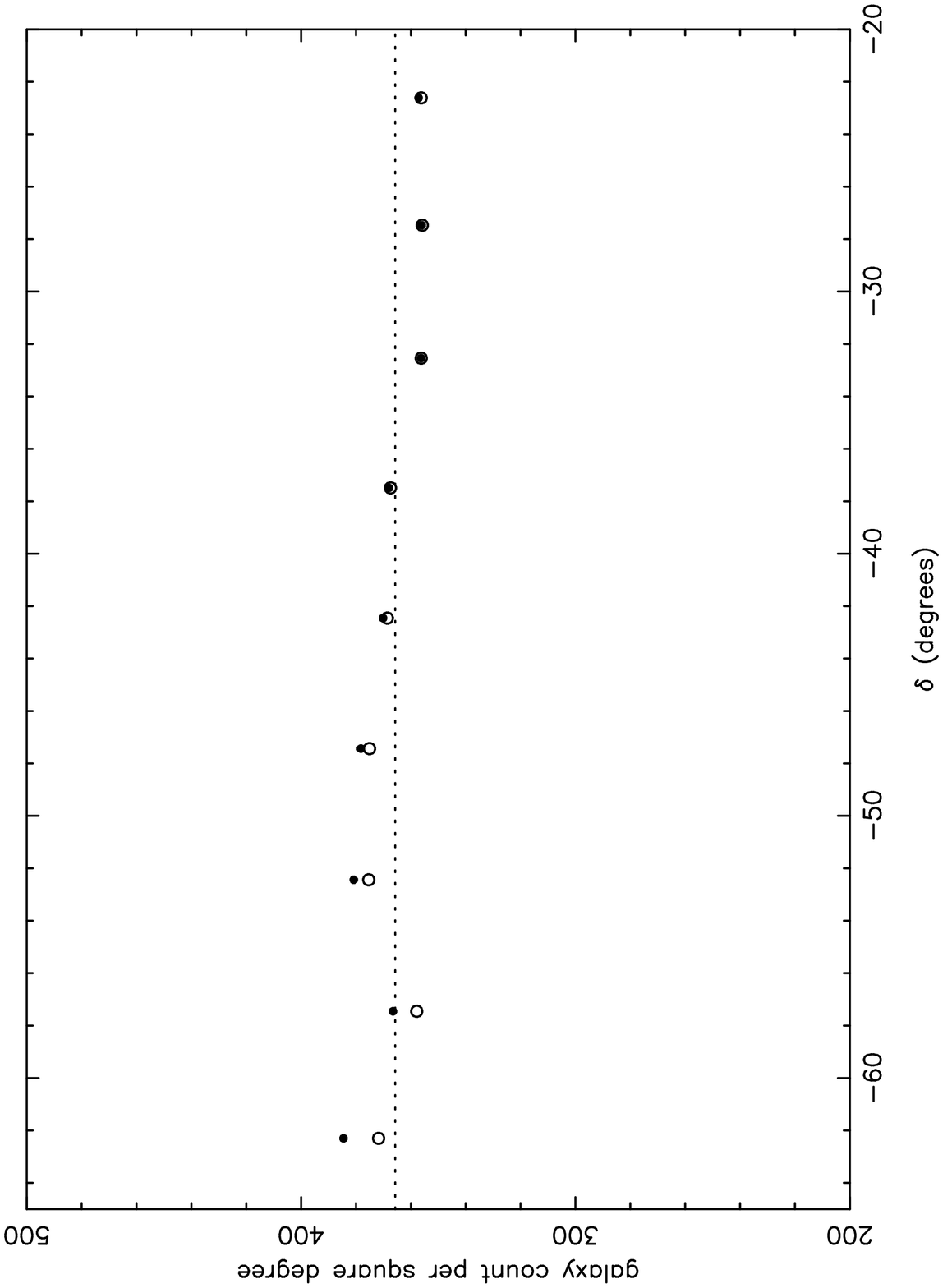}

\noindent
{\bf Figure 18}. The mean galaxy surface density 
in the magnitude range $17 \le b_J \le 20$ plotted as a
function of declination.  
The open symbols show the density before applying the correction for
atmospheric extinction, and the filled symbols show the density after
correction. 
The changes in density are smaller than the variations caused by real
structure in the galaxy distribution. 
The dotted line shows the mean galaxy count averaged over the entire
survey.

\vfill\eject

\noindent
$\;$

\vskip 6 truein

\includegraphics{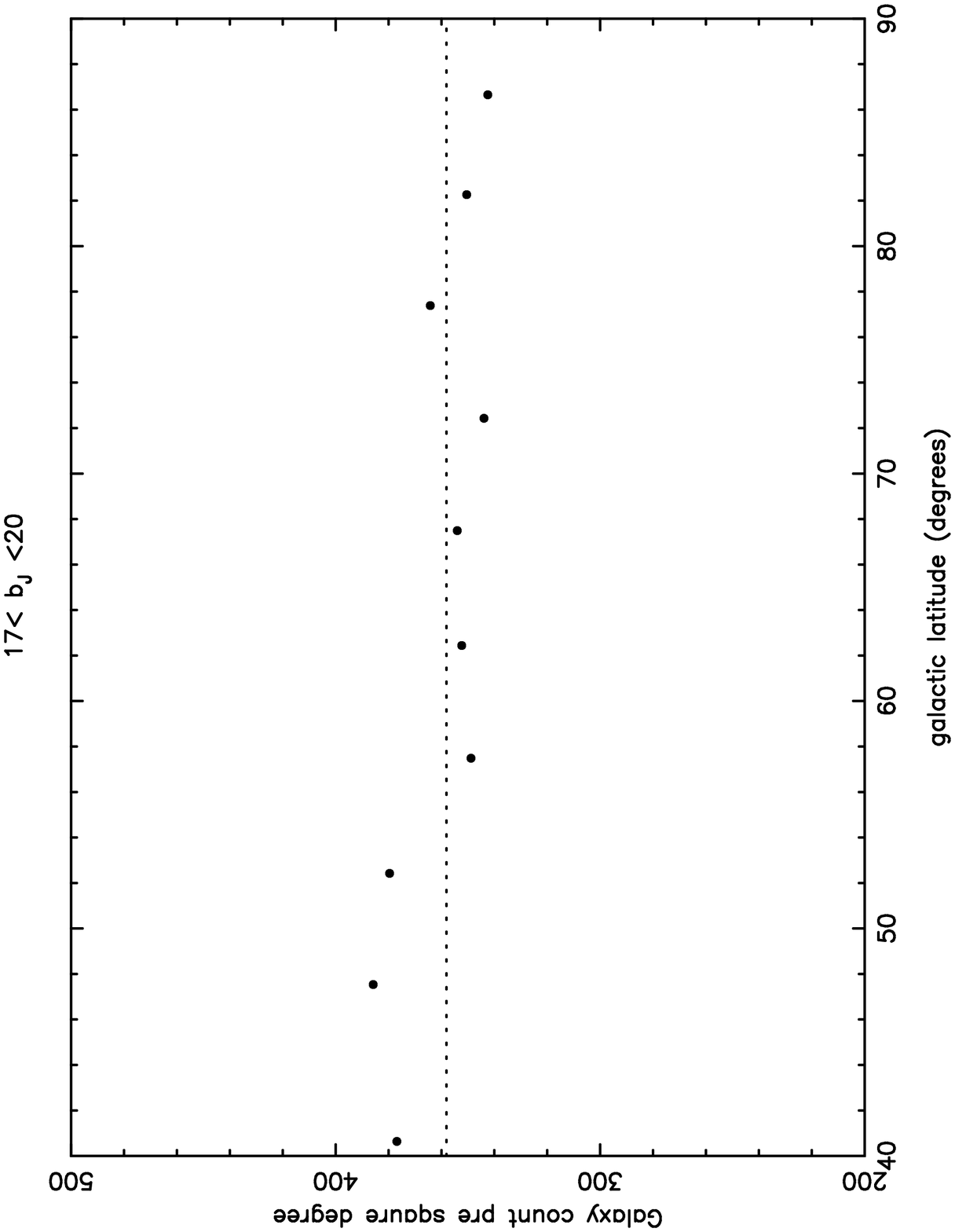}

\noindent
{\bf Figure 19}. 
The mean surface density of images in the galaxy sample 
in the magnitude range $17 \le b_J \le 20$ plotted as a
function of galactic latitude after correcting for galactic extinction.   
The dotted line shows the mean surface density. 
A small gradient can be seen, amounting to about $10\%$ in density
from the high to low latitude limits of the survey.
This gradient arises because of  the increase in the number of
stellar blends at low latitudes that are misclassified as galaxies.

\vfill\eject

\noindent
$\;$

\vskip 6 truein

\includegraphics{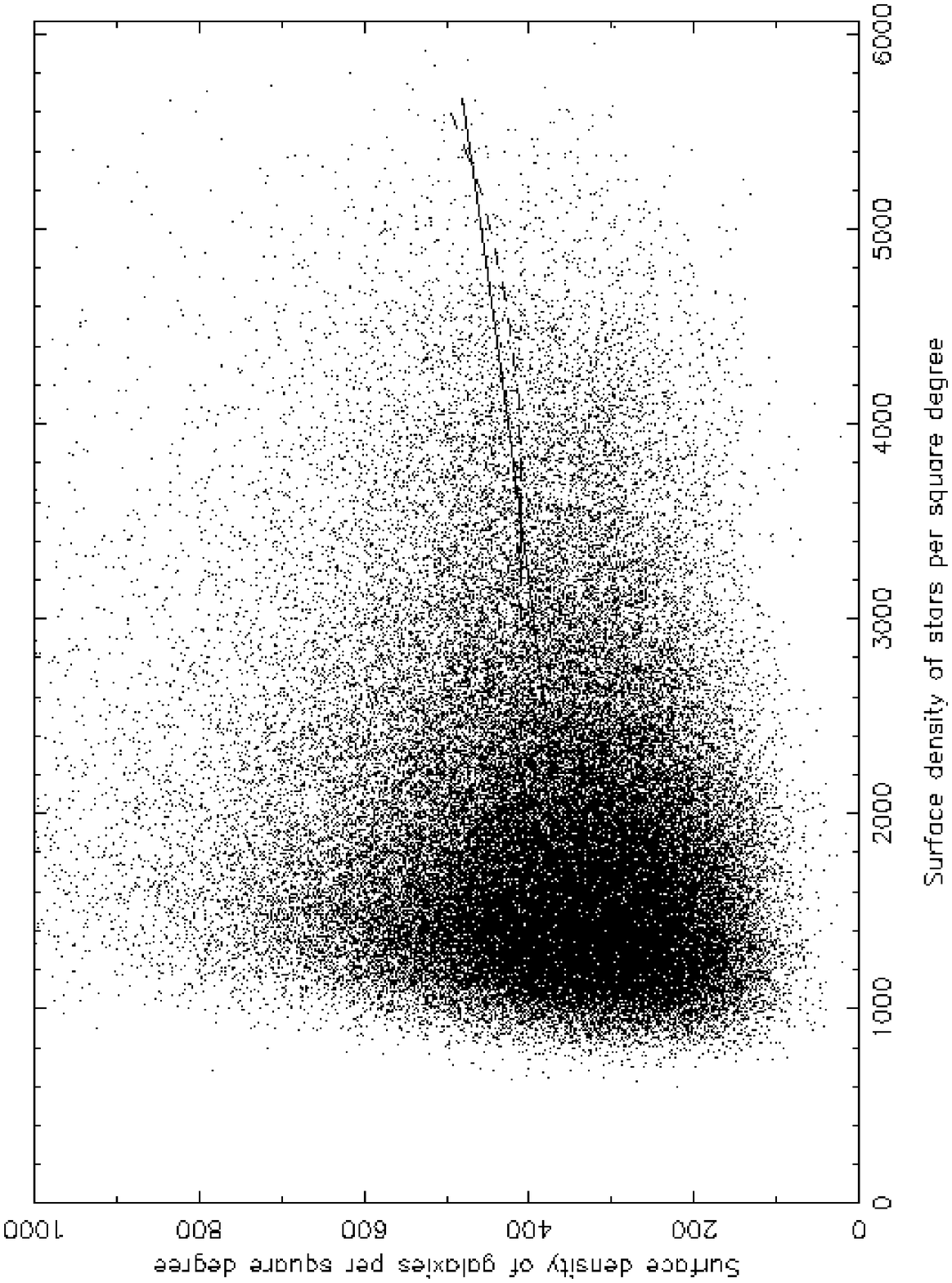}

\noindent
{\bf Figure 20}. A scatter-plot of the surface density of galaxies
with $20.0 > \bj > 17.0 $ against the surface density of stars
brighter than $\bj = 20.5$ .  A point has been plotted for each
$0.23^\circ \times 0.23^\circ $ cell of equal area maps as described
in Section 2.3.  The dashed line shows the mean galaxy density binned
as a function of the stellar density, and the solid line is the best
fit quadratic.

\vfill\eject

\noindent
$\;$

\vskip 6 truein

\includegraphics{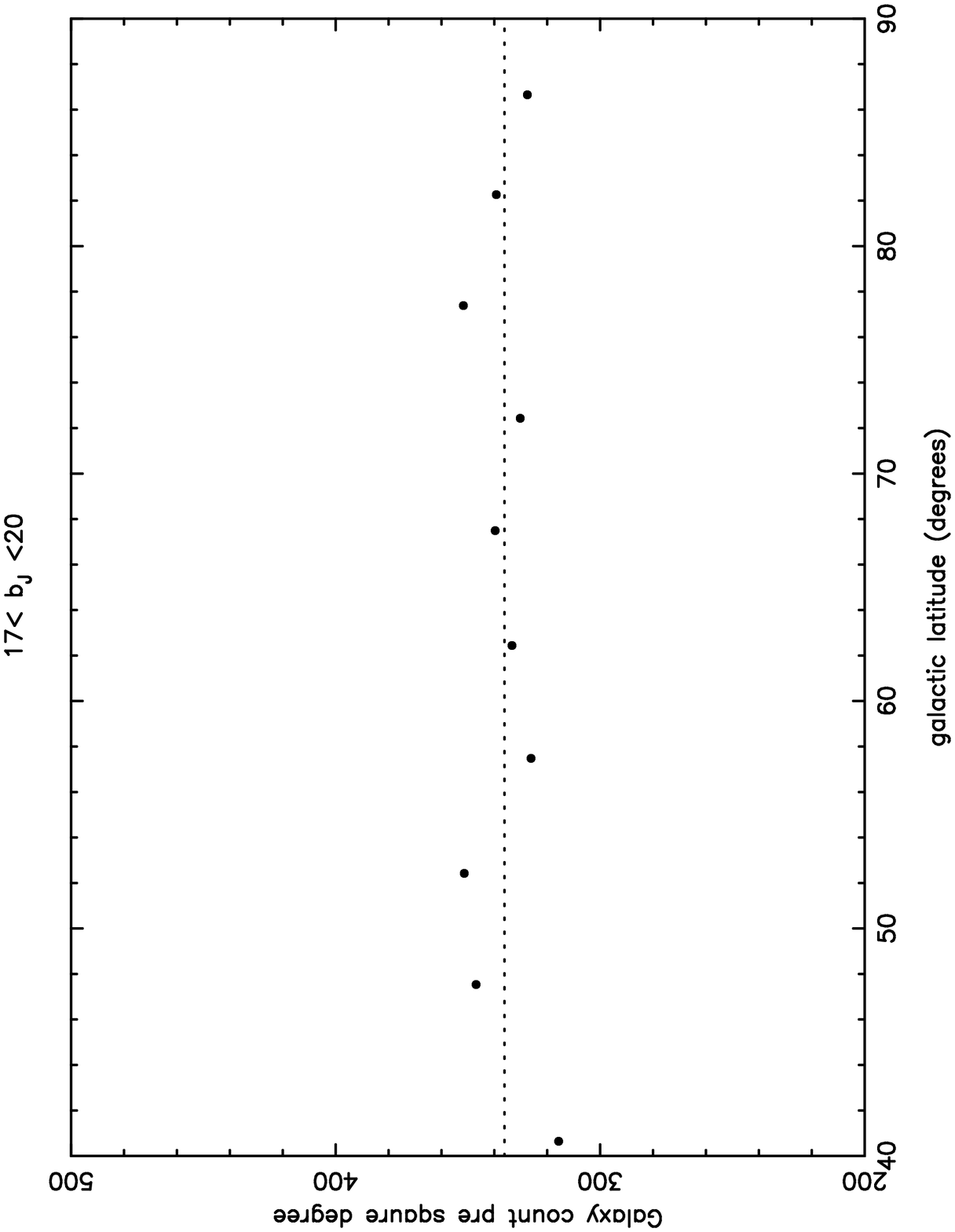}

\noindent
{\bf Figure 21}. The mean surface density of galaxies 
as a function of galactic latitude after applying the correction for
the variations in contaminating fraction of stellar blends. 
The dotted line shows the overall mean surface density. 
There is no significant gradient as a function of galactic latitude.

\vfill\eject

\noindent
$\;$

\vskip 6 truein

\includegraphics{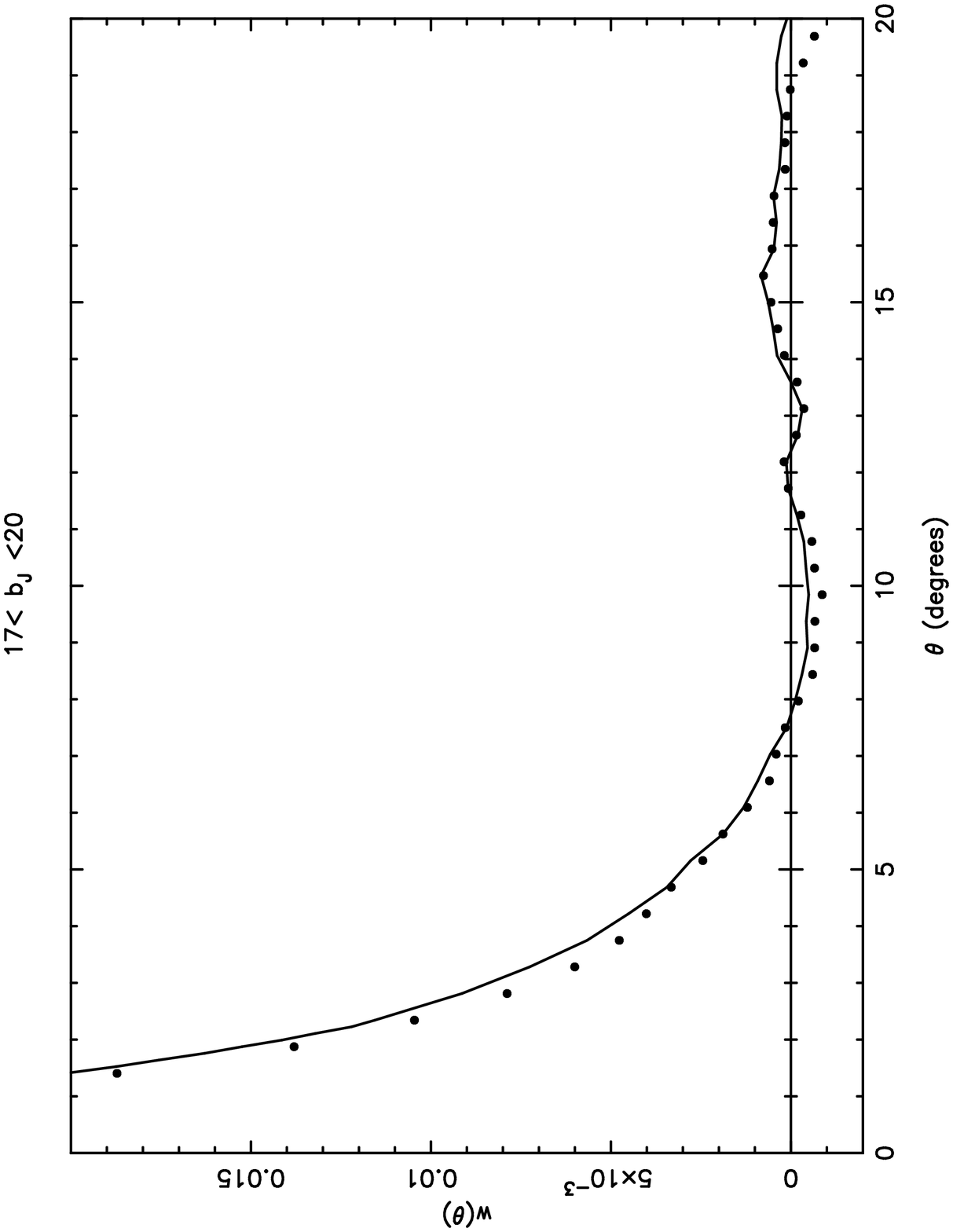}

\noindent
{\bf Figure 22}. The solid points show $\w$ estimated from the fully
corrected map of galaxies with $ 17< b_J < 20$, including corrections
for galactic obscuration, atmospheric extinction and variations in the
fraction of contaminating images.
The solid line is the estimate from the uncorrected galaxy map, as
plotted in Figure~2. 
It can be seen that these corrections have a very small effect.

\vfill\eject

\noindent
$\;$

\vskip 6 truein

\includegraphics{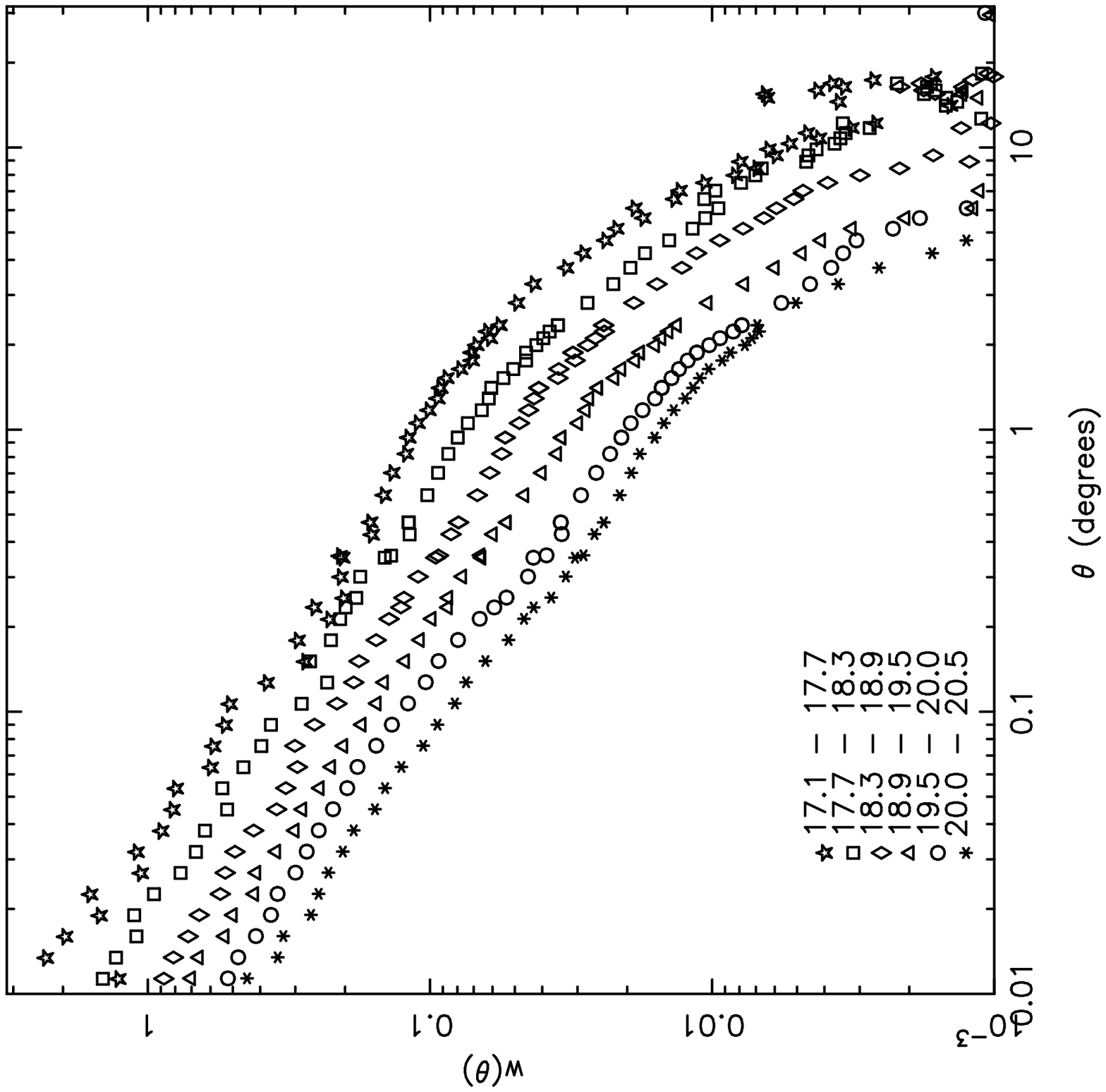}

\noindent
{\bf Figure 23}. Estimates of $\w$ for six disjoint magnitude slices in
the range $ 17 < b_J < 20.5$ as in Figure 3 except here we have
corrected the estimated of $\w$ for the various systematic 
errors described in the text.

\vfill\eject

\noindent
$\;$

\vskip 6 truein

\includegraphics{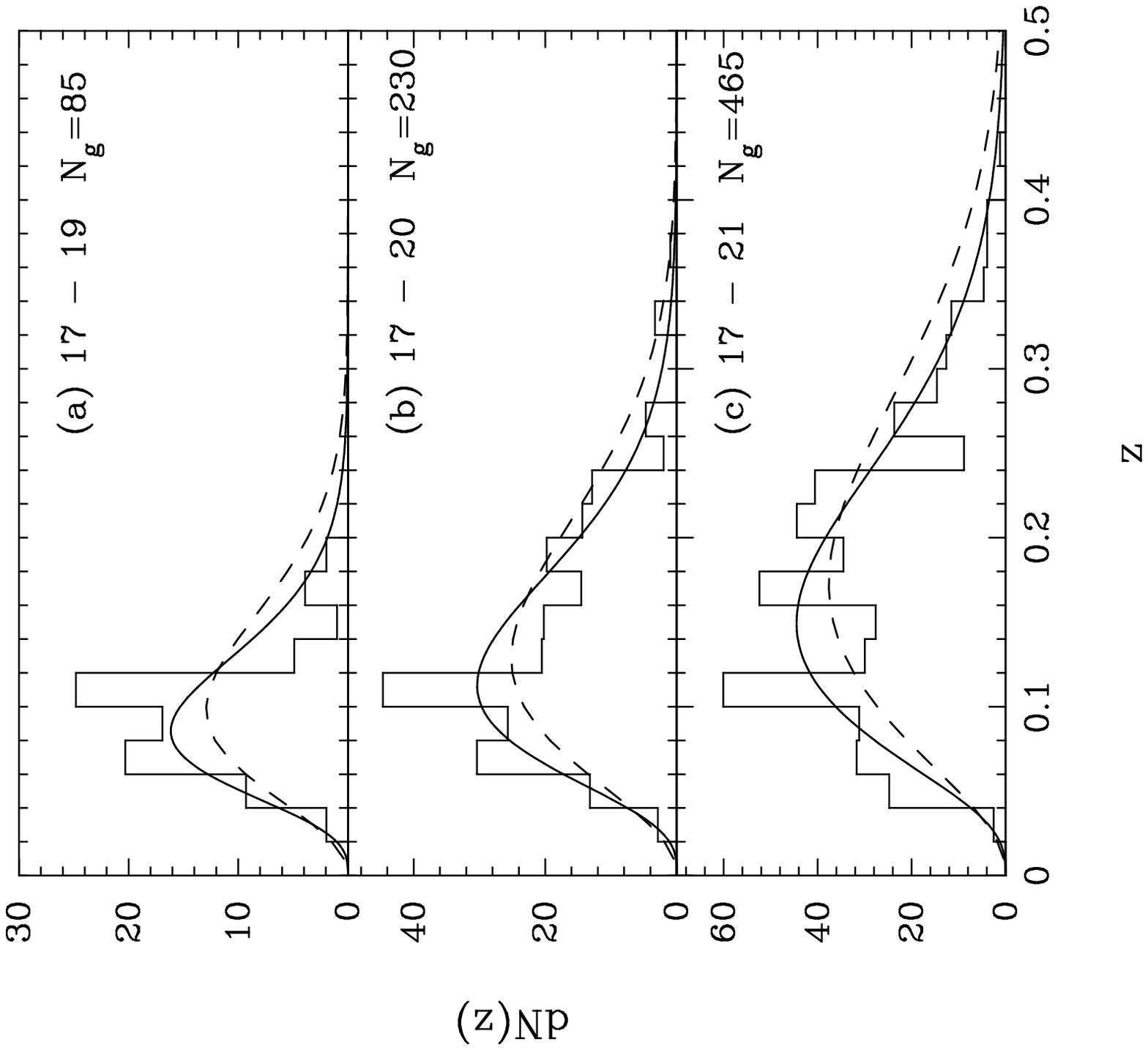}

\noindent
{\bf Figure 24}. The redshift distributions in three
magnitude slices determined from deep redshift surveys.
The magnitude ranges and the number of galaxies, $N_g$,
in each histogram are given in each panel. The solid
lines show the redshift distributions derived from
equations (38) and the dashed lines show the redshift
distribution calculated from the model (equation 37) 
of Maddox \etal (1990c).

\vfill\eject

\noindent
$\;$

\vskip 6 truein

\includegraphics{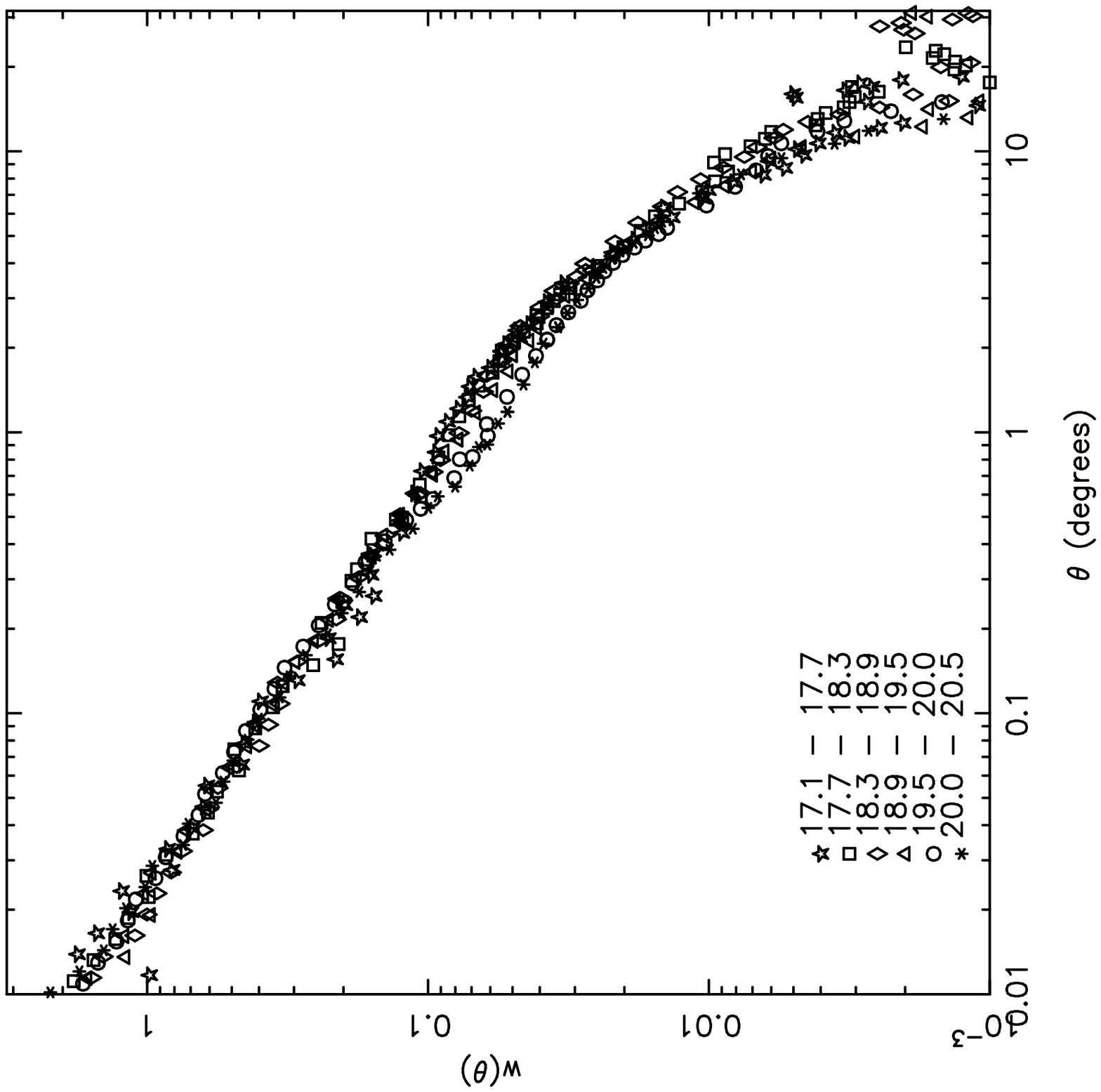}

\noindent
{\bf Figure 25}. Corrected estimates of $\w$ for the six disjoint magnitude
slices from Figure 23 scaled to a magnitude limit
$b_J = 18.4$ (approximately the depth of the Lick catalogue)
using the scaling factors given by equation 38 with $\epsilon = 0 $
and $\Omega = 1$ (columns 4 and 5 of Table 5).

\vfill\eject

\noindent
$\;$

\vskip 7.5 truein

\includegraphics{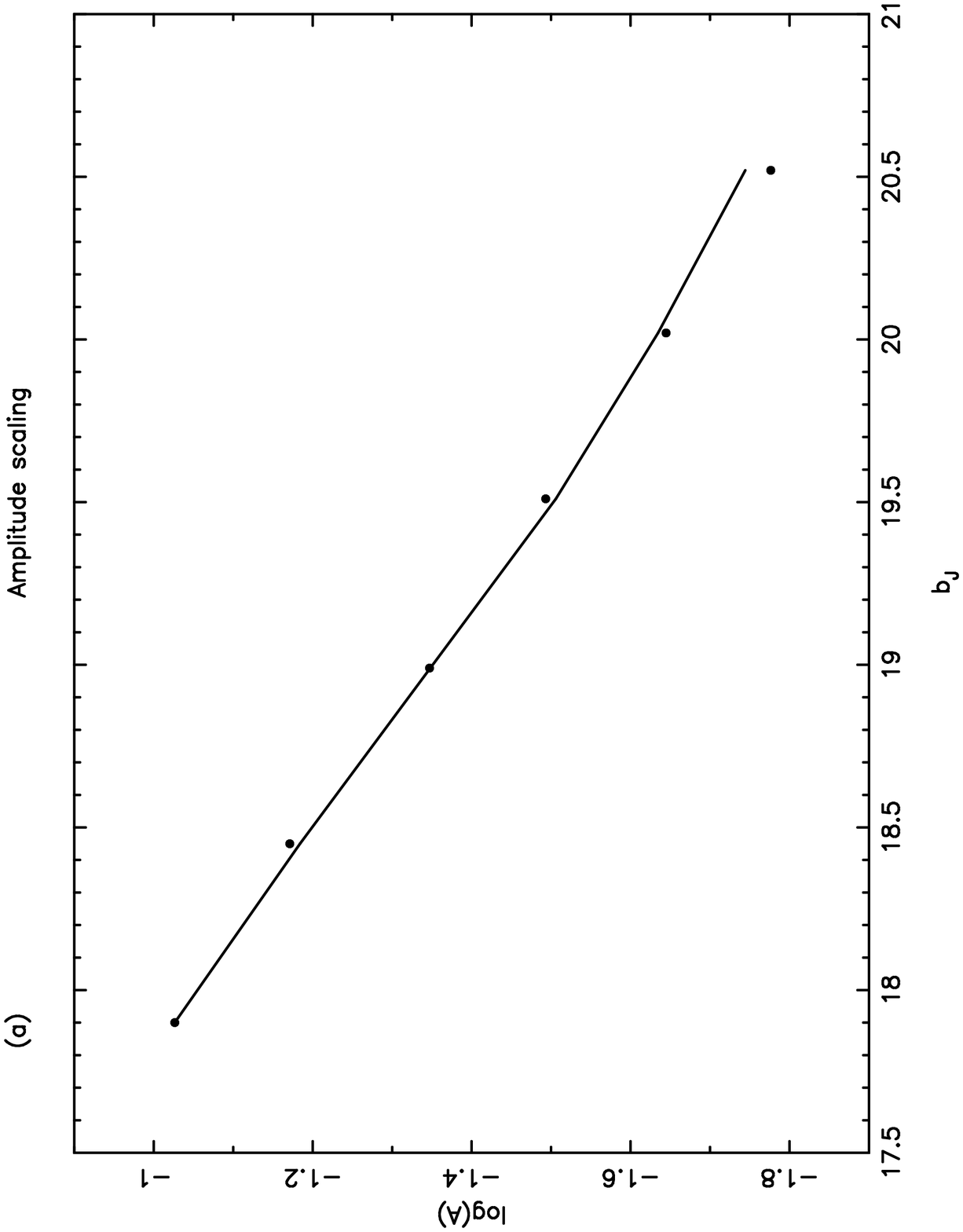}
\includegraphics{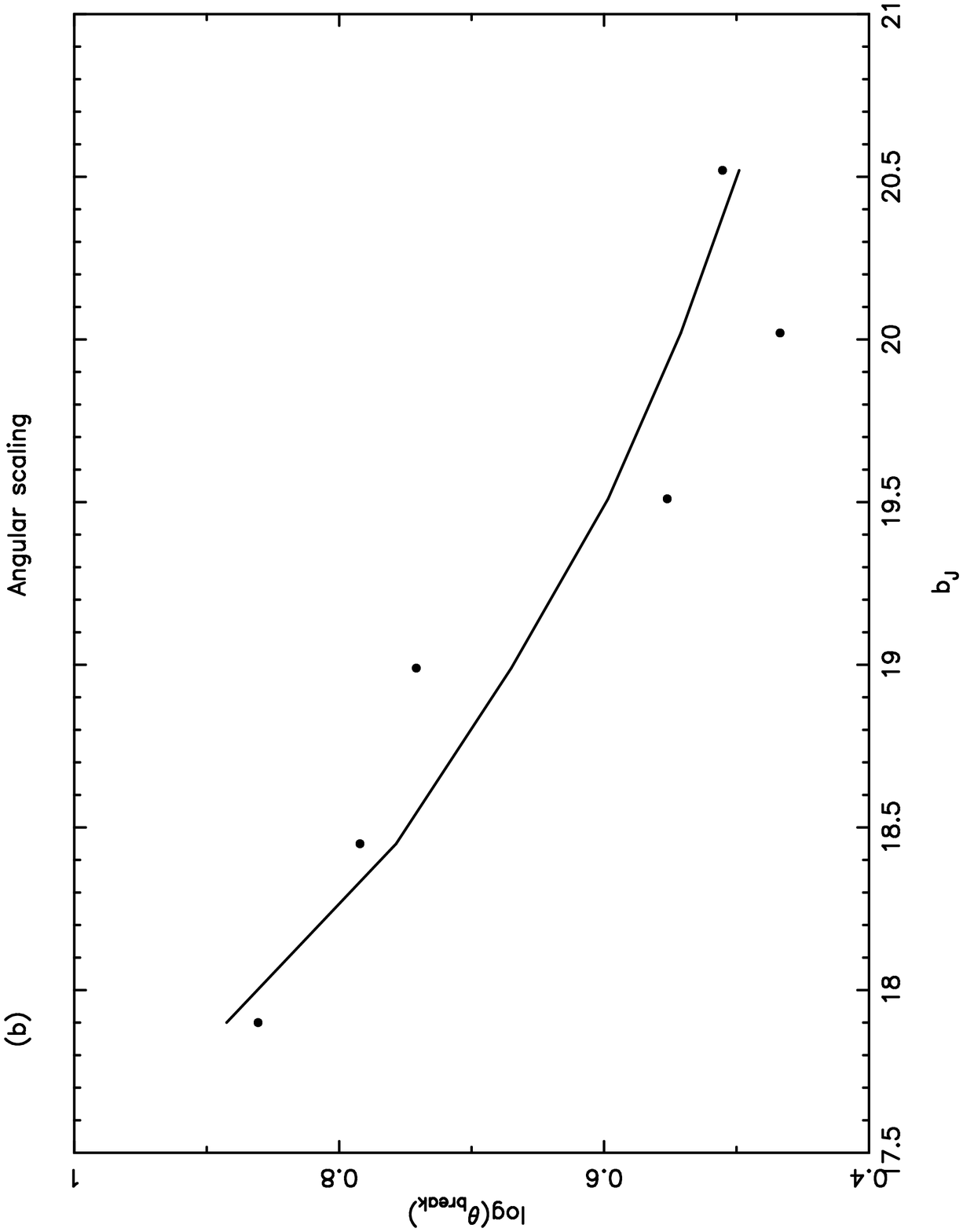}
%\special{psfile=pg_20a.ps hscale=40 vscale=40 angle=-90 hoffset=30
%voffset=490}
%\special{psfile=pg_20b.ps hscale=40 vscale=40 angle=-90 hoffset=30
%voffset=250}

\noindent
{\bf Figure 26}. Figure (26a) shows the amplitude of 
$\w$ at $\theta=1^\circ$ plotted as a function of magnitude
limit for the six magnitude slices shown in Figure 23. The
amplitude $A$ was determined by least squares power law
fits as described in the text. Figure (26b) shows the
break angle $\theta_{break}$, as defined in the text,
plotted against magnitude limit for the six slices. 
The solid lines show the predictions of the scaling relation
based on equation (38) with $\epsilon = 0$ and $\Omega_0 =1$.

\vfill\eject

\noindent
$\;$

\vskip 6 truein

\includegraphics{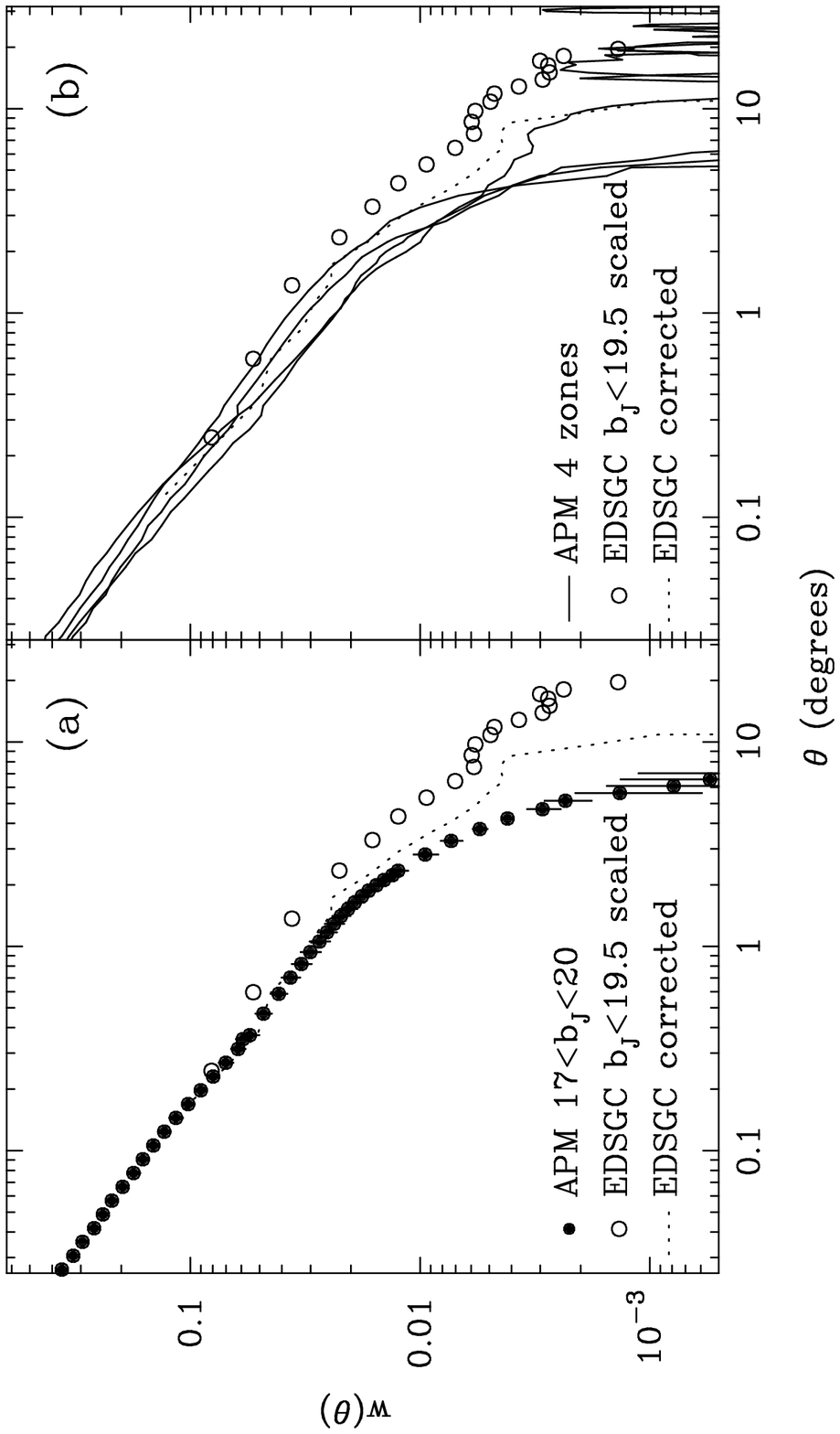}

\noindent
{\bf Figure 27}. The solid points in Figure (27a) show the APM
$\w$ with $1\sigma$ errors for the magnitude slice 
$17 \le b_J \le 20$, as plotted in  Figure (2a). The open
circles show $\w$ estimated from the EDSGC limited at $b_J=19.5$
from Figure 2 of Collins \etal (1992) scaled to the APM results
as described in the text. The dotted lines show the corrected
EDSGC $\w$ from Figure 7 of Nichol and Collins (1993) scaled
to the APM results. In Figure (27b) we plot separately the $\w$ 
estimates
for the four zones of the APM Survey, each of which has about the
same area as the EDSGC.

\vfill\eject

\noindent
$\;$

\vskip 6 truein

\includegraphics{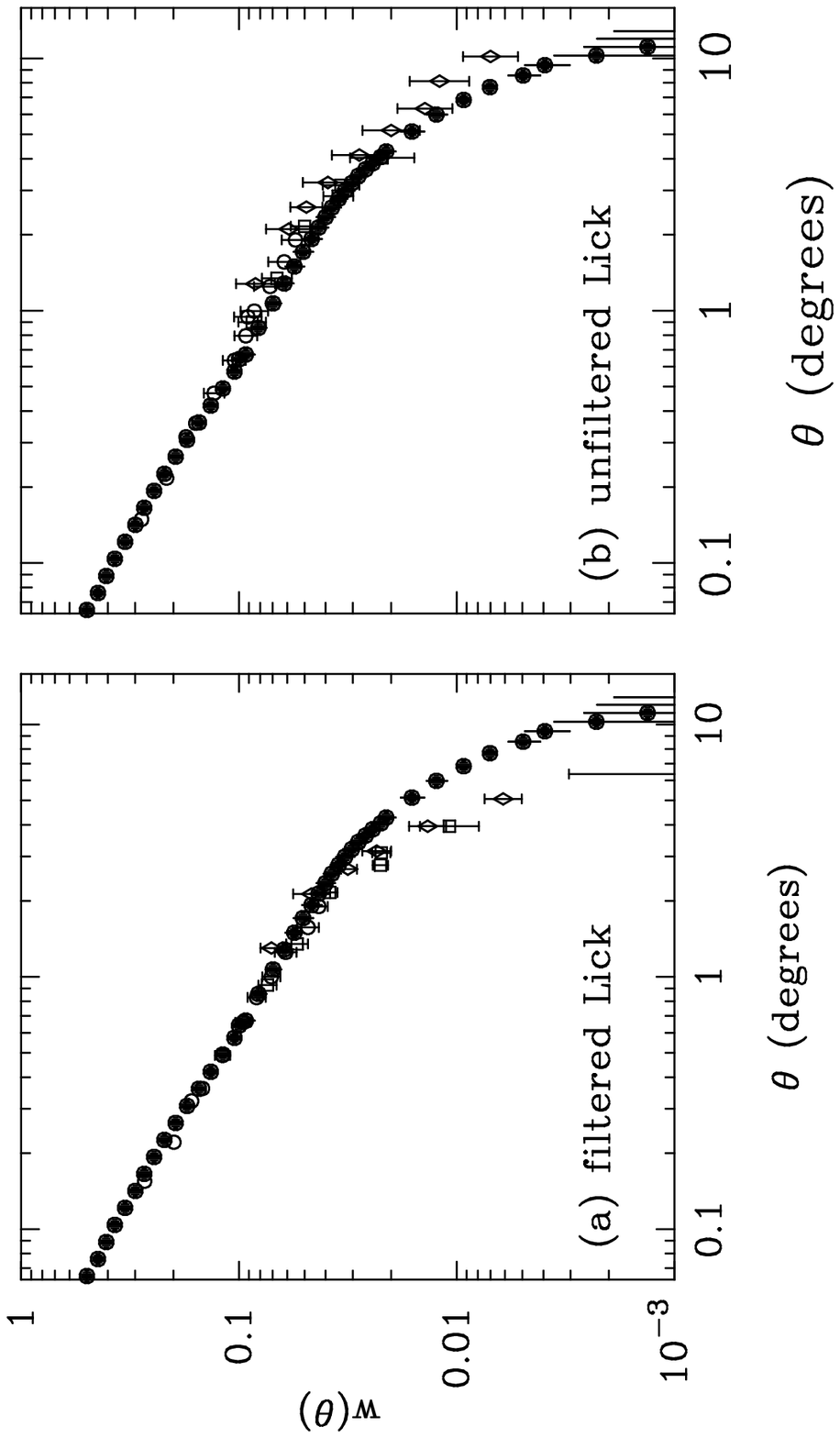}

\noindent
{\bf Figure 28}. Estimates of $w(\theta)$ (solid points) for 
APM galaxies in the magnitude slice $17 < b_J < 20$
scaled to the depth of the Lick Survey. The open symbols
show the GP77 estimates of $w(\theta)$ for the Lick
survey after filtering the Lick counts (Figure 28a) and
before filtering (Figure 28b).

\vfill\eject

\noindent
$\;$

\vskip 6 truein

\includegraphics{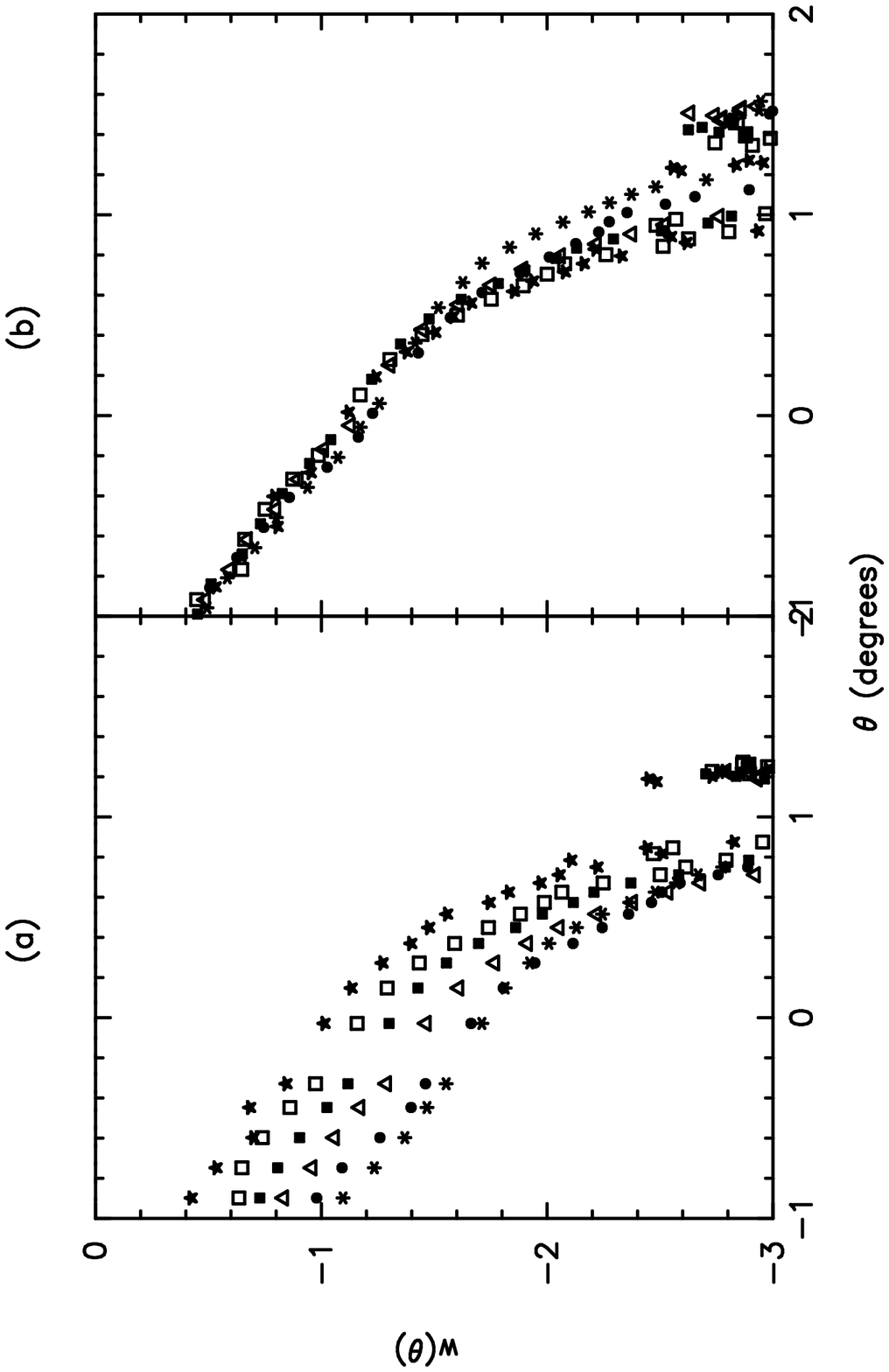}

\noindent
{\bf Figure 29.} The effect on $w(\theta)$ of filtering the APM maps.
For each magnitude slice 
(shown with the same symbols as in Figures 3 and 23)
we fitted a bi-cubic spline to the 
map of galaxy density and then multiplied by the inverse of the fit before
calculating the plotted $w(\theta)$. Figure (29a) shows the
angular correlations of the six magnitude slices after filtering.
For  the brighter slices, the break in $w(\theta)$ is dramatically steepened
compared to the  unfiltered results plotted in Figure (23).
The measurements for the fainter slices are hardly changed by the 
filtering. Figure (29b) shows the scaling test applied to the
filtered estimates of $w(\theta)$, as in Figure 25.

\vfill\eject

\noindent
$\;$

\vskip 6 truein

\includegraphics{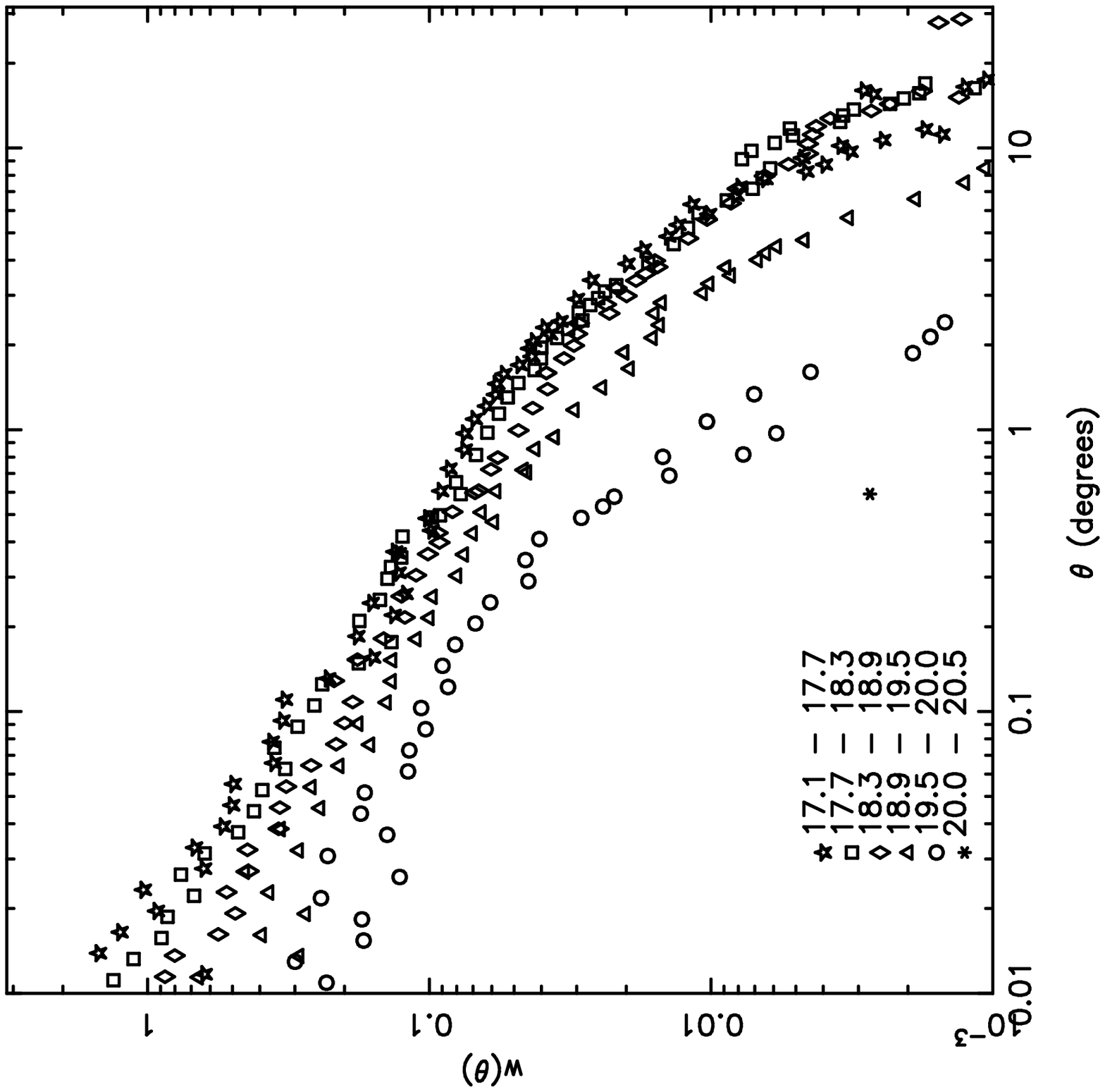}

\noindent
{\bf Figure 30}. The correlation functions from the five brighter
slices after subtracting the correlation function of the faintest
slice.  The correlation functions have been scaled to a common depth
as in Figure 25. This correction provides a gross overestimate of the
errors in $\w$ since it requires that {\it all} of the observed
correlations of the faint galaxies are produced by systematic errors.
However, even in this extreme case, the scaled estimates of $\w$ for
the brightest three slices are in reasonable agreement, showing
significant clustering between scaled angles of $5^\circ$ and
$10^\circ$.

\vfill\eject

\noindent
$\;$

\vskip 6 truein

\includegraphics{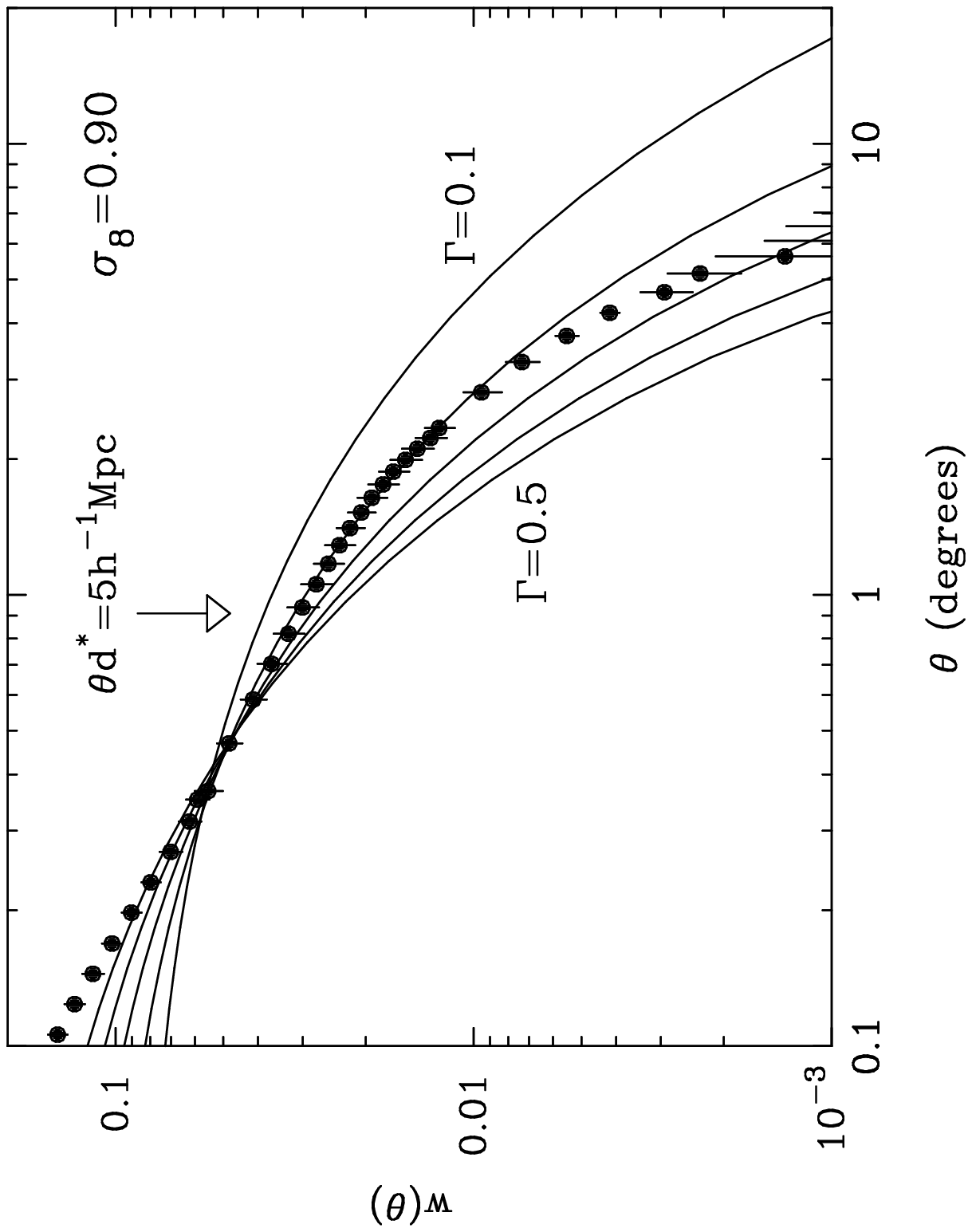}

\noindent
{\bf Figure 31}. The APM $w(\theta)$ estimates for the magnitude
slice $17 < b_J < 20$ compared with the predictions of a
family of CDM-like models. The theoretical predictions are
computed from the linear theory power spectrum (equation 40),
where the parameter $\Gamma = \Omega_0 h$. The power spectra
have been normalized so that the {\it rms} fluctuations in
spheres of radius $8 \hmpc$ is $\sigma_8 = 0.9$. We plot
curves for $\Gamma = 0.1$, $0.2$, $0.3$, $0.4$ and $0.5$
(from top to bottom). The arrow marks the angular scale corresponding
to a physical separation of $\approx 5 h^{-1}{\rm Mpc}$ at the
median depth of the survey.

\vfill\eject

\noindent
$\;$

\vskip 7.2 truein

\includegraphics{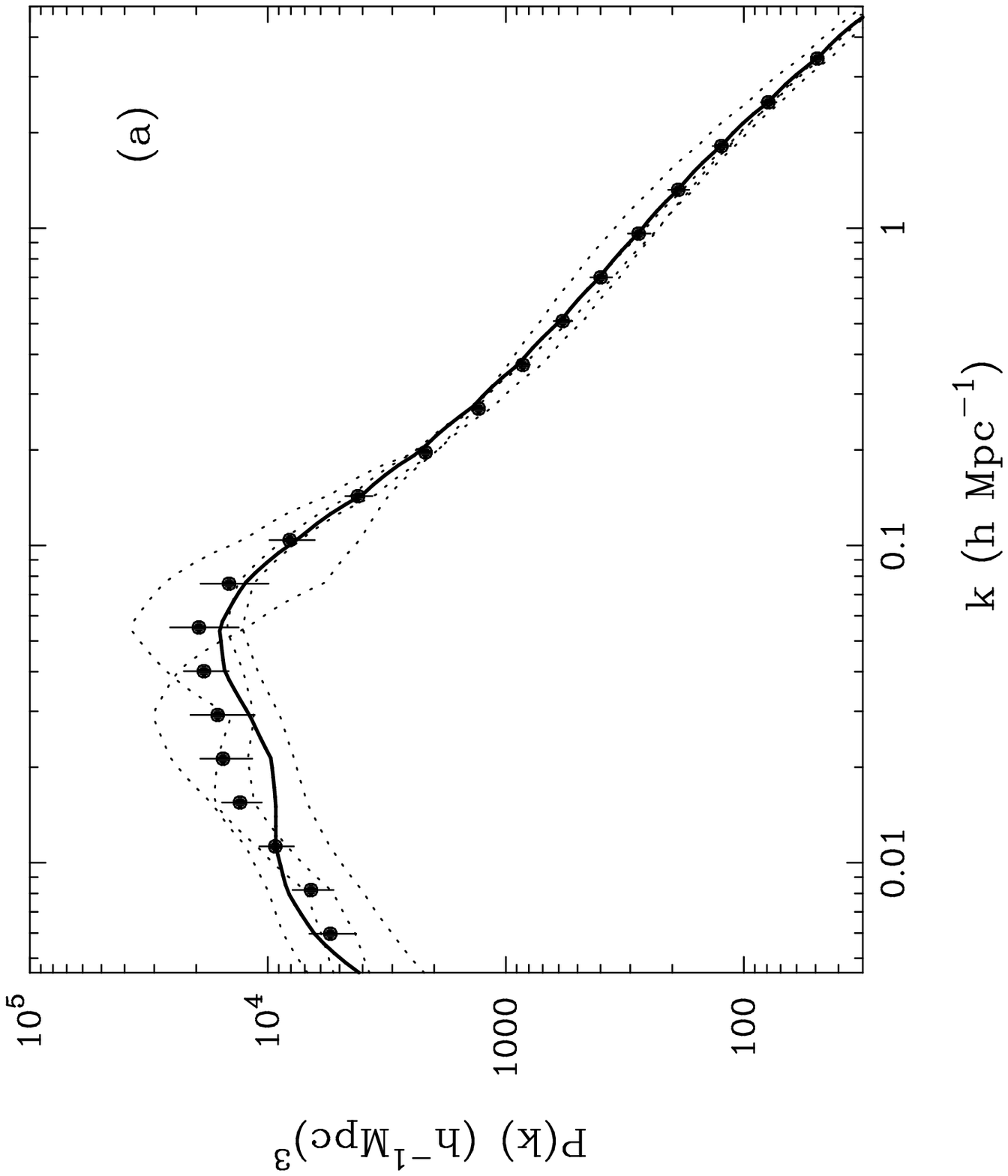}

\includegraphics{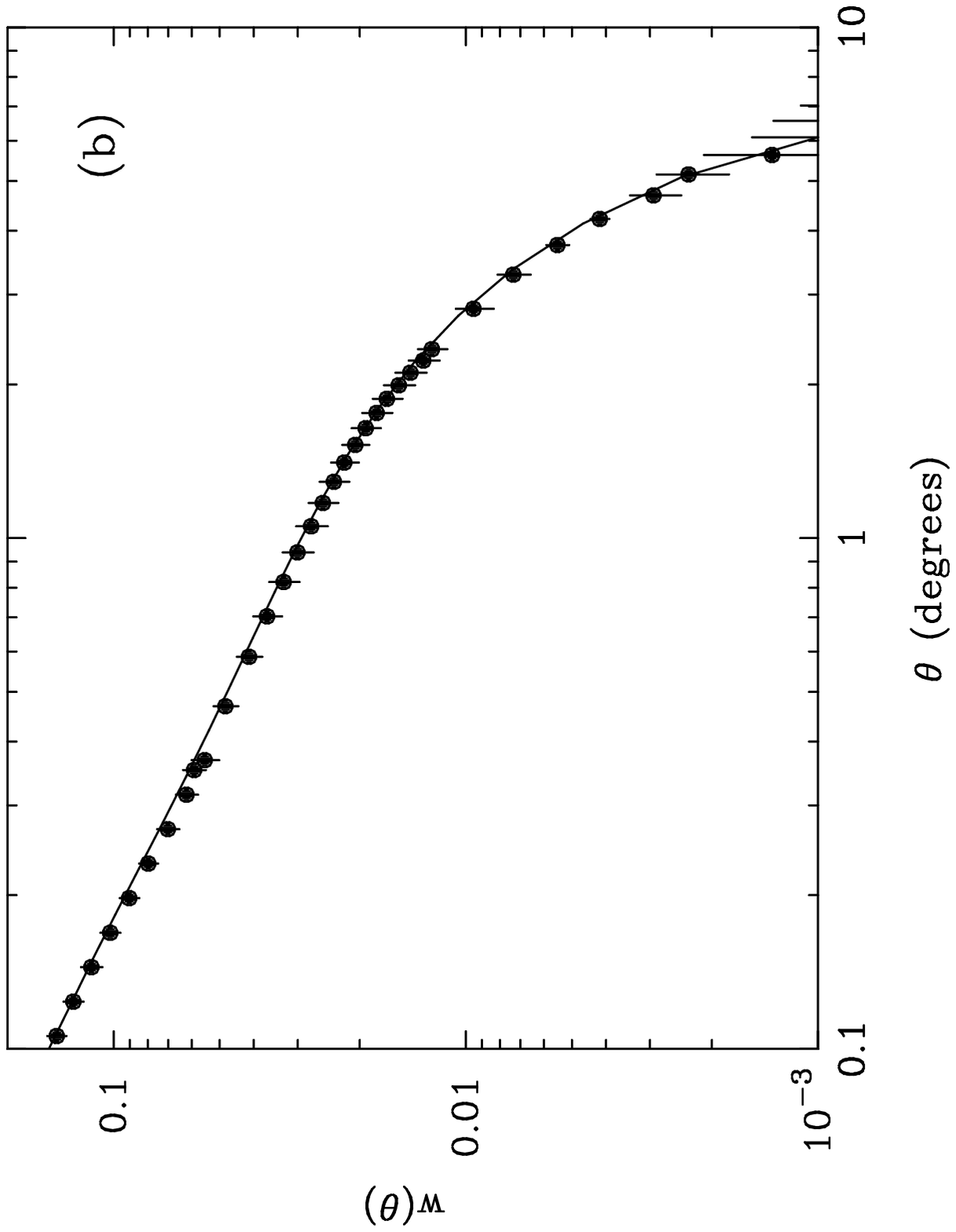}

\noindent
{\bf Figure 32}. Figure~32a shows the three-dimensional power
spectrum derived by inverting the APM angular correlation function
as described in the text. The dotted lines show the inversion
of $P(k)$ for 4 nearly equal area zones. The filled circles
show the means of the these estimates together with $1\sigma$
error bars. The thick solid line shows the power spectrum
by inverting $w(\theta)$ estimated from the full survey area.
Figure 32b shows a check of the inversion. The solid line
is computed from equation (42a) using the power spectrum 
plotted as the thick solid line in Figure~32a and
reproduces the angular correlation function (shown by 
the filled symbols) used in the inversion.

\vfill\eject

\noindent
$\;$

\vskip 6 truein

\includegraphics{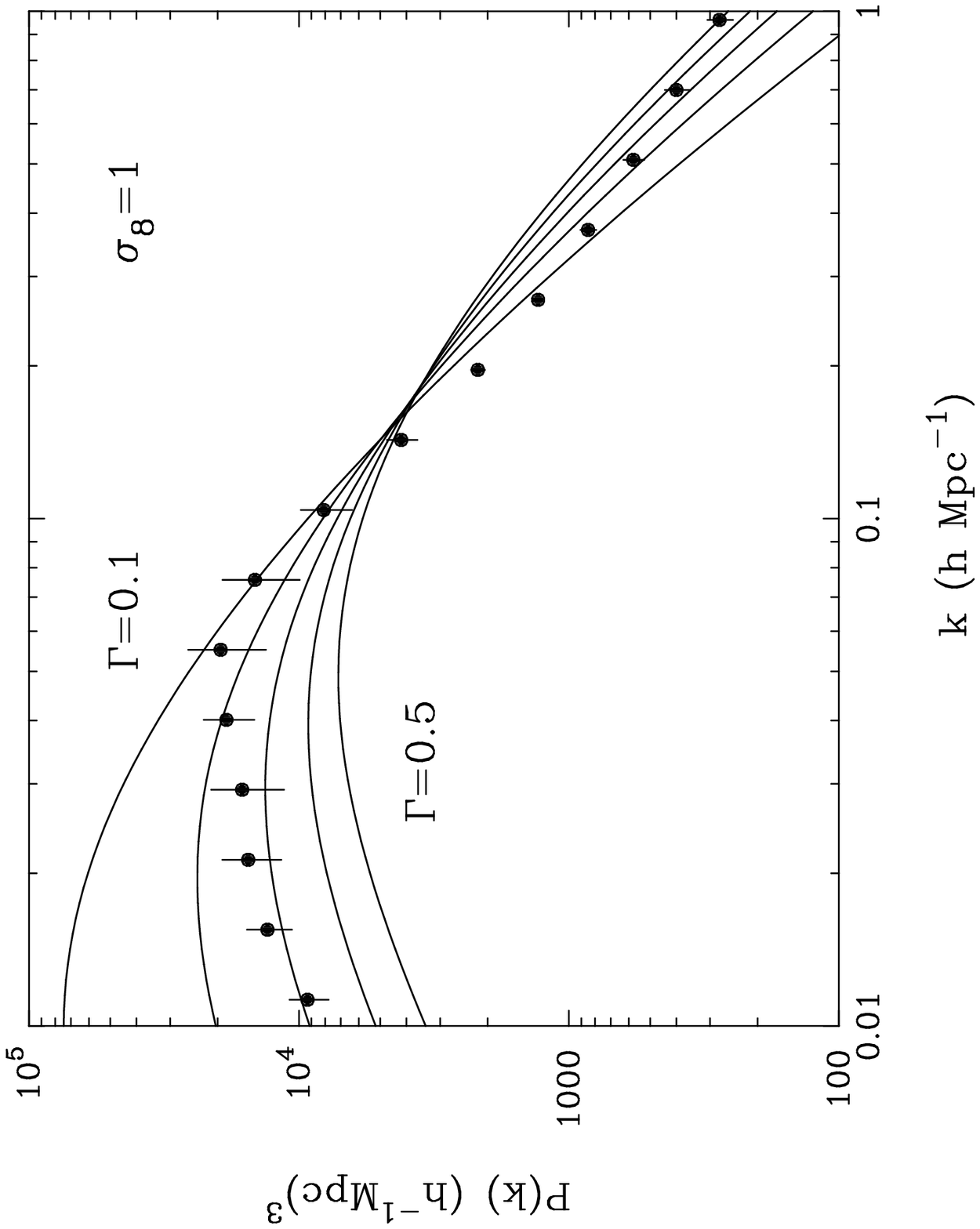}

\noindent
{\bf Figure 33}. The points show the three-dimensional power spectrum
from the APM survey together with $1\sigma$ errors, as in Figure~32.
The curves show the family of linear theory CDM power spectra
(equation 40) with $\Gamma = 0.1$, $0.2$, $0.3$, $0.4$ and $0.5$
from top to bottom. The CDM power spectra have been normalized to
$\sigma_8=1$.

\vfill\eject

\noindent
$\;$

\vskip 6 truein

\includegraphics{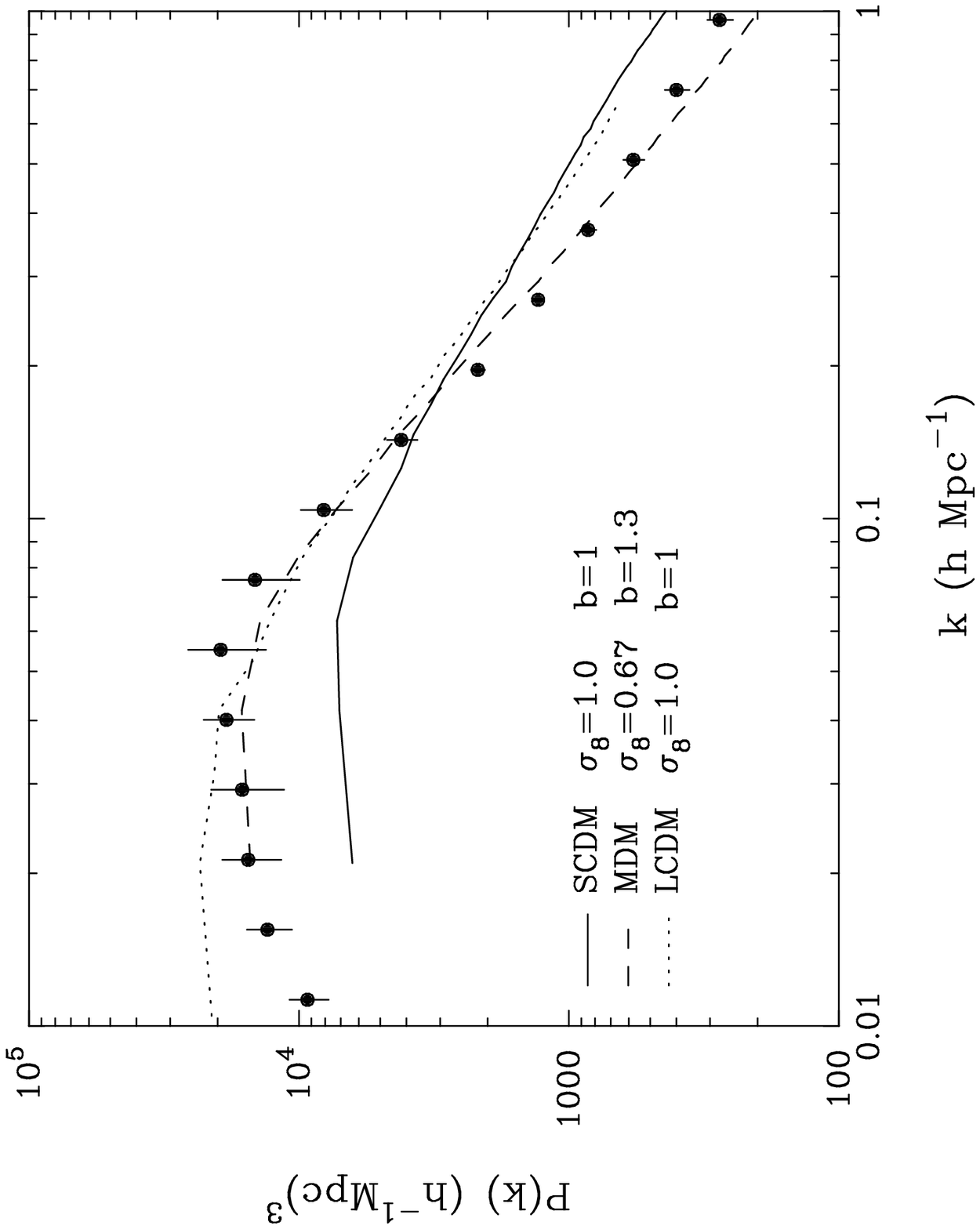}

\noindent
{\bf Figure 34}. The APM power spectrum compared with the power spectra
determined from N-body simulations as described in the text. The
simulations are normalized to match approximately the amplitude
of the microwave background anistropies detected by COBE;  the values
of $\sigma_8$ for the mass fluctuations in each model are given
in the figure. The curves show the power spectra of the mass
fluctuations; for the MDM model we have multiplied the power
spectrum by a biasing factor $b^2$, with $b=1.3$, to match the
observations.

\vfill\eject

\noindent
$\;$

\vskip 6 truein

\includegraphics{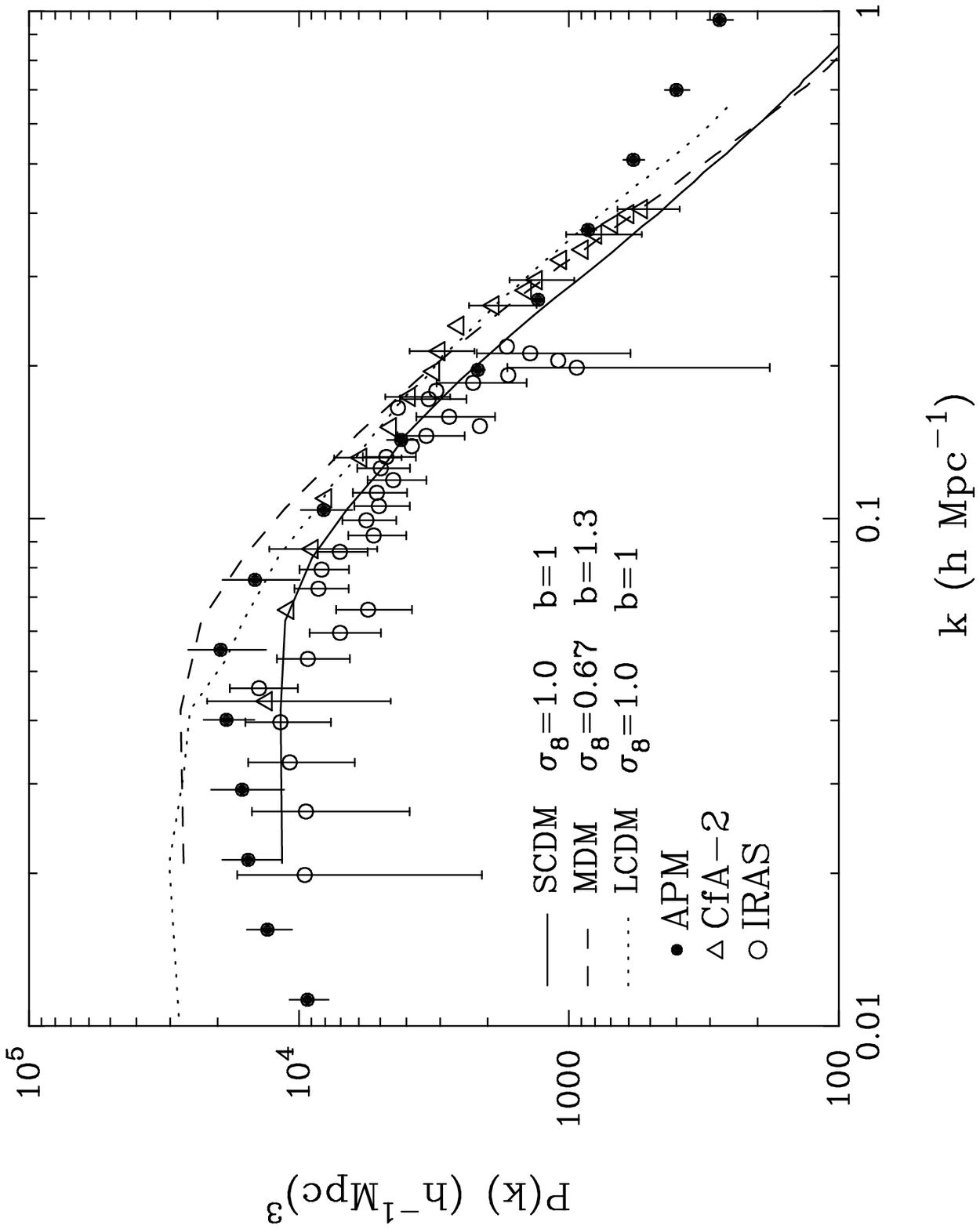}

\noindent
{\bf Figure 35}. The APM power spectrum compared with 
the power spectra determined from redshift surveys.
The open circles shows the power spectrum of IRAS
galaxies, as determined by Tadros \& Efstathiou (1995a).
The open triangles show the power spectrum of optically
selected galaxies determined by Park \etal (1994)
from a volume limited subset of the CfA-2 survey
(see text for details). The error bars on the 
points show $1\sigma$ errors. The curves show the power
spectra of the N-body simulations of Figure~33
determined {\it in redshift space}.

\vfill\eject

\noindent $\;$

\vskip 8. truein

\includegraphics{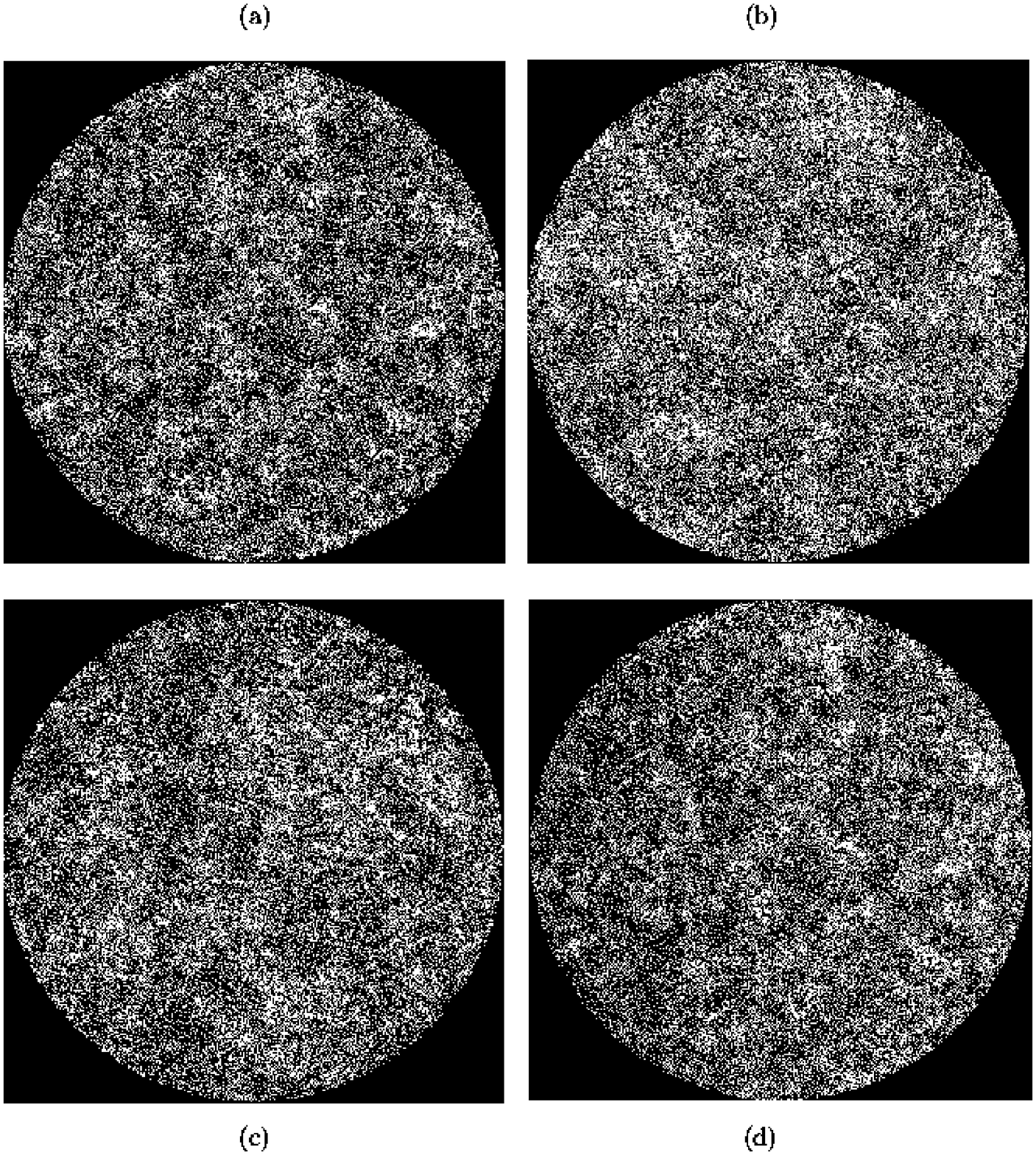}

\noindent
{\bf Figure A1}. Greyscale maps showing the galaxy density in a
variety of simulated galaxy catalogues constructed using the
Soneira-Peebles  prescription. The simulated cones are $60^\circ$ in
diameter. (a) SP0; (b) SP2; (c) SP4; (d) SPS.  
Details of the model parameters are given in the text. 

\vfill\eject

\noindent
$\;$

\vskip 7.5 truein

\includegraphics{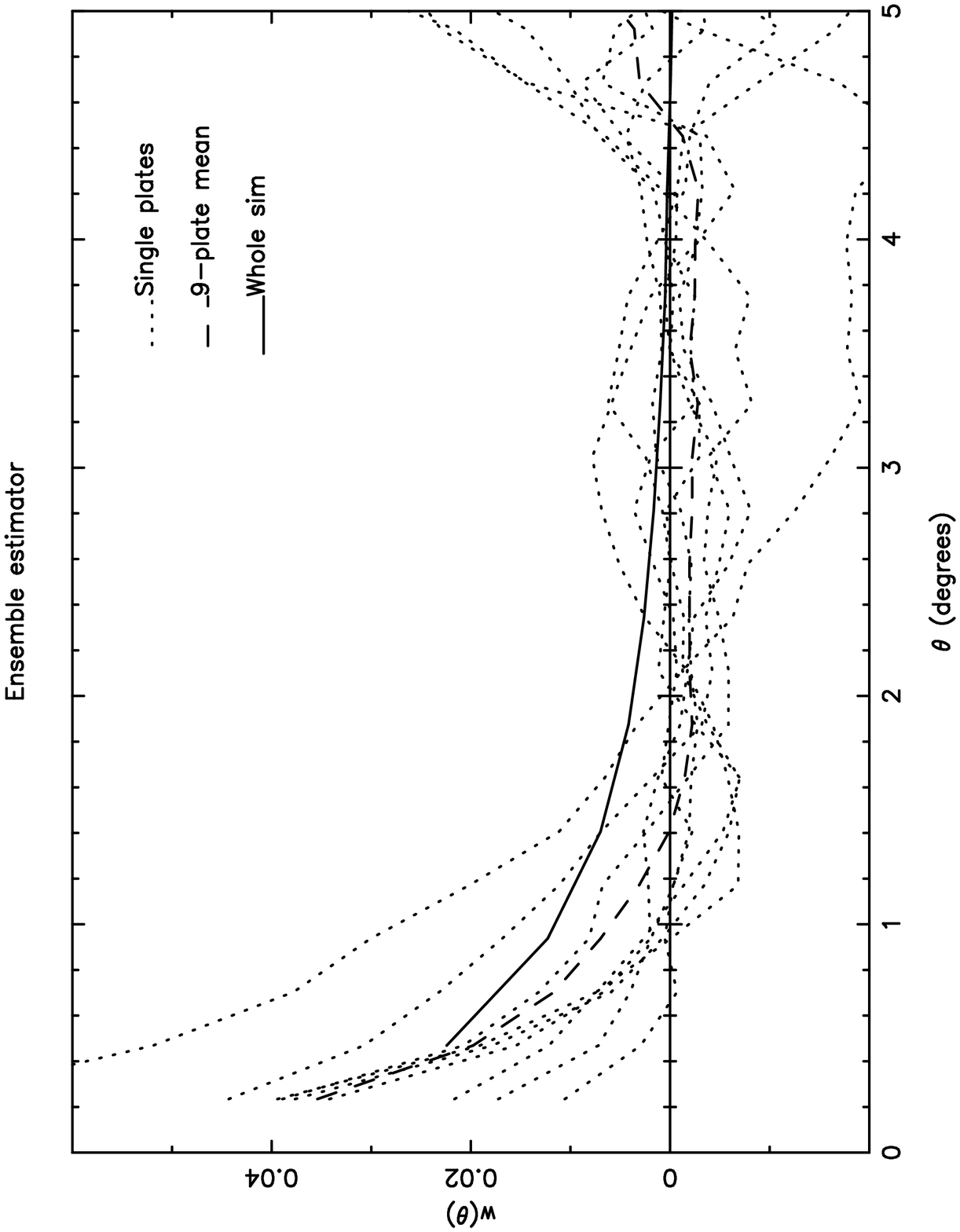}
\includegraphics{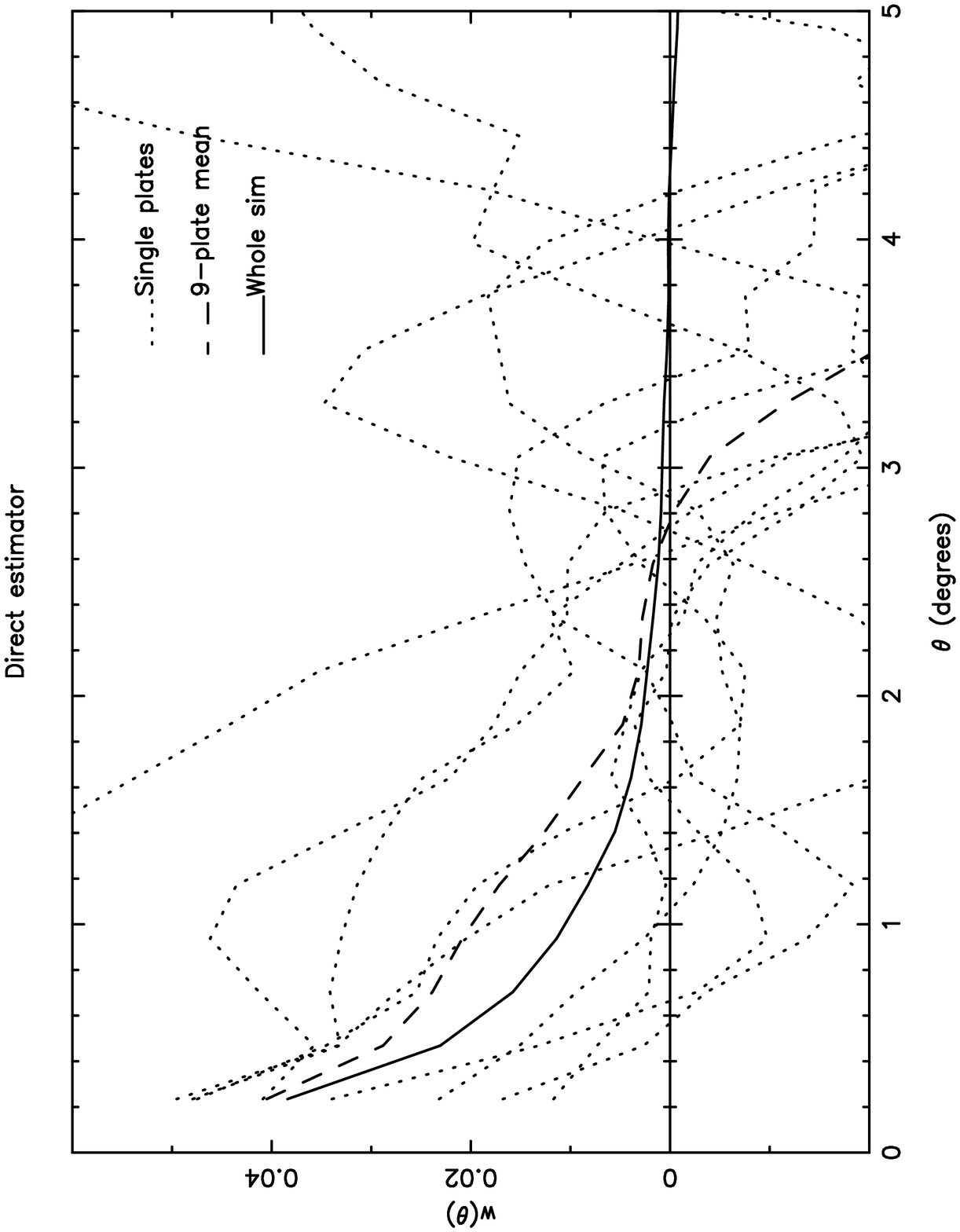}

\noindent
{\bf Figure A2}. ``Single plate estimates'' from the SP2 simulations
(a) using the ensemble estimator; (b) using the direct estimator. 

\vfill\eject

\noindent
$\;$

\includegraphics{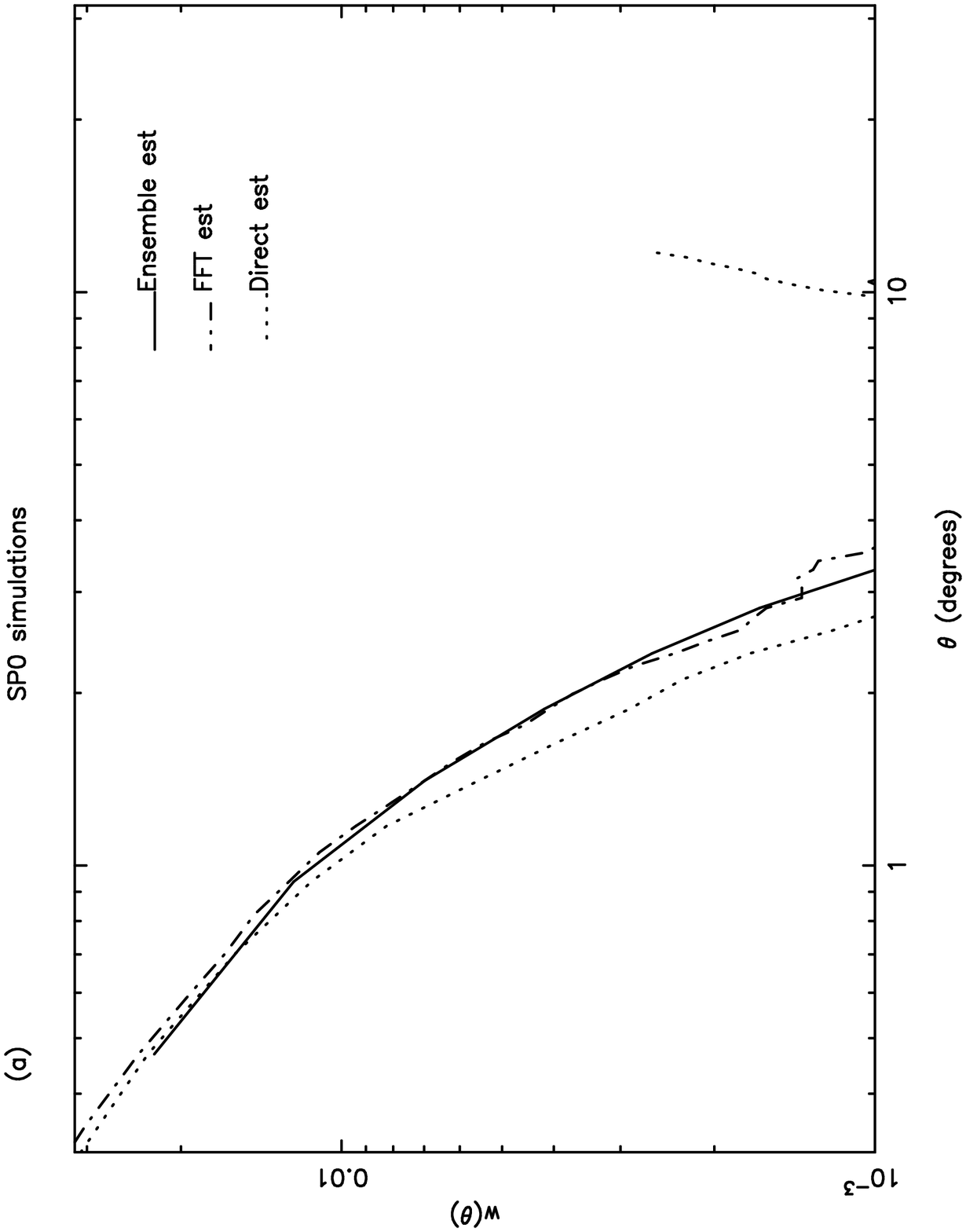}
\includegraphics{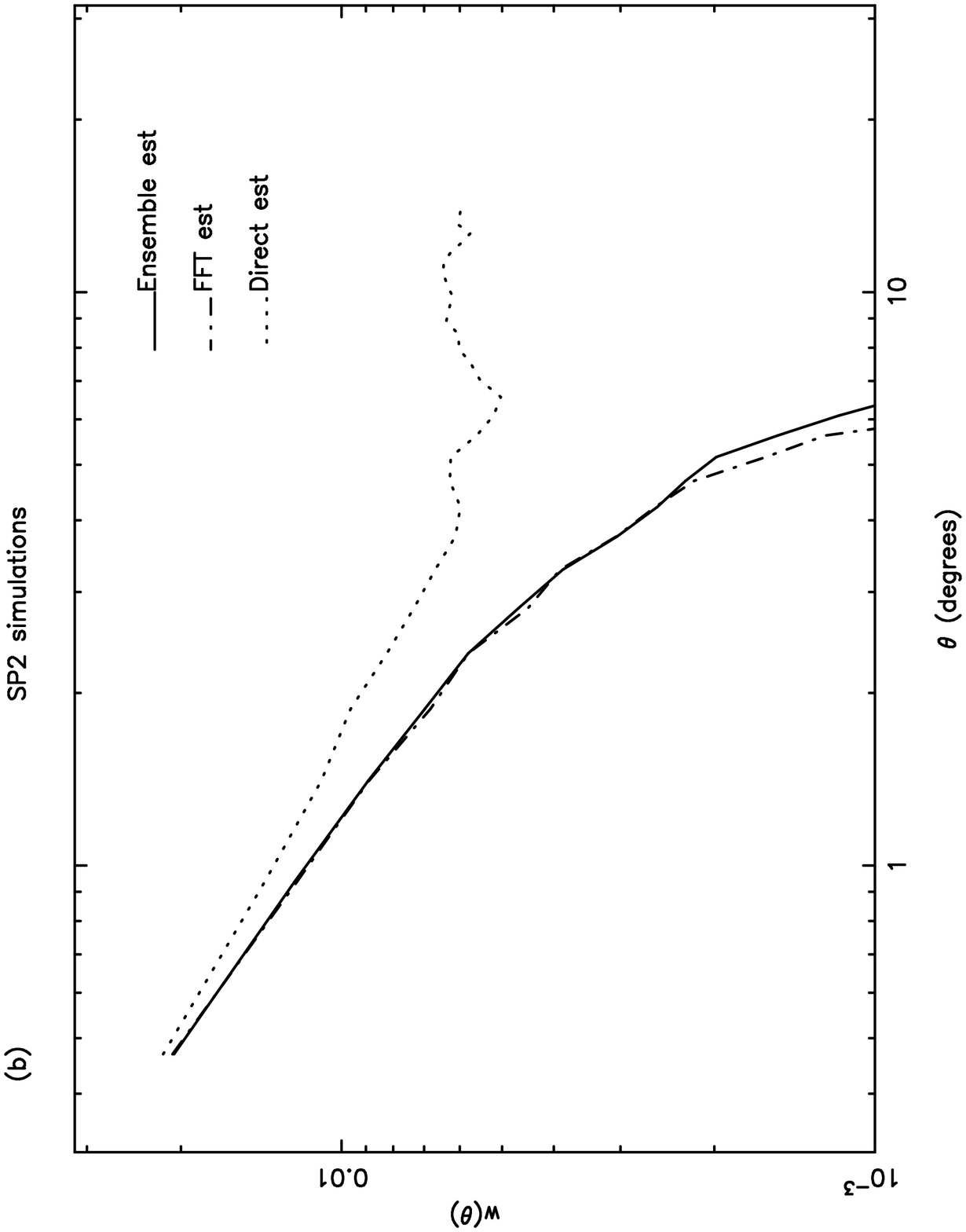}
\includegraphics{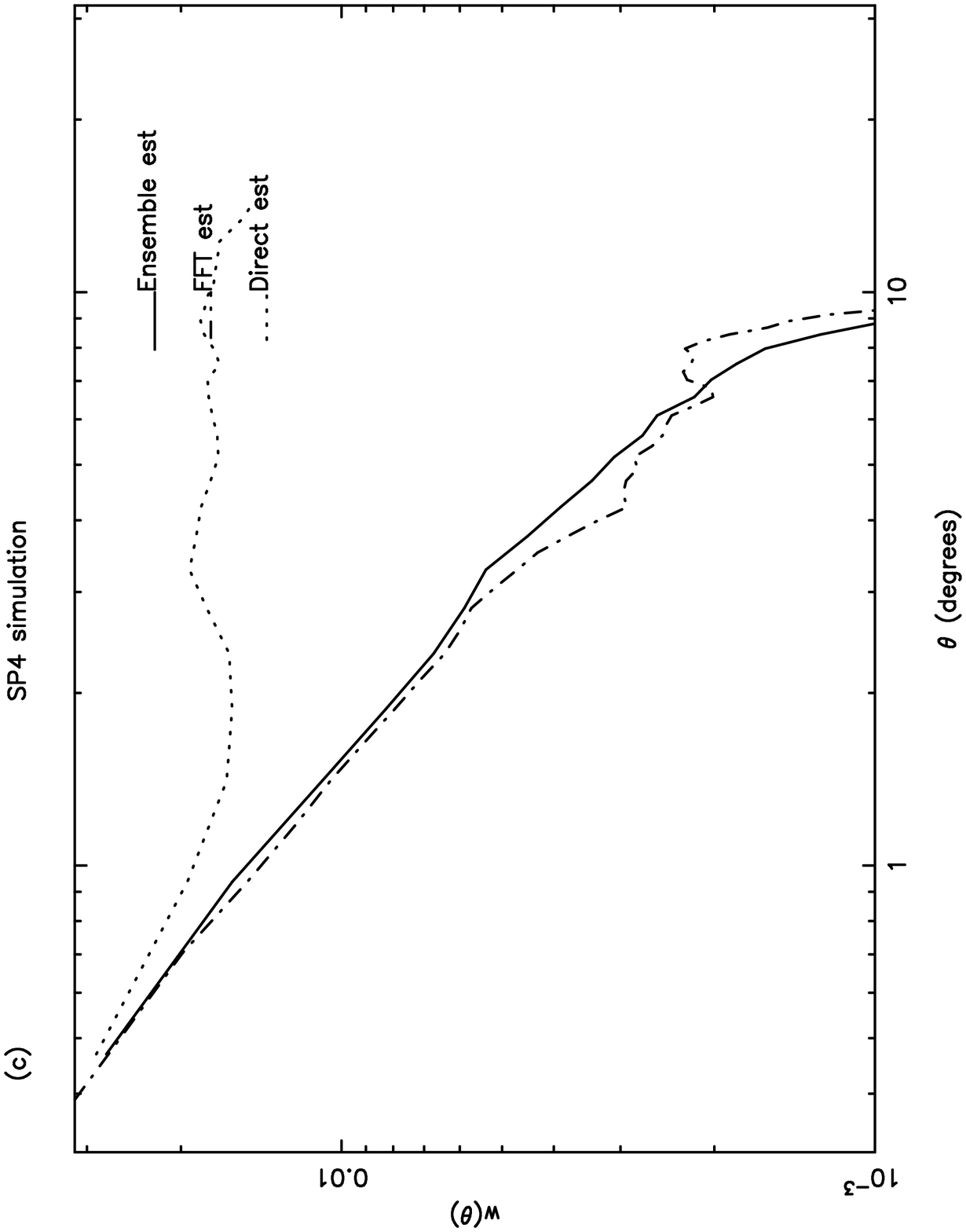}

\vskip 8.5 truein

\noindent
{\bf Figure A3}. Measurements of $\w$ using the ensemble estimator,
the FFT estimator and the direct estimator for three different types
of models (a) SP0; (b) SP2; and (c) SP4. 
Details of the models are given in the text.

\vfill\eject

\noindent
$\;$
\vskip 1 truein 

\includegraphics{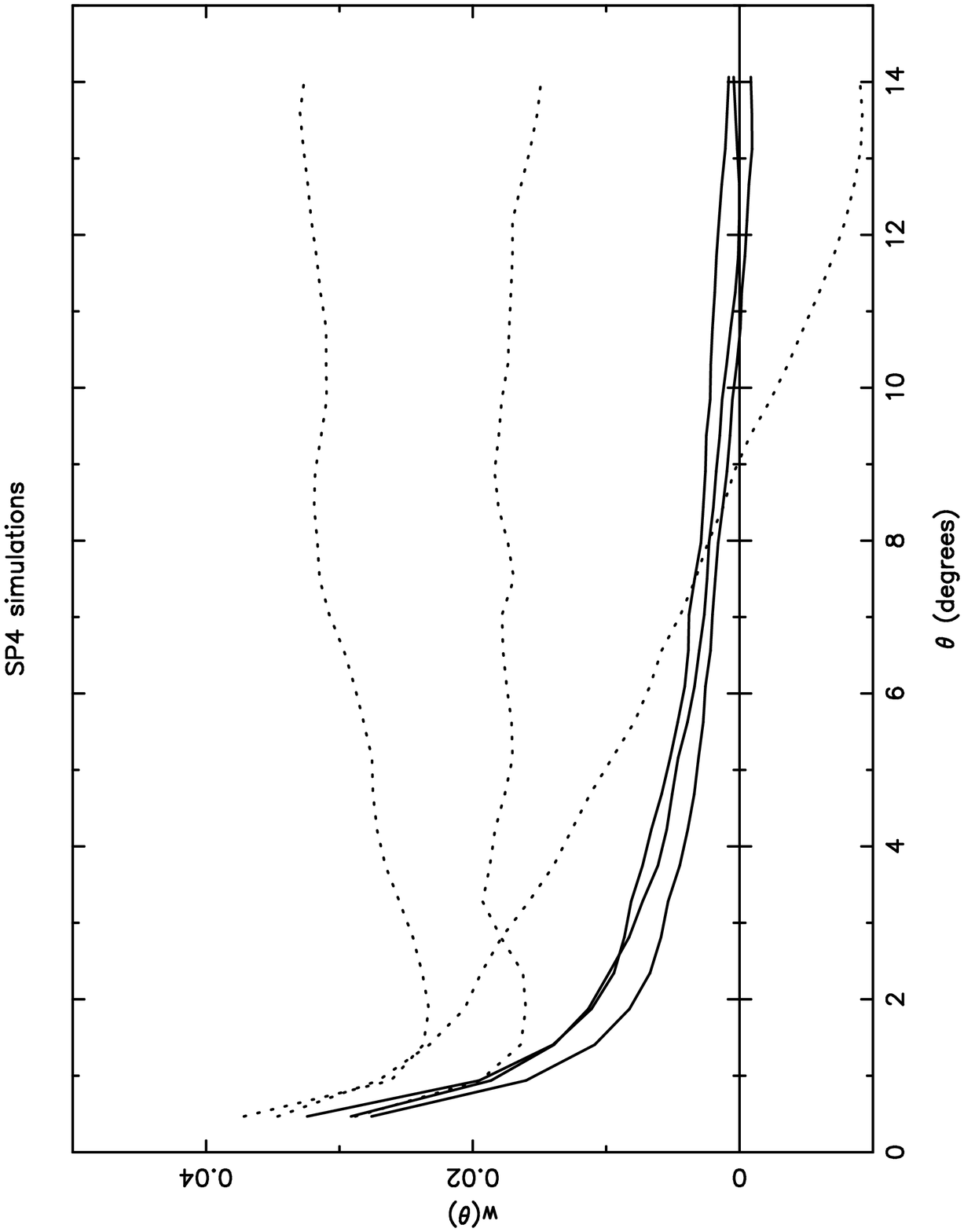}

\vskip 6 truein

\noindent
{\bf Figure A4}. Measurements of $\w$ from three different SP4
realizations using the direct estimator (dotted lines) and the
ensemble estimator (solid lines). 

\vfill\eject

\noindent
$\;$

\includegraphics{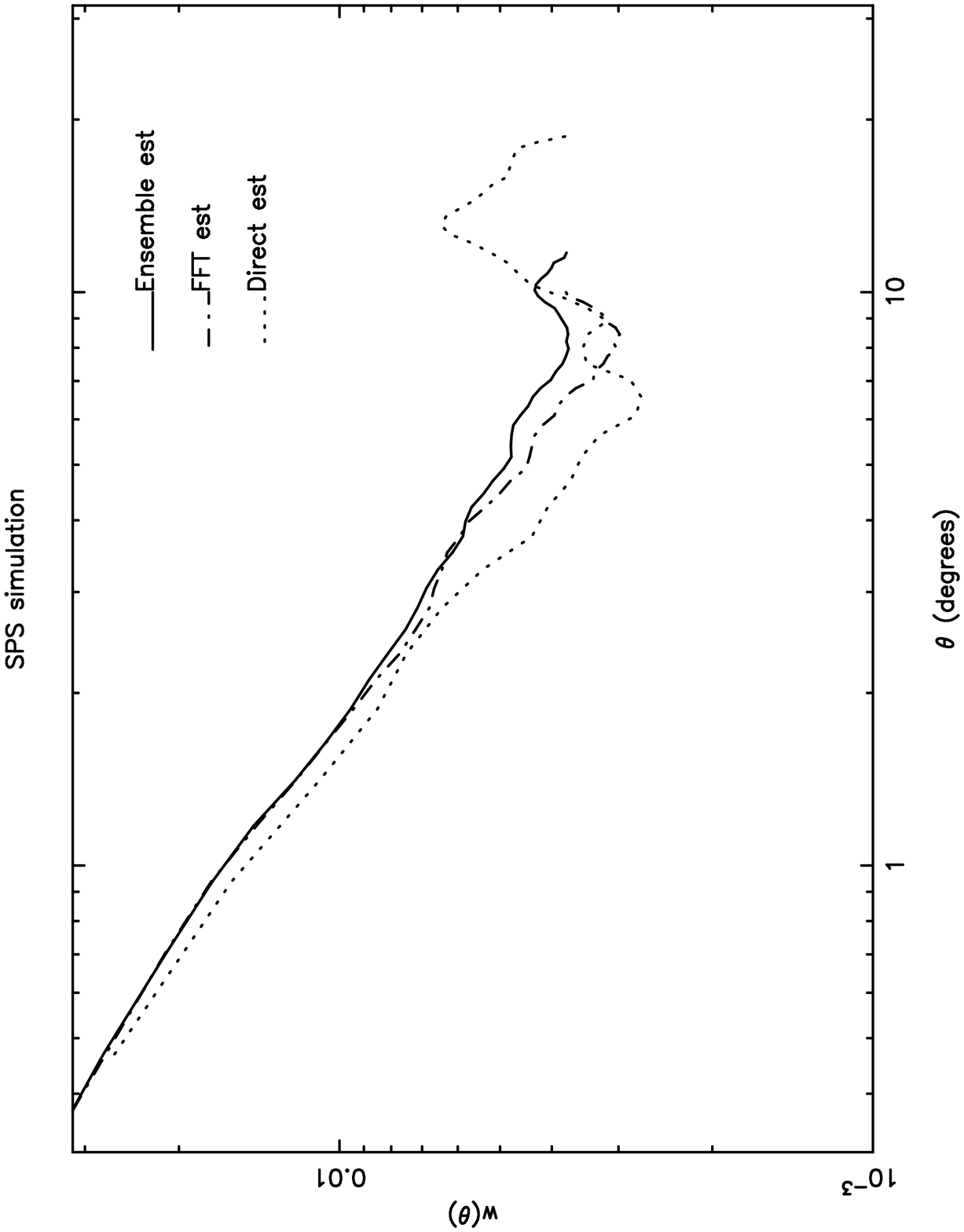}

\vskip 6 truein

\noindent
{\bf Figure A5}. The results of using different estimators for
simulations including a linear gradient.  The lines show the direct,
ensemble, and FFT estimates of $\w$ from an SPS simulation, as
described in the text.

\vfill\eject

\noindent
$\;$

\includegraphics{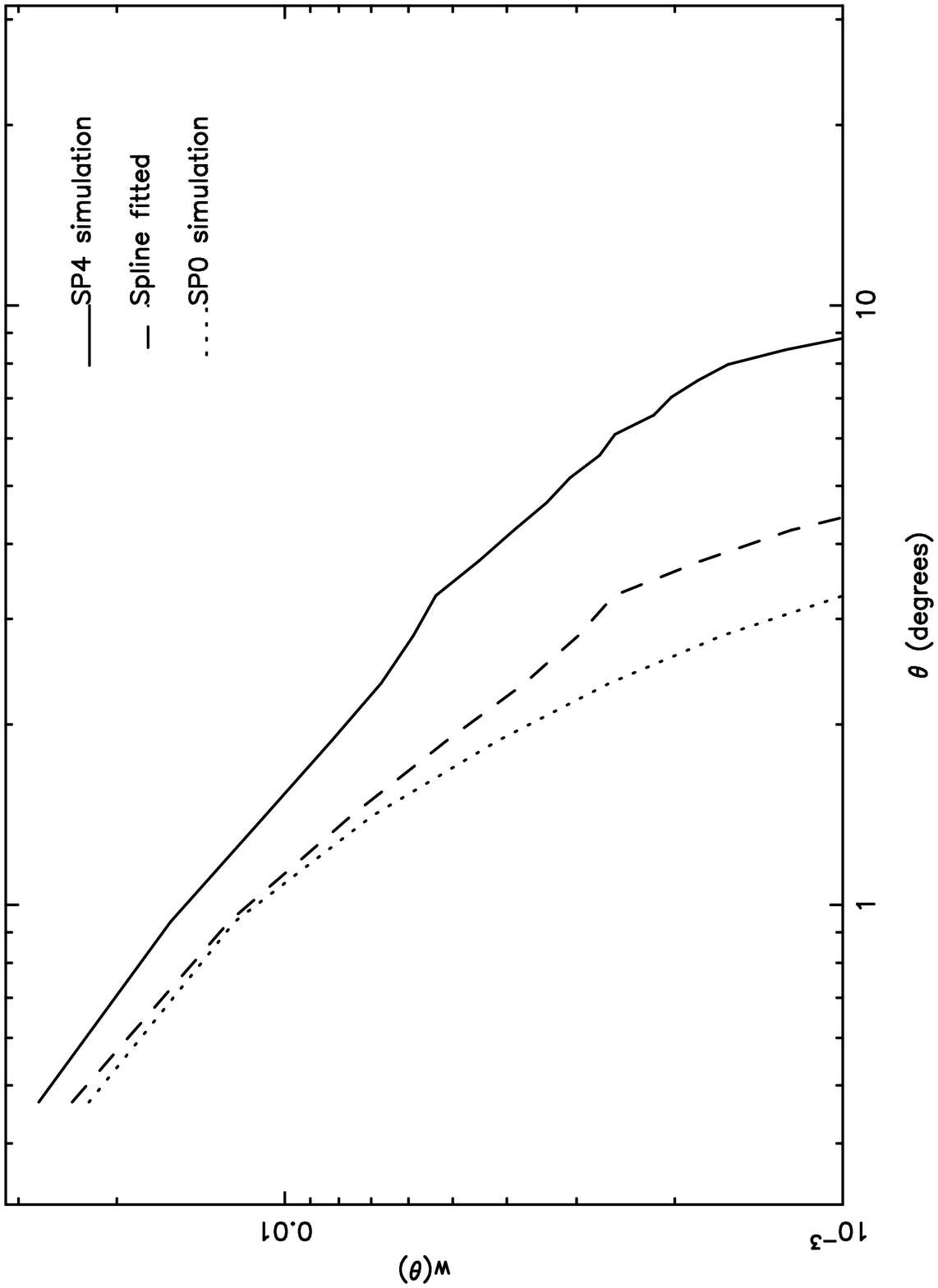}

\vskip 6 truein

\noindent
{\bf Figure A6}. The effects of removing large-scale structure using
spline fits. The solid line shows $\w$ from the SP4
simulation. The dashed line shows $\w$ from the same simulation after 
subtracting a bicubic spline fitted to the original
simulation. For comparison the dotted line shows $\w$ for an SP0 simulation. 

\bye